\documentclass[a4paper,12pt]{article}

\usepackage{epsfig}
\usepackage{amstex}
\usepackage{amssymb}

\def\ppb{p\bar{p}}
\def\as{\alpha_s}
\def\gl{\tilde{g}}
\def\sq{\tilde{q}}
\def\sqb{\bar{\tilde{q}}}
\def\qb{\bar{q}}
\def\sqL{\tilde{q}_{_L}}
\def\sqR{\tilde{q}_{_R}}
\def\ms{m_{\tilde q}}
\def\mg{m_{\tilde g}}
\def\Gs{\Gamma_{\tilde q}}
\def\Gg{\Gamma_{\tilde g}}
\def\md{m_{-}}
\def\eps{\varepsilon}

\def\DR{$\overline{DR}$\,\,}
\def\MS{$\overline{MS}$\,\,}

\def\MSm{\overline{MS}}
\def\ghat{\hat{g}_s}

\def\sihat{\hat{\sigma}}

\def\bs{\beta_{\sq}}

\def\bg{\beta_{\gl}}

\def\lsp{\tilde{\chi}_1^0}

\newcommand{\Lg}[2]{\log\left(\frac{#1}{#2}\right)}

\begin{document}

\thispagestyle{empty}

\hfill\vbox{\hbox{\bf DESY 96-150}
  \hbox{\bf CERN-TH/96-215}
  \hbox{October 1996}
  }
\vspace{1.in}
\begin{center}
\renewcommand{\thefootnote}{\fnsymbol{footnote}}
{\large\bf \sc Squark and Gluino Production \\[2mm]
  at Hadron Colliders}\\
\vspace{0.5in}
{\sc W.~Beenakker}$^{1}$\footnote{Research supported by a fellowship of the 
Royal Dutch Academy of Arts and Sciences.}, {\sc R.~H\"opker$^2$, M.~Spira$^3$
and P.~M.~Zerwas}$^2$ \\ 
\vspace{0.5in}

\noindent
$^1$ \textit{Instituut-Lorentz, P.O.~Box 9506, NL-2300 RA Leiden, 
  The Netherlands}\\ 
$^2$ \textit{Deutsches Elektronen-Synchrotron DESY, D-22603 Hamburg,
  Germany}\\ 
$^3$ \textit{TH Division, CERN, CH-1211 Geneva 23, Switzerland}\\

\vspace{2cm}

ABSTRACT \\
\end{center}
We have determined the theoretical predictions for the cross-sections
of squark and gluino production at $\ppb$ and $pp$ colliders (Tevatron
and LHC) in next-to-leading order of supersymmetric QCD. By reducing
the dependence on the renormalization/factorization scale
considerably, the theoretically predicted values for the
cross-sections are much more stable if these higher-order corrections
are implemented. If squarks and gluinos are discovered, this improved
stability translates into a reduced error on the masses, as extracted
experimentally from the size of the production cross-sections.  The
cross-sections increase significantly if the next-to-leading order
corrections are included at a renormalization/factorization scale near
the average mass of the produced massive particles. This rise results
in improved lower bounds on squark and gluino masses.  By contrast,
the shape of the transverse-momentum and rapidity distributions
remains nearly unchanged when the next-to-leading order corrections
are included.

\newpage

\section{Introduction}

The supersymmetric extension of the Standard Model \cite{susy1} is a
well-motivated step. In supersymmetric theories the hierarchy problem
of the Higgs sector can be solved \cite{hier}. Even in the context of
very high energy scales, as required by grand unification, it is
possible to retain fundamental scalar Higgs particles with low masses.
This is a consequence of pairing bosons with fermions in
supersymmetric multiplets, which removes the quadratic divergences due
to these high scales from the quantum fluctuations. Moreover, the
electroweak Higgs mechanism can be generated radiatively \cite{rsb}.
For a top mass in the experimentally observed range, the theory can
evolve from a symmetric phase at the grand-unification scale to a
phase of broken electroweak symmetry at low energies, while leaving
the electromagnetic and colour gauge symmetry unbroken. Strong
supporting evidence for supersymmetry is provided by the successful
theoretical prediction of the electroweak mixing angle \cite{swth},
$\sin^2\theta_w= 0.2334 \pm 0.0035$, based on the particle spectrum of
the Minimal Supersymmetric extension of the Standard Model (MSSM).
This prediction is matched quite well by the measured value
\cite{swex}, $\sin^2\theta_w^{\text{exp}}= 0.2317 \pm 0.0004$.
Supersymmetric extensions offer solutions for many other problems that
cannot be solved within the Standard Model [see
{\emph{e.g.}~Ref.~\cite{10re} for a comprehensive discussion].

In the MSSM \cite{susy2} quarks and leptons are paired with squarks
and sleptons, gauge and Higgs particles with gauginos and higgsinos.
Supersymmetric QCD (SUSY-QCD) is based on the coloured particles of
this spectrum: quarks and spin-0 squarks ($\sq=\sq_L,\sq_R$), gluons
and spin-1/2 gluinos ($\gl$).  The magnitude of the SUSY-QCD
interactions is set by the gauge and Yukawa couplings $g_s$ and
$\ghat=g_s$, respectively; the two couplings are required by
supersymmetry to be equal. If supersymmetry were an exact symmetry,
squarks and quarks would have equal masses and gluinos would be
massless. However, supersymmetry is a broken symmetry, and the masses
of the supersymmetric partners must exceed the masses of the
Standard-Model particles considerably.  Even though the mechanism for
the breaking of supersymmetry is not identified yet at the fundamental
level, requiring that no quadratic divergences be reintroduced into
the theory by breaking the symmetry, provides a powerful guiding
principle \cite{ssb}. In this approach, supersymmetry is broken within
SUSY-QCD by introducing heavy masses for squarks and gluinos, lifting
the mass degeneracy with quarks and gluons. In order not to ruin the
solution of the hierarchy problem, the masses of these particles
should not exceed limits of ${\cal O}(1~\text{TeV})$.

In the present analysis we will assume that the scalar partners
$\sq=(\sq_L,\sq_R)$ of the five light quark flavours are mass
degenerate. We defer the discussion of final-state stop particles,
with potentially significant L--R mixing due to the large top--Higgs
Yukawa coupling, to a subsequent paper \cite{stop}. Stop masses in
loop diagrams can be identified with the other squark masses since
their impact is small. The masses of the five light quarks are
neglected, while the top-quark mass is set to $m_t=175~\text{GeV}$.
[The set of Feynman rules in SUSY-QCD, including an extensive
discussion of the Majorana gluinos, is summarized in
Appendix~\ref{feynman_rules}; more details are given in
Ref.~\cite{phdrh}.]

\medskip %
The search for supersymmetric particles ranks among the most important
experimental endeavours of high-energy physics. The coloured
particles, squarks and gluinos, can be searched for most efficiently
at hadron colliders. As R parity is conserved in the QCD sector of
$N=1$ supersymmetric theories, these particles are always produced in
pairs.  Squarks and gluinos decay primarily into cascades of jets plus
the lightest supersymmetric particle (LSP), which escapes undetected.
This particle is in general identified with the lightest neutralino
$\lsp$. Missing momentum is therefore one of the classical
characteristics of SUSY events. Pairs of gluinos can decay into
like-sign dileptons plus two LSPs, providing an almost background-free
signal.

Squarks and gluinos can at present be searched for at the Fermilab
Tevatron, a $\ppb$ collider with a centre-of-mass energy of
$1.8~\text{TeV}$, which will be upgraded to $2~\text{TeV}$ in the
near future.  Lower bounds on the squark and gluino masses have been
set by both Tevatron experiments, CDF and D0 \cite{cdf,d0}. At the
95\% confidence level, the lower bound for the gluino mass is
$173~\text{GeV}$, independent of the value of the squark mass. If
squarks and gluinos have the same mass, the lower limit is given as
$225~\text{GeV}$. Within the theoretical set-up of the experimental
analysis, no limit for squark masses can be derived if the gluino mass
exceeds $550~\text{GeV}$.  [The lower limits on squark masses
obtained at LEP are independent of this assumption.]  This set of
experimental bounds on squark and gluino production has been obtained
in the framework of supergravity models, in which gluinos cannot be
much heavier than squarks. If the supergravity relations are relaxed
to the supersymmetric GUT relations between the gaugino masses, the
excluded range in the gluino/squark-mass plane can be extended
slightly.  For gluino masses above $550~\text{GeV}$, no bound on the
squark masses can be obtained any more, since the LSP $\lsp$ becomes
so heavy that the missing transverse momentum is insufficient to
generate an observable signal.  [Very light gluinos, which may have
escaped detection at hadron colliders, are improbable in the light of
the observed topologies of hadronic $Z$ decays; see
\emph{e.g.}~Ref.~\cite{ligl}].  In the near future the search for squarks
and gluinos can be extended at the upgraded Tevatron to masses between
$300~\text{GeV}$ and $400~\text{GeV}$. At the $pp$ collider LHC the
mass range up to $1.5$--$2~\text{TeV}$ can be sweeped, which covers
the canonical range of the coloured supersymmetric particles, yet may
not exhaust the parameter space entirely.

\medskip %
The cross-sections for the production of squarks and gluinos in hadron
collisions were calculated at the Born level already quite some time
ago \cite{born}. Only recently have the predictions been improved by
next-to-leading order (NLO) SUSY-QCD corrections for squark--antisquark
\cite{bhsz1} and gluino-pair production \cite{bhsz2} in $\ppb$
collisions\footnote{ The \emph{gluonic} radiative corrections to
  gluino-pair production in gluon fusion are closely related to the
  gluonic corrections for heavy-quark production \cite{ggtt}, just
  requiring the appropriate change of the Casimir invariants.}. In the
present paper we give a systematic analysis of the next-to-leading
order SUSY-QCD corrections of all possible supersymmetric pair
channels,
\begin{equation}
  \ppb/pp \longrightarrow \sq\sq, \sq\sqb, \sq\gl, \gl\gl +X
  \quad(\sq \neq \tilde{t})
\end{equation}
in the proton--antiproton collisions at the Tevatron and the
proton--proton collisions at the LHC.

Several arguments demand the NLO SUSY-QCD analysis in order to 
obtain adequate theoretical predictions for the cross-sections:

\noindent %
\emph{(i)} The lowest-order cross-sections depend strongly on the
\emph{a priori} unknown renormalization/factorization scale. As a
result, the theoretical predictions are uncertain within factors of 2.
By implementing the NLO corrections, this scale dependence is expected
to be reduced significantly.

\noindent %
\emph{(ii)} Drawing from the experience with similar hadronic
processes [\emph{e.g.}~hadroproduction of top quarks], the NLO corrections
are expected to be positive and large, thus enhancing the production
cross-sections and raising the presently available (conservative)
bounds on squark and gluino masses.

\noindent %
\emph{(iii)} When squarks and gluinos will be discovered, the
comparison of the measured total cross-sections with the theoretical
predictions can be used to determine the masses of the particles. Due
to the two escaping LSPs that are produced in the final-state decay
cascades, the masses of squarks and gluinos cannot be determined by
reconstructing the original squark and gluino states.
[Transverse-momentum spectra can eventually be exploited to evaluate
squark and gluino masses from the final-state distributions
\cite{paige}].

Because of the large number of mechanisms, the calculation of the NLO
corrections is tedious but straightforward. The only
non-straightforward component of the theoretical set-up is one element
of the renormalization program. To make maximal use of the common
infrastructure developed earlier for top-quark production \cite{ggtt},
we have chosen the \MS renormalization scheme. However, this scheme
leads in $n \neq 4$ dimensions to a mismatch between the 2 fermionic
gluino degrees of freedom and the $(n-2)$ transverse gluon degrees of
freedom. As a consequence, the gauge coupling $g_s(\MSm)$ and the
Yukawa coupling $\ghat(\MSm)$ of the \MS scheme differ in higher
orders by a finite amount, even in exact supersymmetric theories. The
problem, however, can be solved by introducing a proper counter term
that restores the supersymmetry also in higher orders \cite{ghatg}.

\medskip %
The paper is organized as follows. In the next section we recapitulate
the lowest-order cross-sections for the partonic subprocesses of
squark and gluino production for the sake of completeness and to
define the notation. In Section 3 we carry out the calculation of the
virtual SUSY-QCD corrections, followed by real-gluon radiation and the
discussion of final states including light quarks. In Section 4 we
present the overall corrections at the parton level and at the
hadronic level for the total $\ppb$ and $pp$ cross-sections.
Furthermore, the transverse-momentum and rapidity distributions for
semi-inclusive squark/gluino final states will be discussed briefly.
We conclude the paper with an assessment of the results. Some useful
technical details are collected in several Appendices.

\newpage
\section{Squark and Gluino Production in Leading Order}

To set the stage for the subsequent discussion of higher-order
effects, it is useful to recapitulate the lowest-order processes of
squark and gluino production in quark and gluon collisions
\cite{born}. Moreover, the main features of the production mechanisms
will be briefly described.

\medskip %
The hadroproduction of squarks and gluinos in leading order (LO) of the
perturbative expansion proceeds through the following partonic
reactions:
\begin{alignat}{4}
\sq\sqb \,\text{ production:}\quad  &q_i &+&\qb_j &\,\longrightarrow\,& 
  \sq_k &+&\sqb_l \label{qbtosb}\\[0.2cm]
  &g   &+&g     &\,\longrightarrow\,& 
  \sq_i &+&\sqb_i \label{ggtosb}\\[0.2cm]
\sq\sq \,\text{ production:}\quad  &q_i &+&q_j   &\,\longrightarrow\,& 
  \sq_i &+&\sq_j \,\text{ and } c.c.\label{qqtoss}\\[0.2cm]
\gl\gl \,\text{ production:}\quad  &q_i &+&\qb_i &\,\longrightarrow\,& 
  \gl   &+&\gl \label{qbtogg}\\[0.2cm]
  &g   &+&g     &\,\longrightarrow\,& 
  \gl   &+&\gl \label{ggtogg}\\[0.2cm]
\sq\gl \,\text{ production:}\quad  &q_i &+&g     &\,\longrightarrow\,& 
  \sq_i &+&\gl \,\text{ and } c.c.\label{qgtosg}
\end{alignat}
The momenta of the two partons in the initial states are denoted by
$k_1$ and $k_2$, those of the particles in the final states by $p_1$ and
$p_2$.

The chiralities of the squarks $\sq=(\sq_L,\sq_R)$ are not noted
explicitly. The indices $i$--$l$ indicate the flavours of the quarks
and squarks. Also charge-conjugated processes ($c.c.$) are possible,
related to the reactions (\ref{qqtoss}) and (\ref{qgtosg}); for the
sake of simplicity they are not given explicitly.  They will be taken
into account properly when the hadronic cross-sections are calculated.
The Feynman diagrams corresponding to these partonic reactions are
displayed in Fig.~\ref{fig:feynborn}.
\begin{figure}[p]
  \begin{center}
  \vspace*{-2.3cm}
  \hspace*{-0cm}
  \epsfig{file=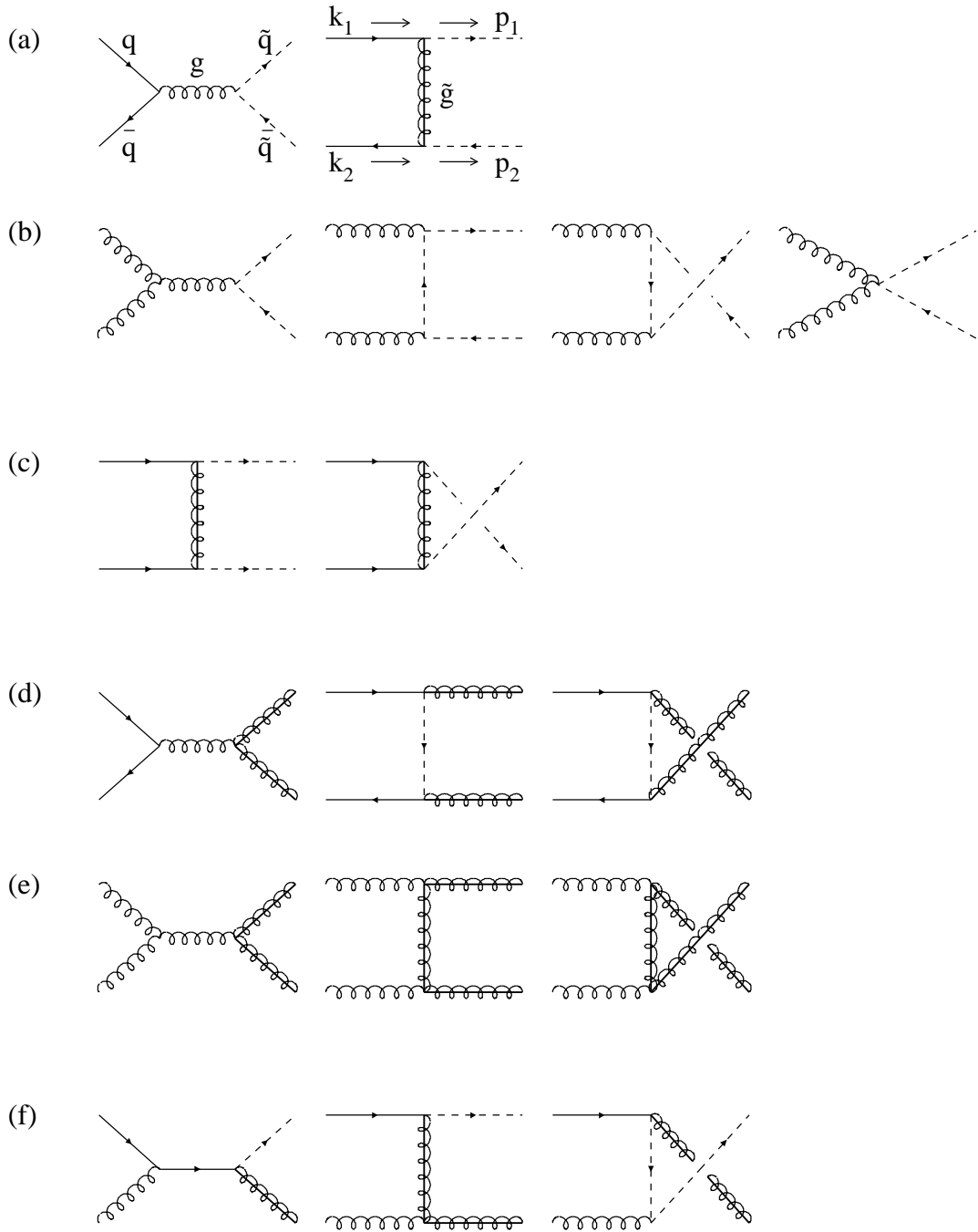,width=17cm}
  \vspace*{-2.5cm}
  \end{center}
  \caption[]{Feynman diagrams for the production of squarks and
    gluinos in lowest order. The diagrams without and with crossed
    final-state lines [\emph{e.g.}~in (b)] represent $t$- and $u$-channel
    diagrams, respectively. The diagrams in (c) and the last diagram
    in (d) are a result of the Majorana nature of gluinos.  Note that
    some of the above diagrams contribute only for specific flavours
    and chiralities of the squarks.}
  \label{fig:feynborn}
\end{figure}
The production of squark--antisquark final states requires
quark--antiquark (a) or gluon--gluon (b) initial states. Squark pairs
can only be produced from quark-pair (c) initial states. Gluino pairs
are produced from quark--antiquark (d) and gluon--gluon (e) initial
states. The squark--gluino final states can only be produced in
quark--gluon (f) collisions.

\medskip %
We use the Feynman gauge for the internal gluon propagators. For the
external gluons only the transverse polarization states are
generated. In the axial gauge the polarization sum for the external
gluons is given by
\begin{equation}
  P_i^{\mu\nu} = \sum_T \epsilon^{\mu\ast}_T(k_i) \,\epsilon^\nu_T(k_i)
  = -g^{\mu\nu} + \frac{n_i^\mu k_i^\nu + k_i^\mu n_i^\nu}{(n_i\, k_i)}
  -\frac{n_i^2 \,k_i^\mu k_i^\nu}{(n_i\, k_i)^2}.
\end{equation}
Here $n_i\neq k_i$ is an arbitrary vector. This polarization sum obeys
the transversality relations
\begin{displaymath}
  k_{i\mu} \,P_i^{\mu\nu} = P_i^{\mu\nu} k_{i\nu} = 
  n_{i\mu} \,P_i^{\mu\nu} = P_i^{\mu\nu} n_{i\nu} = 0.
\end{displaymath}
Combining these transversality relations with the Slavnov--Taylor
identities for on-shell external particles, \emph{e.g.}
\begin{equation}
  \label{slavnov}
  k_i^\mu M_{\mu\rho} = M^{ghost}_\rho \propto k_{j\rho}
\end{equation}
for two external gluons [labelled $i$ and $j$], allows us to perform
ghost subtraction. Because of the transversality relations all terms
proportional to $k_i^\mu$ and $k_j^\rho$ can be removed from the LO
matrix elements, resulting in the nullification of the right-hand side
of Eq.~(\ref{slavnov}) [ghost subtraction]. As a consequence, the
polarization sum can effectively be replaced by $P_i^{\mu\nu} =
-g^{\mu\nu}$.

\medskip %
The $n_f=5$ light quark flavours and the gluons are treated as massless
particles. Since we have excluded the top-squarks from the final
state, all squark-flavour and chirality states are considered to be
mass degenerate with mass $\ms$. The gluino mass is denoted by $\mg$.
The set of kinematical invariants used for the description of the
reactions (\ref{qbtosb})--(\ref{qgtosg}) is given by
\begin{alignat}{3}
  \\[-0.7cm]
  s & = (k_1 +k_2)^2 \\
  t & = (k_2 -p_2)^2 &\qquad t_1 & = (k_2 -p_2)^2 -\ms^2
  &\qquad  t_{g} &= (k_2 -p_2)^2 -\mg^2 \notag\\
  u & = (k_1 -p_2)^2 &\qquad u_1 & = (k_1 -p_2)^2 -\ms^2 
  &\qquad  u_{g} &= (k_1 -p_2)^2 -\mg^2. \notag
\end{alignat}
The Mandelstam invariants are related by $s+t+u = p_1^2+p_2^2$. All
in- and outgoing particles are assumed to be on their respective mass
shells, \emph{i.e.}~$k_i^2=0$, $p_i^2 = \ms^2$ for squarks, and $p_i^2
= \mg^2$ for gluinos.

\medskip %
Applying the Feynman rules given in Appendix~\ref{feynman_rules}, we
obtain the following squared lowest-order matrix elements $|{\cal
  M}^B|^2$ in $n=4-2\eps$ dimensions\footnote{The generalization to
  $n\neq 4$ dimensions is anticipated here in view of the
  renormalization and mass factorization that will have to be performed 
  in higher orders later on.}:
\begin{small}
\begin{eqnarray*}
  \sum |{\cal M}^B|^2 (q_i\qb_j\to\sq\sqb) & = & 
  \delta_{ij}\left[8 n_f g_s^4 \,N C_F \,\frac{t_1 u_1 -\ms^2 s}{s^2} 
  + 4 \ghat^4 \,N C_F \,\frac{t_1 u_1 -(\ms^2 -\mg^2) s}{t_g^2}
\right. \label{born1} \\ 
  & &\hphantom{\delta_{ij}a} \left. {} \hspace*{-1cm}
  - 8 g_s^2 \ghat^2 \,C_F \,\frac{t_1 u_1 -\ms^2 s}{s t_g}
\right] 
  + (1-\delta_{ij}) 
  \left[ 4 \ghat^4 \,N C_F \,\frac{t_1 u_1 -(\ms^2 -\mg^2)s}{t_g^2}\right]
    \nonumber\\[0.2cm] 
    \sum |{\cal M}^B|^2 (gg\to\sq\sqb) & = & \!\!4 n_f g_s^4
    \left[C_O\!\left(1 -2\,\frac{t_1 u_1}{s^2}\right) -\!C_K\right]\! \left[ 1
      -\!\eps -\!2\,\frac{s\ms^2}{t_1 u_1}\left(1 -\!\frac{s\ms^2}{t_1
        u_1}\right)\right] \quad\qquad\\[0.2cm] 
  \label{squarkborn}
  \sum |{\cal M}^B|^2 (q_i q_j\to\sq\sq) & = & 
  \delta_{ij}\Bigg[ 2\ghat^4 \,N C_F \left(t_1 u_1 -\ms^2 s\right)
  \left(\frac{1}{t_g^2} +\frac{1}{u_g^2}\right)  \\ 
  & &{} \hphantom{\delta_{ij}a} + 4\ghat^4 \,\mg^2  s \left( N C_F 
  \left(\frac{1}{t_g^2} +\frac{1}{u_g^2}\right)
  - 2 C_F \,\frac{1}{t_g u_g}\right) \Bigg]\nonumber\\
  & &{} + (1-\delta_{ij}) 
  \left[ 4 \ghat^4 \,N C_F \,\frac{t_1 u_1 -(\ms^2-\mg^2) s}{t_g^2}\right]
    \nonumber 
\end{eqnarray*}
\begin{eqnarray*}
  \sum |{\cal M}^B|^2 (q\qb\to\gl\gl) & = & 
  4 g_s^4 \,C_O \left[\frac{2 \mg^2 s +t_g^2 +u_g^2}{s^2} -\eps \right]\\
  & &{} + 4 g_s^2 \ghat^2 \,C_O \left[\frac{\mg^2 s +t_g^2}{s t_1}
  +\frac{\mg^2 s +u_g^2}{s u_1} +\eps\left(\frac{t_g}{t_1}
  +\frac{u_g}{u_1}\right) \right]\nonumber\\ 
  & &{} + 2\ghat^4 \left[ C_O \left(\frac{t_g^2}{t_1^2} +
  \frac{u_g^2}{u_1^2}\right) +C_K \left(2\,\frac{\mg^2 s}{t_1\,u_1} -
  \frac{t_g^2}{t_1^2} - \frac{u_g^2}{u_1^2} \right)\right]\nonumber\\[0.2cm] 
  \sum |{\cal M}^B|^2 (gg\to\gl\gl) & = & 8g_s^4 \,N C_O \left(1
  -\frac{t_g u_g}{s^2}\right) \left[\frac{s^2}{t_g\,u_g}
  \left(1-\eps\right)^2 -2 \left(1-\eps\right)
  +4\,\frac{\mg^2 s}{t_g u_g} \left (1-\frac{\mg^2 s}{t_g u_g}\right)
  \right]\\[0.2cm]    
  \sum |{\cal M}^B|^2 (qg\to\sq\gl) & = & 
  2  g_s^2 \ghat^2 \left[ C_O \left( 1 -2\,\frac{su_1}{t_g^2}
  \right) - C_K \right] 
  \Bigg[(-1+\eps) \,\frac{t_g}{s} \label{born6} \\
  & &{} + \frac{2(\mg^2-\ms^2)\,t_g}{su_1} \left(1 +\frac{\ms^2}{u_1}
  +\frac{\mg^2}{t_g} \right) \Bigg]. \nonumber
\end{eqnarray*}
\end{small}

\noindent The QCD gauge coupling ($qqg$) is denoted by $g_s$ and the Yukawa
coupling ($q\sq\gl$) by $\ghat$. These couplings are identical. For
squarks all chiralities and non-stop flavours are summed
over\footnote{The first term of $\sum |{\cal M}^B|^2 (q_i
  q_j\to\sq\sq)$ corresponds to the production of squarks with
  different chiralities, whereas the second term corresponds to equal
  chiralities. When calculating cross-sections, the second term will
  be weighted by a statistical factor of $1/2$, since the squarks in
  the final state are identical.}.  As mentioned before, the
charge-conjugate final states that are not denoted explicitly in the
above listing, will be included in the hadronic cross-sections. The
$SU(3)$ colour factors are given by $N=3$, $C_O=N(N^2-1)=24$,
$C_K=(N^2-1)/N=8/3$, and $C_F=(N^2-1)/(2N)=4/3$.

After performing the $n$-dimensional phase-space integration and
taking into account colour and spin averaging, we find for the
lowest-order double-differential distributions:
\begin{eqnarray}
  s^2 \,\frac{d^2\sigma^B}{dt\,du} &=& K_{ij} \,\frac{\pi
    S_\eps}{\Gamma(1-\eps)} \left[\frac{(t-p_2^2)(u-p_2^2)-p_2^2
    s}{\mu^2 s}\right]^{-\eps} \Theta(\,[t-p_2^2][u-p_2^2]-p_2^2 s\,)
    \nonumber \\[1mm]
  & & \times \,\Theta(s-4m^2)\,\delta(s+t+u-p_1^2 -p_2^2)\sum |{\cal M}^B|^2.
\end{eqnarray}
Here $m$ denotes the average mass of the
produced particles, \emph{i.e.}~$m = (\sqrt{p_1^2}+\sqrt{p_2^2}\,)/2$.
The averaging of the initial-state colours and spins is incorporated in
the factor $K_{ij}$:
\begin{displaymath}
  K_{qq} = K_{q\qb} = \frac{1}{4 N^2}, \qquad 
  K_{gg} = \frac{1}{4(1-\eps)^2 (N^2-1)^2}, \qquad 
  K_{qg} = \frac{1}{4(1-\eps) N (N^2-1)}. 
\end{displaymath}
The gluons have $(n-2)$ degrees of freedom in $n$ dimensions. The
scale parameter $\mu$ accounts for the correct dimension of the
coupling in $n$ dimensions.  The term $S_\eps=(4\pi)^{-2+\eps}$
follows from the angular integrations.

\newpage
The subsequent integration over the remaining invariants\footnote{The
  explicit boundaries of these integrations are given in
  Appendix~\ref{phasespace}.}
yields the total lowest-order partonic cross-sections \cite{born}:
\begin{small}
\begin{eqnarray*}
  \sigma^B(q_i\qb_j\to\sq\sqb) & = &
  \delta_{ij} \,\frac{n_f\pi\as^2}{s}\,\bs\left[\frac{4}{27}
  -\frac{16\ms^2}{27s} \right] \\
  &&{} +\delta_{ij}\,\frac{\pi\as\hat{\alpha}_s}{s}\left[\bs\left(\frac{4}{27}
  +\frac{8\md^2}{27s}\right) +\left(\frac{8\mg^2}{27s}
  +\frac{8\md^4}{27s^2} \right) L_1\right] \nonumber \\ 
  &&{}  +\frac{\pi\hat{\alpha}_s^2}{s}\left[\bs\left(
  -\frac{4}{9}-\frac{4\md^4}{9(\mg^2s+\md^4)}\right)
  +\left(-\frac{4}{9} -\frac{8\md^2}{9s}\right) L_1 \right]\nonumber \\[0.2cm]
  \sigma^B(gg\to\sq\sqb) & = &
  \frac{n_f\pi\as^2}{s}\left[
  \bs \left(\frac{5}{24} +\frac{31\ms^2}{12s} \right)
  +\left(\frac{4\ms^2}{3s}+ \frac{\ms^4}{3s^2}\right)
  \log\left(\frac{1-\bs}{1+\bs}\right) \right]\\[0.2cm]
  \sigma^B(q_i q_j\to\sq\sq) & = &
  \frac{\pi\hat{\alpha}_s^2}{s}\left[\bs\left(
  -\frac{4}{9}-\frac{4\md^4}{9(\mg^2s+\md^4)}\right)
  +\left(-\frac{4}{9} -\frac{8\md^2}{9s}\right) L_1 \right] \\ 
  & &{} + \delta_{ij}\,\frac{\pi\hat{\alpha}_s^2}{s}
  \left[ \frac{8\mg^2}{27(s+2\md^2)} L_1 \right] \nonumber\\[0.2cm]
  \sigma^B(q\qb\to\gl\gl) & = &
  \frac{\pi\as^2}{s}\, \bg\left(\frac{8}{9} +\frac{16\mg^2}{9s}\right)
  \\
  & &{} +\frac{\pi\as\hat{\alpha}_s}{s}\left[
  \bg\left(-\frac{4}{3}-\frac{8\md^2}{3s} \right)
  +\left(\frac{8\mg^2}{3s} +\frac{8\md^4}{3s^2}\right)L_2\right] \nonumber\\
  & &{} +\frac{\pi\hat{\alpha}_s^2}{s}\left[
  \bg\left(\frac{32}{27}+\frac{32\md^4}{27(\ms^2s+\md^4)}\right)
  +\left(-\frac{64\md^2}{27s} -\frac{8\mg^2}{27(s-2\md^2)}\right)
  L_2\right] \nonumber\\[0.2cm]
  \sigma^B(gg\to\gl\gl) & = &
  \frac{\pi\as^2}{s}\left[
  \bg\left(-3-\frac{51\mg^2}{4s}\right)
  + \left(-\frac{9}{4}
  -\frac{9\mg^2}{s} +\frac{9\mg^4}{s^2} \right)
  \log\left(\frac{1-\bg}{1+\bg} \right) \right] \\[0.2cm]
  \sigma^B(qg\to\sq\gl) & = &
  \frac{\pi\as\hat{\alpha}_s}{s}\,\left[
  \frac{\kappa}{s}\left(-\frac{7}{9} -\frac{32\md^2}{9s}\right)
  + \left(-\frac{8\md^2}{9s}+\frac{2\ms^2\md^2}{s^2}
  +\frac{8\md^4}{9s^2} \right) L_3 \right. \\
  & &\hphantom{\frac{\pi\as\hat{\alpha}_s}{s}a} \left. {}
  + \left(-1-\frac{2\md^2}{s}+\frac{2\ms^2\md^2}{s^2}
  \right) L_4 \right], \nonumber 
\end{eqnarray*}
with
\begin{alignat*}{2}
  L_1 & = \Lg{s+2\md^2-s\bs}{s+2\md^2+s\bs} &\qquad 
  L_2 & = \Lg{s-2\md^2-s\bg}{s-2\md^2+s\bg} \\
  L_3 & = \Lg{s-\md^2-\kappa}{s-\md^2+\kappa} &\qquad
  L_4 & = \Lg{s+\md^2-\kappa}{s+\md^2+\kappa} \\
  \bs & = \sqrt{1-\frac{4\ms^2}{s}} &\qquad 
  \bg & = \sqrt{1-\frac{4\mg^2}{s}} \\
  \md^2 & = \mg^2 -\ms^2 &\qquad
  \kappa & = \sqrt{(s-\mg^2-\ms^2)^2-4\mg^2\ms^2} \\
  \as & = g_s^2/4\pi &\qquad
  \hat{\alpha}_s & = \ghat^2/4\pi. 
\end{alignat*}
\end{small}

\noindent
Note that we have suppressed the theta-functions $\Theta(s-4m^2)$ for the 
production thresholds. For identical particles in the final state
[gluino-pair production or production of squarks with identical
chirality and flavour] a statistical factor $1/2$ has been taken into
account. 

\medskip%
For the production of squark pairs or squark--gluino pairs only one
initial state contributes at lowest order. For squark--antisquark and
gluino pairs both gluon--gluon and quark--antiquark initial states are
possible.

\medskip%
The total hadronic cross-sections are obtained by integrating the
parton cross-sections in the usual way over the parton distributions
$f_i$ in the proton/antiproton:
\begin{equation}
  \sigma(ij\to\sq,\gl) = \int \,dx_1 \,dx_2 \, f_i(x_1)\,f_j(x_2)\,
  \sigma^B(ij\to\sq,\gl; s=x_1 x_2 S),
\end{equation}
where the total centre-of-mass energy of the collider is denoted by $\sqrt{S}$.

To assess the relative weights of $\sq\sqb$, $\sq\sq$, $\gl\gl$ and
$\sq\gl$ final states in $\ppb$ collisions at the Tevatron and in $pp$
collisions at the LHC, the relative yields are shown for a typical set
of mass parameters in Fig.~\ref{fig:bornratio1} and
Fig.~\ref{fig:bornratio2}.  The relative yields of squarks and
gluinos in the final states depend strongly on the mass ratio
$\ms/\mg$, for which we have chosen two representative values, $0.8$ and
$1.6$. If squarks are lighter than gluinos, the valence partons give
the dominant yield of squark--antisquark pairs/squark pairs at the
Tevatron/LHC. By contrast, if the gluinos are the lighter of the two
species, their production is the most copious.

\begin{figure}[h]
  \begin{center}
  \vspace*{-1.5cm}
  \hspace*{-1cm}
  \epsfig{file=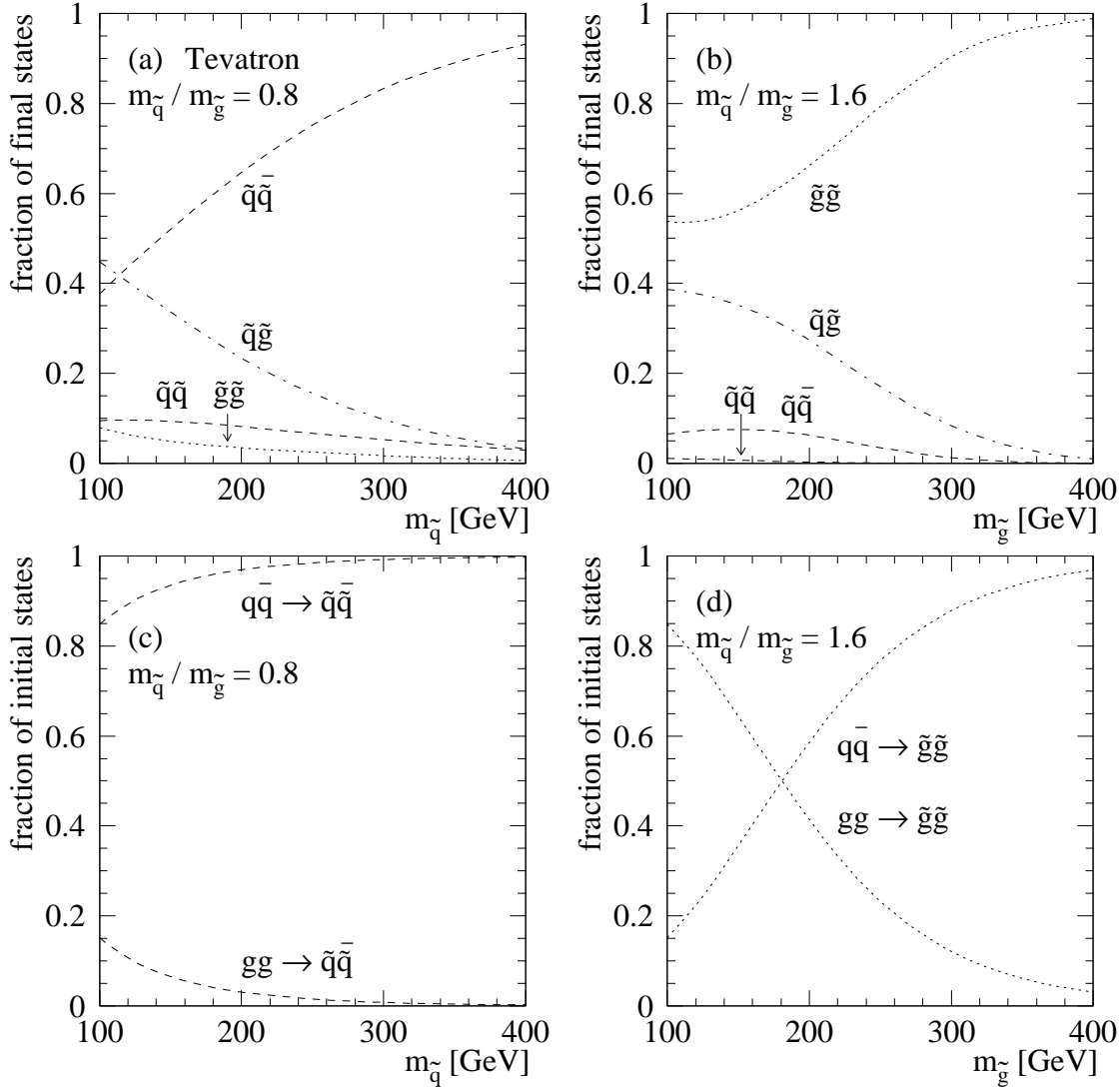,width=17cm}
  \vspace*{-2.2cm}
  \end{center}
  \caption[]{The relative yields of squarks and gluinos in the final
    states at the Tevatron. The mass ratio $\ms/\mg$ is chosen to be
    (a) $0.8$ and (b) $1.6$. Also shown are the leading parton
    contributions for (c) $\sq\sqb$ and (d) $\gl\gl$ final states.
    Parton densities: GRV 94 \cite{GRV}; renormalization and
    factorization scale $Q=\ms$ for squarks, $Q=\mg$ for gluinos, and
    $Q=(\ms+\mg)/2$ for squark--gluino pairs.}
  \label{fig:bornratio1}
\end{figure}

\begin{figure}[h]
  \begin{center}
  \vspace*{-1.5cm}
  \hspace*{-1cm}
  \epsfig{file=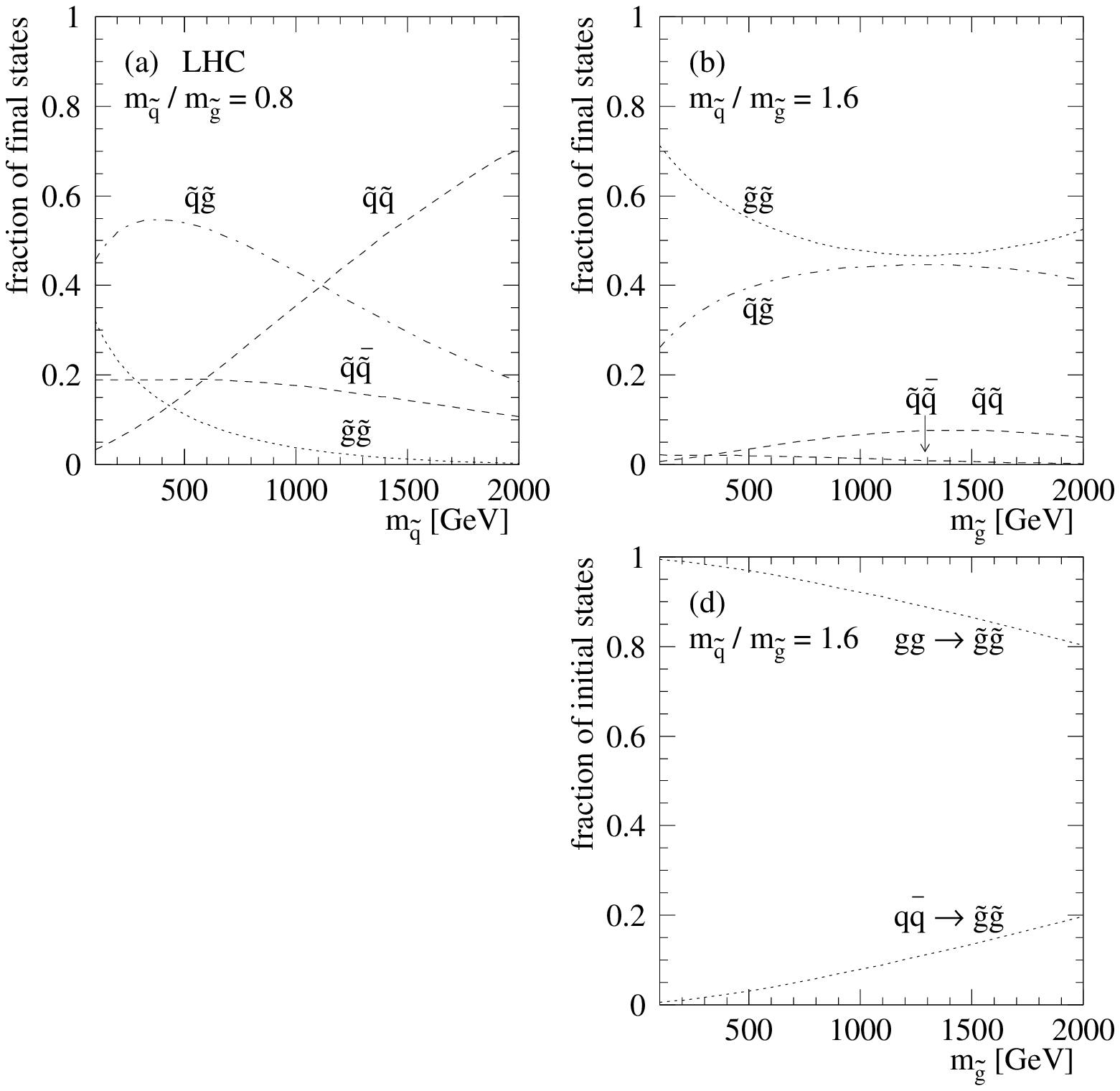,width=17cm}
  \vspace*{-2.2cm}
  \end{center}
  \caption[]{The relative yields of squarks and gluinos in the final
    states at the LHC. The mass ratio $\ms/\mg$ is chosen to be (a)
    $0.8$ and (b) $1.6$. Also shown are the leading parton
    contributions for (d) $\gl\gl$ final states. Parton densities: GRV
    94 \cite{GRV}; renormalization and factorization scale $Q=\ms$ for
    squarks, $Q=\mg$ for gluinos, and $Q=(\ms+\mg)/2$ for
    squark--gluino pairs. [Note that $\sq\sq/\sq\gl$ final states can
    only be generated by $qq/qg$ initial states so that diagram (c) is
    trivial and not shown.]}
  \label{fig:bornratio2}
\end{figure}

\section{SUSY-QCD Corrections}
\subsection{Virtual Corrections}

In this subsection we will present the virtual QCD corrections to the
partonic production cross-sections of squarks and gluinos. The
technical set-up of the calculation will be defined, including the
renormalization of the ultraviolet (UV) divergences. The calculations
are carried out in the \MS renormalization scheme, which requires a
careful analysis of the Yukawa ($q\sq\gl$) coupling $\ghat$ in higher
orders.

\subsubsection{Technical Set-Up}
 
The calculation of the ${\cal O}(\as)$ corrections to the reactions
(\ref{qbtosb})--(\ref{qgtosg}) involves the evaluation of the virtual
corrections, \emph{i.e.}~the interference of the Born matrix element
[after ghost subtraction] with the one-loop amplitudes. The
corresponding differential cross-section is given by
\begin{eqnarray}
  s^2 \,\frac{d^2\sigma^V}{dt\,du} &=& K_{ij} \,\frac{\pi
    S_\eps}{\Gamma(1-\eps)} \left[\frac{(t-p_2^2)(u-p_2^2)-p_2^2
    s}{\mu^2 s}\right]^{-\eps} \Theta(\,[t-p_2^2][u-p_2^2]-p_2^2 s\,)
    \nonumber \\[1mm]
  & & \times \,\Theta(s-4m^2)\,\delta(s+t+u-p_1^2 -p_2^2)\sum \left({\cal
    M}^B {\cal M}^{V\ast} + {\cal M}^V {\cal M}^{B\ast} \right).
\end{eqnarray}
The virtual (one-loop) amplitudes include self-energy corrections,
vertex corrections, and box diagrams. For the virtual particles inside
loops we use the complete supersymmetric QCD spectrum: gluons,
gluinos, all quarks, and all squarks. In Fig.~\ref{fig:feynvirt} we
present a set of typical virtual corrections.
\begin{figure}[p]
  \begin{center}
  \vspace*{-2.5cm}
  \hspace*{-0cm}
  \epsfig{file=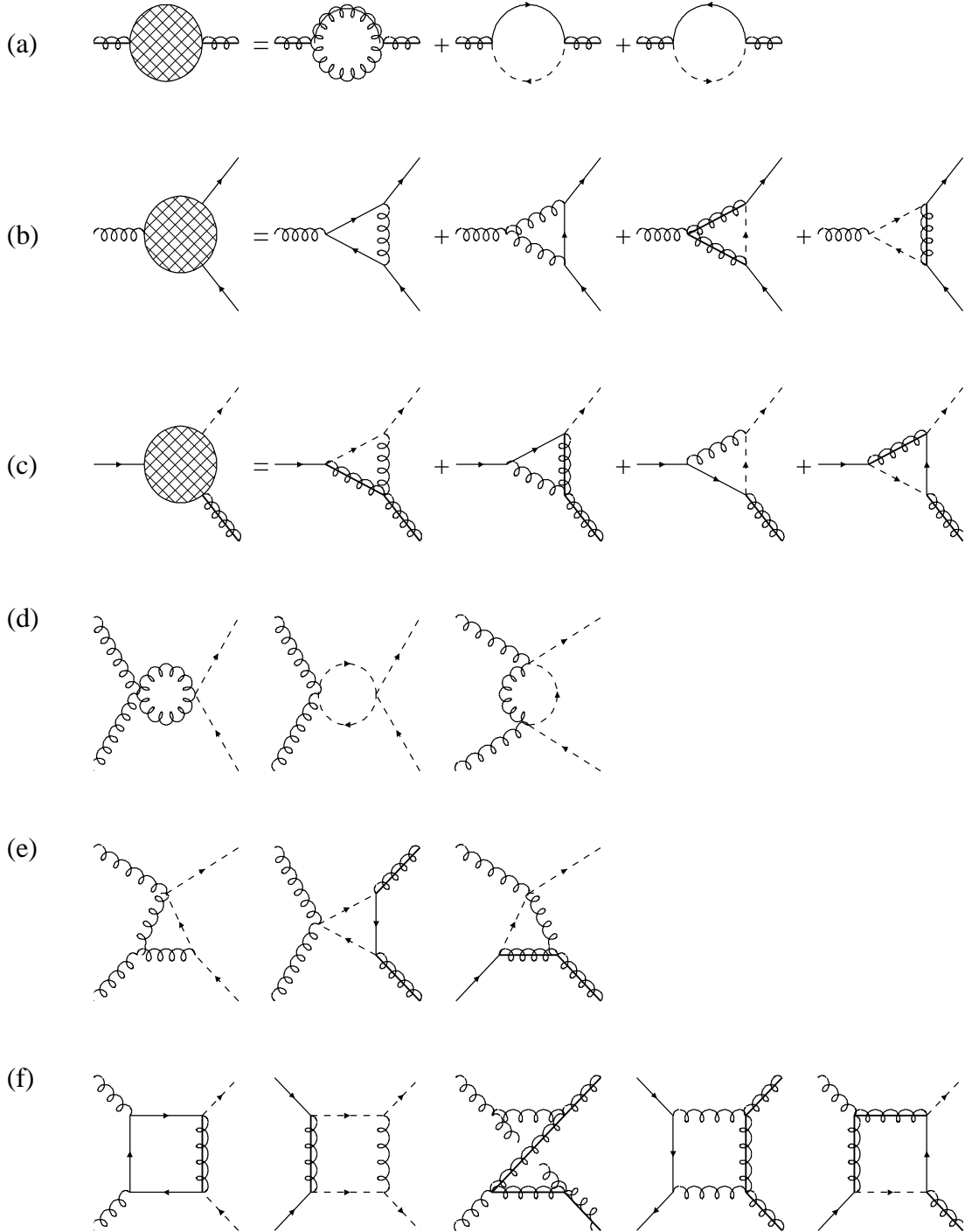,width=17cm}
  \vspace*{-2.5cm}
  \end{center}
  \caption{A selected set of Feynman diagrams for the virtual corrections. 
    (a) Gluino self-energy,
    (b) quark--quark--gluon vertex [gauge coupling],
    (c) quark--squark--gluino vertex [Yukawa coupling],
    (d) two-point boxes,
    (e) three-point boxes, and
    (f) four-point boxes.}
  \label{fig:feynvirt}
\end{figure}
In (a) the gluino self-energy is given. The diagram with reversed
fermion-number flow contributes due to the Majorana nature of the
gluinos. The vertex corrections are exemplified by a gauge vertex (b)
and a Yukawa vertex (c). Typical examples for the different types of
box diagrams are depicted in (d), (e), and (f).

The divergences in the virtual corrections are regularized by
performing the calculations in $n=4-2\eps$ dimensions. These
divergences consist of ultraviolet\footnote{As expected, no
  quadratic UV divergences are generated in softly broken
  supersymmetric models.}, infrared (IR), and collinear divergences
[also called mass singularities], and they show up as poles of the
form $\eps^{-i}$ ($i=1,2$). Since the virtual amplitudes are
contracted with the Born matrix elements, all loop momenta will be
contracted with themselves or external momenta. This leads to a great
simplification of the tensor integrals appearing in the one-loop
corrections. These tensor integrals are evaluated in $n$ dimensions by
means of an adapted version of the standard Passarino--Veltman tensor
integral reduction \cite{pave}, and they are expressed in terms of
scalar integrals.  The coefficients of these scalar integrals are
finite, and they have to be calculated up to ${\cal O}(\eps^2)$. The
divergences are contained in the scalar integrals. We have calculated
these, using two different techniques. One is based on the Feynman
parametrization, the other proceeds via the computation of the
absorptive part by applying the Cutkosky cutting rules in $n$
dimensions, followed by the use of dispersion-integral techniques to
get the real part. All analytical manipulations were performed with
the help of the symbolic computer program FORM \cite{form}.

The case of equal masses for squarks and gluinos is calculated
separately, in view of additional singularities. Divergent terms of
the form $\log[(\mg^2-\ms^2)/\ms^2]$ lead to additional $1/\eps$
poles. It has been checked explicitly that the final cross-sections
are nevertheless continuous at this point of the mass-parameter space.

The singularity structure of the scalar integrals in $n$ dimensions
can be summarized as follows. The non-vanishing scalar one-point and
two-point functions only give rise to UV poles. The derivative of the
on-shell two-point functions\footnote{In the top--stop loop
  contributing to the gluino self-energy we insert widths for top,
  stop, and gluino states.}, and the three- and four-point functions
are UV finite and give rise only to IR and collinear singularities. IR
poles appear when a massless particle is exchanged between two
on-shell particles; collinear poles show up when a massless particle
splits into two massless collinear particles.  Double poles are
generated only when IR and collinear singularities are present at the
same time.

The $\gamma_5$ Dirac matrix, entering the calculation through the
quark--squark--gluino Yukawa couplings, is treated in the `naive'
scheme. In this scheme the $\gamma_5$-matrix anticommutes with the
other $\gamma_\mu$-matrices. This is a legitimate procedure at the
one-loop level for anomaly-free theories.

We have excluded the top-squarks from the final states, as discussed
in the introduction. However, to carry out the NLO calculation
consistently, we have to take into account the top-squarks inside
loops.  For the sake of simplicity we take them to be mass degenerate
with the other squarks. For the top quark we use the mass
$m_t=175~\text{GeV}$.  Thus, the final results will depend on two
free parameters: the squark mass $\ms$ and the gluino mass $\mg$.

The definitions of the invariant energies and momentum transfers, and
the gauge choices for internal and external gluons are the same as in
lowest order. Faddeev--Popov ghost contributions have therefore to be
taken into account in the gluon self-energy and in the three-gluon
vertex corrections.

\subsubsection{Renormalization of UV Divergences}
\label{renorm}

The UV divergences can be removed by renormalizing the coupling
constants and the masses of the heavy particles. The external
self-energies are multiplied by a factor $1/2$ to properly account for
the transition from Green's functions to the S-matrix. For the
renormalization of the QCD coupling constant one usually resorts to
the $\MSm$ scheme, which involves $n$-dimensional regularization,
\emph{i.e.}~fields, phase space, and loop momenta are defined in $n$
dimensions. The UV $1/\eps$ poles are subtracted, together with
specific transcendental constants, at an arbitrary subtraction point
$Q_R$, the charge-renormalization scale.

\medskip %
In supersymmetric theories, however, a complication occurs. In $n\neq
4$ dimensions the \MS scheme introduces a mismatch between the number
of gluon ($n-2$) and gluino (2) degrees of freedom.  Since this ${\cal
  O}(\eps)$ mismatch will result in finite non-zero contributions, the
\MS scheme violates supersymmetry explicitly in higher orders. In
particular, the Yukawa coupling $\ghat$, which by supersymmetry should
coincide with the gauge coupling $g_s$, deviates from it by a finite
amount at the one-loop level.  Requiring the physical amplitudes to
preserve this supersymmetric relation, a shift between the bare Yukawa
coupling and the bare gauge coupling must be introduced in the \MS
scheme:
\begin{equation}
  \hat{g}_s = g_s \left[ 1 + \frac{\as}{4\pi}\left(\frac{2}{3}N
  -\frac{1}{2}C_F\right)\right] = 
  g_s \left[ 1 + \frac{\as}{3\pi}\right],
  \label{finshift}
\end{equation}
which effectively subtracts the contributions of the false,
non-supersymmetric degrees of freedom [also called $\eps$ scalars].

The need for introducing a finite shift is best demonstrated for the
effective [one-loop corrected] Yukawa coupling, which must be equal to
the effective gauge coupling in an exact supersymmetric world with
massless gluons/gluinos and equal-mass quarks/squarks. For the sake of
simplicity we define the effective couplings
${\Gamma}^{\text{eff}}(Q^2)$ and $\hat{\Gamma}^{\text{eff}}(Q^2)$ in
the limit of on-shell quarks/squarks and almost on-shell
gluons/gluinos, with virtuality $Q^2\ll \ms^2=m_q^2$; in this limit
the couplings do not contain gauge-dependent terms.  In the \MS scheme
we find, after charge renormalization:
\begin{eqnarray}
  \hspace*{-0.6cm}\overline{MS}: \, 
  {\Gamma}^{\text{eff}}(Q^2) \!& = &\! g_s \left\{1 +
  \frac{\as}{4\pi}\,N\,\left[ 
  - \frac{1}{\bar{\varepsilon}} 
  - \log\left(\frac{\mu^2}{\ms^2}\right)
  - \frac{1}{2}\,\log\left(\frac{Q^2}{\ms^2}\right) + \frac{7}{6} \right]
  \label{qqgtot} \right\}\\[2mm] 
  \hat{\Gamma}^{\text{eff}}(Q^2) \! & = & \! \ghat \left\{1 +
  \frac{\as}{4\pi}\,\left(
  N\,\left[-\frac{1}{\bar{\varepsilon}} 
  - \log\left(\frac{\mu^2}{\ms^2}\right)
  - \frac{1}{2}\,\log\left(\frac{Q^2}{\ms^2}\right) + \frac{1}{2}
\right] + \frac{C_F}{2} \right) \right\}. \label{qsqgltot}
\end{eqnarray}
The singular term $1/\bar{\eps}$ represents the combination $1/\eps
-\gamma_E +\log(4\pi)$.  The remaining $1/\eps$ poles are IR and
collinear singularities. Inspecting ${\Gamma}^{\text{eff}}$ and
$\hat{\Gamma}^{\text{eff}}$, it is easy to prove that the difference
between the two effective couplings coincides with the shift in
Eq.~(\ref{finshift}). Taking into account this finite shift of the
bare couplings in the \MS scheme, both effective couplings become
identical at the one-loop level. In this way supersymmetry is
preserved and the \MS renormalization becomes a consistent scheme.

An alternative renormalization scheme is the modified Dimensional
Reduction (\DR) scheme in which the fields are treated in four
dimensions, but the phase space and loop momenta in $n$ dimensions. In
this scheme no mismatch between bosonic and fermionic degrees of
freedom is apparently introduced and supersymmetry is preserved
\emph{ab initio}.  At the level of the effective couplings
${\Gamma}^{\text{eff}}$ and $\hat{\Gamma}^{\text{eff}}$, this is
reflected in the equalities
\begin{eqnarray}
  \overline{DR}: \quad {\Gamma}^{\text{eff}}(Q^2) & = &  
  g_s \left\{ 1 + \frac{\as}{4\pi}\,N\,\left[ -\frac{1}{\bar{\varepsilon}} 
  - \log\left(\frac{\mu^2}{\ms^2}\right)
  -\frac{1}{2}\,\log\left(\frac{Q^2}{\ms^2}\right) + 1 \right] \right\}
  \\
  & = & \hat{\Gamma}^{\text{eff}}(Q^2).
\end{eqnarray}
As a result, both couplings are identical order by order. [It should
be noted that the transition from the effective gauge coupling in \MS
to the gauge coupling in \DR involves a well-known finite
renormalization $\as N/(24\pi)=\as/(8\pi)$ \cite{ssk}.]

\medskip %
In the following we use the \MS renormalization scheme, supplemented
by the finite shift of the Yukawa coupling. In this way supersymmetry
is preserved on the one hand, while on the other hand the definition
of the strong gauge coupling corresponds to the usual Standard-Model
measurements.

Below we list the various renormalizations needed for the production
of squarks and gluinos. In order to preserve the form of the Ward
identity given in Eq.~(\ref{slavnov}), non-zero particle masses 
have to be renormalized in an on-shell scheme. We have opted for a
real mass renormalization, involving the subtraction of the real part
of the on-shell self-energies at the real-valued pole masses. In the
case of squarks and gluinos, this is equivalent to replacing the bare
masses in the lowest-order propagators by
\begin{small}
\begin{eqnarray*}
  \left(\ms^2\right)^{bare}
    & \to & \ms^2 \left\{ 1 + \frac{\as}{4\pi}\,C_F\left[ 
  \left(-\frac{1}{\bar\eps} 
    -\log\left(\frac{\mu^2}{\ms^2}\right)
  \right)\left(\,4\frac{\mg^2}{\ms^2}\right)
 \right.\right.\\
 & & \left.\left.  -2 -6\,\frac{\mg^2}{\ms^2} 
+\left(2-4\,\frac{\mg^2}{\ms^2}\right)
 \log\left(\frac{\ms^2}{\mg^2}\right) +\left(-2 + 4\,\frac{\mg^2}{\ms^2}
 -2\,\frac{\mg^4}{\ms^4}\right)\log\left|1 -\frac{\ms^2}{\mg^2}\right|\,
\right] \right\} \nonumber\\
  \left(\mg\right)^{bare} 
    & \to & \mg \left\{ 1 + \frac{\as}{4\pi}\left[ 
  \left(-\frac{1}{\bar\eps} 
    -\log\left(\frac{\mu^2}{\mg^2}\right)\right)
    \left(\,3 N - n_f - 1\right) \right.\right.\\
 & & -4N + \frac{\ms^2}{\mg^2}
 -\frac{m_t^2}{\mg^2} + n_f\left(2 -\frac{\ms^2}{\mg^2}\right)
 + \left(-n_f -\frac{\ms^2}{\mg^2}\right)
 \log\left(\frac{\ms^2}{\mg^2}\right)
+ \frac{m_t^2}{\mg^2}  \log\left(\frac{m_t^2}{\mg^2}\right)
 \nonumber\\
 & &  \left.\left. 
 + n_f\left(-1 +2\,\frac{\ms^2}{\mg^2} -\frac{\ms^4}{\mg^4}\right)
 \log\left|1 -\frac{\mg^2}{\ms^2}\right|
 + \left(1 -\frac{\ms^2}{\mg^2} + \frac{m_t^2}{\mg^2}\right) B_0\,
\right] \right\}, \nonumber
\end{eqnarray*}
with
\begin{eqnarray*}
   B_0 &=& {\cal\text{Re}}\,\left[ \vphantom{\frac{1}{2}} 2
   -\log\left(\frac{\ms^2}{\mg^2}\right) 
   +x_1\log(1-1/x_1) +x_2\log(1-1/x_2)\right] \\
   x_{1,2} &=& \frac{1}{2\mg^2}\left[ \mg^2 +m_t^2 -\ms^2\pm 
  \sqrt{(\mg^2-m_t^2 -\ms^2)^2 -4 m_t^2\ms^2} \,\right].
\end{eqnarray*}
\end{small}
The parameters $\ms$ and $\mg$ are the pole masses.

As discussed above, the couplings are renormalized in the \MS scheme,
including the finite shift of the bare Yukawa coupling given by
Eq.~(\ref{finshift}). This leads to the following replacements of the
bare couplings in the LO expressions:
\begin{eqnarray*}
  \left(g_s \right)^{bare} 
    & \to & g_s(Q_R^2)   \left\{ 1 + \frac{\as(Q_R^2)}{4\pi}\left[\left(
  -\frac{1}{\bar\eps} 
  +\log\left(\frac{Q_R^2}{\mu^2}\right)\right)\frac{\beta_0}{2} 
  \right.\right.\\
  & & \hspace*{1.5cm}\left.\left.
  -\frac{N}{3} \log\left(\frac{\mg^2}{Q_R^2}\right)
  -\frac{n_f+1}{6} \log\left(\frac{\ms^2}{Q_R^2}\right)
  -\frac{1}{3}  \log\left(\frac{m_t^2}{Q_R^2}\right)
  \right] \right\} \nonumber\\
  \left(\ghat\right)^{bare}
    & \to & g_s(Q_R^2) \left\{ 1 + \frac{\as(Q_R^2)}{4\pi}\left[\left(
  -\frac{1}{\bar\eps} 
  +\log\left(\frac{Q_R^2}{\mu^2}\right)\right)\frac{\beta_0}{2} 
  \right.\right.\\
  & & \hspace*{1.5cm}\left.\left.
  -\frac{N}{3} \log\left(\frac{\mg^2}{Q_R^2}\right)
  -\frac{n_f+1}{6} \log\left(\frac{\ms^2}{Q_R^2}\right)
  -\frac{1}{3}  \log\left(\frac{m_t^2}{Q_R^2}\right)
  +\frac{2N}{3} - \frac{C_F}{2}
  \right] \right\}. \nonumber
\end{eqnarray*}
The first coefficient $\beta_0$ of the SUSY-QCD $\beta$ function can
be decomposed into a sum of contributions from light and heavy
particles:
\begin{equation}
  \beta_0 = \left[ \frac{11}{3}N -\frac{2}{3}n_f\right]
  + \left[-\frac{2}{3}N -\frac{2}{3} -\frac{1}{3}(n_f+1)\right] = 
  \beta_0^L + \beta_0^H.
\end{equation}
In addition to the poles, also some logarithms are subtracted in order
to decouple the heavy particles [top quark, squarks, gluinos] from the
running of $\as(Q_R^2)$. In this decoupling scheme the $Q_R^2$
evolution of the strong coupling is determined completely by the
light-particle spectrum [gluons and $n_f=5$ massless quarks]:
\begin{equation}
  \frac{\partial \,g_s^2(Q_R^2)}{\partial\log(Q_R^2)} = g_s(Q_R^2)
  \beta(g_s) = -\as^2(Q_R^2) \,\beta_0^L.
\end{equation}

The methods described above to renormalize the UV divergences lead to
cross-sections that are UV-finite. Nevertheless, there are still
divergences left. The IR divergences will cancel against the
contribution from soft-gluon radiation. The collinear singularities,
finally, will be removed by applying mass factorization. These steps
will be discussed in detail in the next subsections.

\subsection{Real-Gluon Radiation}

Two important aspects of the real-gluon radiation, the split-up of the
phase space in soft- and hard-gluon regimes, and the renormalization
of collinear divergences by means of mass factorization, will be
discussed in detail in the following subsections.

\subsubsection{Matrix Elements and Phase Space}

In order to complete the NLO evaluation of squark and gluino
production we need, in addition to the afore-mentioned virtual
SUSY-QCD corrections, also the corrections from real-gluon radiation.
They are obtained from the LO partonic reactions by adding a gluon to
the final state:
\begin{alignat}{5}
  &q_i &+&\qb_j &\,\longrightarrow\,& \sq_k &+&\sqb_l 
  &+& g\label{hard1}\\[0.2cm]
  &g   &+&g     &\,\longrightarrow\,& \sq_i &+&\sqb_i
  &+& g\label{hard2}\\[0.2cm]
  &q_i &+&q_j   &\,\longrightarrow\,& \sq_i &+&\sq_j 
  &+& g\label{hard3}\\[0.2cm]
  &q_i &+&\qb_i &\,\longrightarrow\,& \gl   &+&\gl 
  &+& g\label{hard4}\\[0.2cm]
  &g   &+&g     &\,\longrightarrow\,& \gl   &+&\gl 
  &+& g\label{hard5}\\[0.2cm]
  &q_i &+&g     &\,\longrightarrow\,& \sq_i &+&\gl 
  &+& g.\label{hard6}
\end{alignat}
Again, the momenta of the initial-state partons are denoted by
$k_1,k_2$, while the particles in the final states carry momenta $p_1,
p_2$, and $k_3$.  The charge-conjugate final states, which are not
given here explicitly, will be taken into account when the hadronic
cross-sections are calculated.  A representative set of Feynman
diagrams, contributing to the real-gluon amplitude ${\cal M}^R$, is
given in Fig.~\ref{fig:feynhard}.
\begin{figure}[h]
  \begin{center}
    \vspace*{-1.5cm}
    \hspace*{-0.5cm}
    \epsfig{file=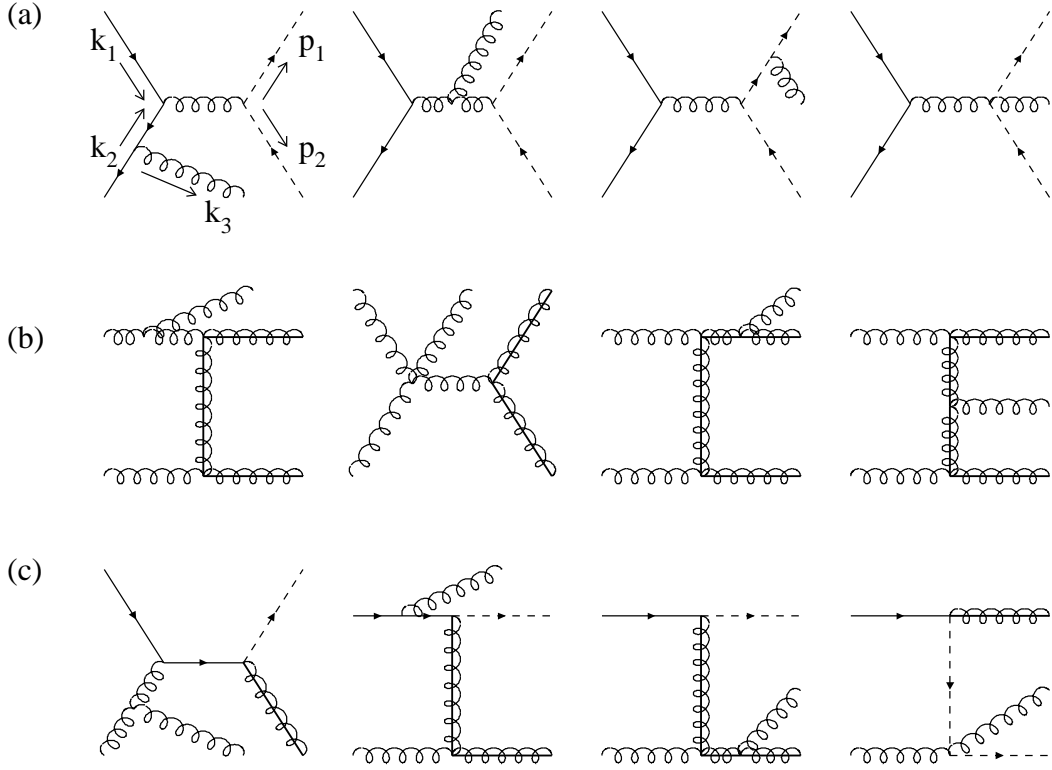,width=17cm}
    \vspace*{-2.5cm}
  \end{center}
  \caption{A representative set of Feynman diagrams corresponding to
    real-gluon radiation: squark--antisquark production (a),
    gluino-pair production (b), and squark--gluino production (c).}
  \label{fig:feynhard}
\end{figure}
We display some diagrams for (a) squark--antisquark production in
quark--antiquark annihilation, (b) gluino-pair production in gluon
fusion, and (c) squark--gluino production in quark--gluon collisions.
The internal and external gluon lines, and the $\gamma_5$
Dirac matrix, are treated in the same way as before.

\medskip %
To evaluate the squared real-gluon matrix elements $|{\cal M}^R|^2$ 
we define the following kinematical invariants \cite{ggtt}: 
\begin{alignat}{2}
  & \\[-0.7cm]
  s   &= (k_1 + k_2)^2 \qquad\quad & s_5 &= (p_1 +p_2)^2 \label{hardkin}\\
  s_3 &= (k_3 +p_2)^2 -\ms^2 \qquad\quad & s_4 &= (k_3 +p_1)^2 -\ms^2
  \nonumber\\ 
  t & = (k_2 -p_2)^2 \qquad\quad &  t' &= (k_2 -k_3)^2 \nonumber \\
  u & = (k_1 -p_2)^2 \qquad\quad & u' &= (k_1 -k_3)^2 \nonumber\\
  u_6 &= (k_2 -p_1)^2 -\ms^2 &\qquad\quad u_7 &= (k_1 -p_1)^2 -\ms^2.
  \nonumber 
\end{alignat}
For the sake of convenience we will use the following additional
invariants:  
\begin{alignat*}{2}
  s_{3g}& = s_3 +\ms^2 -\mg^2\qquad\quad & 
  s_{4g}& = s_4 +\ms^2 -\mg^2 \nonumber\\
  s_{4x} &= s_4 +\ms^2 -p_1^2 =  2(k_3\,p_1) \qquad \nonumber\\
  t_1 & = t -\ms^2 \qquad\quad & t_g & = t -\mg^2 \nonumber \\
  u_1 & = u -\ms^2 \qquad\quad & u_g & = u -\mg^2 \nonumber \\
  u_{6g} &=  u_6 +\ms^2 -\mg^2 \qquad\quad &   u_{7g} &=  u_7 +\ms^2 -\mg^2.
  \nonumber 
\end{alignat*}
The squared matrix elements must be evaluated in $n=4-2\eps$
dimensions up to ${\cal O}(\eps^2)$.

\medskip %
After performing the $n$-dimensional three-particle phase-space
integration, we find for the double-differential distributions
\begin{eqnarray}
  \label{sighard}
  s^2 \,\frac{d^2\sigma^R}{dt\,du} &=& K_{ij} \,\frac{ S_\eps^2
    \mu^{2\eps}}{2\Gamma(1-2\eps)}
    \left[\frac{(t-p_2^2)(u-p_2^2)-p_2^2s}{\mu^2 s}\right]^{-\eps}
    \Theta(\,[t-p_2^2][u-p_2^2]-p_2^2 s\,)
    \nonumber\\[1mm]
   & & \times \,\Theta(s-4m^2)\,
   \frac{(2k_3p_1)^{1-2\eps}}{(2k_3p_1 +p_1^2)^{1-\eps}}\,\Theta(2k_3p_1) \int
   d\Omega_n \sum |{\cal M}^R|^2,     \label{defhard}
\end{eqnarray}
with $ 2(k_3 p_1) = s+t+u-p_1^2 -p_2^2 \ge 0$. The $n$-dimensional angular
integration is given explicitly by $\int d\Omega_n
=\int_0^{\pi}\sin^{1-2\eps}(\theta_1)\, d\theta_1\,
\int_0^{\pi}\sin^{-2\eps}(\theta_2)\,d\theta_2$ [see
Appendix~\ref{phasespace}]. To perform the angular integrations, we
isolate the 
coefficients of the form $(s')^k (s'')^l$.  The variables $s'$ and
$s''$ denote kinematical invariants of the list in
Eq.~(\ref{hardkin}), and $k$ and $l$ are integer numbers.  One of
those invariants should contain both integration variables
($\theta_1,\theta_2$), whereas the second should depend only on
$\theta_1$. This can be achieved by means of partial fractioning,
exploiting the fact that only five of the kinematical invariants are
in fact independent\footnote{The relevant relations between the
  various invariants are given in Appendix~\ref{phasespace} for the
  process of squark--gluino production, which represents the most
  general case.}. The angular integrals are therefore of the form
\begin{equation}
  \label{defangint}
  I_n^{(k,l)} = \int_0^\pi \sin^{1-2\eps}(\theta_1)\, d\theta_1\,
  \int_0^\pi \sin^{-2\eps}(\theta_2) \,d\theta_2 (a\!+\!b \cos\theta_1)^{-k}
  ( A \!+\! B\cos\theta_1 \!+\! C\sin\theta_1\cos\theta_2)^{-l}.
\end{equation}
Explicit analytical expressions for these angular integrals can be
found in Ref.~\cite{ggtt}. In Appendix~\ref{phasespace} we demonstrate how
the kinematical invariants can be expressed in terms of the angular
variables.

\subsubsection{Soft- and Hard-Gluon Radiation}

To identify the IR singularities, the phase space for gluon radiation
is split into two distinct regimes, one describing soft gluons and the
other describing hard gluons. This separation can be defined by
introducing a cut-off parameter $\Delta$ in the invariant mass
$s_{4x}=2(k_3p_1)$, corresponding to the radiated gluon and one of the
heavy particles in the final state.  The cut-off parameter is chosen
so small that it can be neglected with respect to any other mass scale
in the process. In terms of the single-differential distribution
$d\sigma/dt$ the split-up takes the form
\begin{equation}
  \label{softhard}
  \frac{d\sigma^R}{dt} = 
  \int_0^{s_{4x}^{max}} ds_{4x} \,\frac{d^2\sigma^R}{dt\, du} = 
  \int_0^{\Delta} ds_{4x} \,\frac{d^2\sigma^S}{dt\, du} + 
  \int_\Delta^{s_{4x}^{max}} ds_{4x} \,\frac{d^2\sigma^H}{dt\, du}.
\end{equation}
The first term on the right-hand side of Eq.~(\ref{softhard})
represents the regime of soft-gluon radiation. In this regime the
momentum of the radiated gluon, $k_3$, tends to zero and an eikonal
approximation can be applied, \emph{i.e.}~neglecting $k_3$ whenever
possible.  In the limit $k_3 \to 0$, the $(2\to 3)$ kinematics is
reduced to $(2\to 2)$ kinematics, and the kinematical invariants of
Eq.~(\ref{hardkin}) take the form
\begin{alignat}{4}
  \\[-0.7cm]
  s_5 &\to s&\qquad 2(k_3p_1) &\to 0 &\qquad 2(k_3p_2) &\to 0 &\qquad
  t' &\to  0 \\ 
  u' & \to 0 &\qquad u_6/u_{6g} &\to u_1/u_g &\qquad u_7/u_{7g}
  &\to t_1/t_g, \nonumber
\end{alignat}
while the remaining invariants are not affected. 

After integration over the angles and over the invariant mass
$s_{4x}$, singular expressions of the form $\eps^{-i}$ $(i=1,2)$ are
generated. The double poles correspond to configurations where IR and
collinear singularities coincide. When the single-differential
soft-gluon distribution $d\sigma^S/dt$ is added to the virtual
corrections, the sum is IR-finite. This sum, however, is not free of
divergences until the collinear singularities are removed by means of
mass factorization.

The second term on the right-hand side of Eq.~(\ref{softhard})
represents the regime of hard-gluon radiation. In this regime only
collinear singularities occur, generated when the radiated
gluon is collinear with one of the initial-state massless particles.
They show up in $|{\cal M}^R|^2$ as terms proportional to $1/t'$ or 
$1/u'$, which behave as $1/[1\pm\cos(\theta_i)]$ \,$(i=1,2$) and lead to 
$1/\eps$ poles after the angular integration. Also these collinear
singularities have to be removed by means of mass factorization.

As a result of the split-up of the phase space, terms of the form
$\log^i(\Delta/m^2)$ $(i=1,2)$ occur in both the soft and hard
cross-sections.  They come from the same terms that generate the IR
singularities. If soft and hard contributions are added up, however,
any $\Delta$ dependence disappears from the cross-sections in the
limit $\Delta \to 0$.

\subsubsection{Mass Factorization}
\label{massfac}
 
The collinear divergences, generated by the radiation of gluons [or massless
quarks], have a universal structure. The partonic cross-sections $\sigma_{ij}$,
which contain the collinear singularities, have the following form,
factorized to all orders of perturbation theory:
\begin{eqnarray}
  s^2\,\frac{d^2\sigma_{ij}(s,t_x,u_x,\mu^2,\eps)}{dt_x\, du_x} 
  &=& \int_0^1 \frac{dx_1}{x_1} \int_0^1 \frac{dx_2}{x_2} \sum_{l,m}
  \Gamma_{li}(x_1,Q_F^2,\mu^2,\eps)  
  \,\Gamma_{mj}(x_2,Q_F^2,\mu^2,\eps) \nonumber\\   
  & & \times
  \,\hat{s}^2\,\frac{d^2\hat{\sigma}_{lm}(
    \hat{s},\hat{t}_x,\hat{u}_x,Q_F^2)  
    }{ d\hat{t}_x\, d\hat{u}_x} 
\end{eqnarray}
\begin{displaymath}
   t_x = -2(k_2 p_2) \quad u_x = -2(k_1p_2)
  \quad\hat{s}=x_1x_2 s \quad
  \hat{t}_x = x_2 t_x \quad \hat{u}_x = x_1 u_x. 
\end{displaymath}
The indices $i$--$m$ characterize the initial-state partons. The
universal splitting functions $\Gamma_{ij}$, representing the
probability of finding, inside the parent particle $j$ at the scale
$Q_F^2$, a particle $i$ with fraction $x$ of the longitudinal
momentum, contain the collinear divergences. They can be absorbed into
a redefinition of the parton densities at NLO \cite{masfac}, in
general called mass factorization. Since the subtraction point of the
mass-factorization procedure is arbitrary, the splitting functions
will depend on the factorization scale $Q_F$. Adopting the \MS
mass-factorization scheme we can write to ${\cal O}(\as)$
\begin{eqnarray}
  \Gamma_{ij}(x,Q_F^2,\mu^2,\eps) = \delta_{ij}\,\delta(1-x)
  +\frac{\as}{2\pi}\left[-\frac{1}{\bar\eps} 
  + \log\left(\frac{Q_F^2}{\mu^2}\right)\right] P_{ij}(x),  \qquad
\end{eqnarray}
with $1/\bar{\eps} = 1/\eps -\gamma_E +\log(4\pi)$ as before.  The
hard-scattering (reduced) cross-sections $\hat{\sigma}_{ij}$ are free
of collinear divergences. They depend, like the splitting functions,
on the scale $Q_F$. In NLO they have the form
\begin{eqnarray}
  \label{defreduced}
  s^2\,\frac{d^2\hat{\sigma}_{ij}(s,t_x,u_x,Q_F^2)}{dt_x\,du_x} &=& 
  s^2\,\frac{d^2{\sigma}_{ij}(s,t_x,u_x,\mu^2,\eps)}{dt_x\,du_x}\\
  & & \hspace*{-3cm}
  -\frac{\as}{2\pi}\int_0^1 \frac{dx_1}{x_1} 
  \left[-\frac{1}{\bar\eps} +
  \log\left(\frac{Q_F^2}{\mu^2}\right)\right] P_{li}(x_1)\, 
  (x_1s)^2\,\frac{d^2{\sigma}^B_{lj}(x_1s,t_x,x_1u_x,\mu^2,\eps)
    }{dt_x\,d(x_1u_x)}  \nonumber\\
  & & \hspace*{-3cm} -\frac{\as}{2\pi}\int_0^1 \frac{dx_2}{x_2} 
  \left[-\frac{1}{\bar\eps} +
  \log\left(\frac{Q_F^2}{\mu^2}\right)\right] P_{kj}(x_2)\,
  (x_2s)^2\,\frac{d^2{\sigma}^B_{ik}(x_2s,x_2t_x,u_x,\mu^2,\eps)
    }{d(x_2t_x)\,du_x}.  \nonumber
\end{eqnarray}
The Altarelli--Parisi kernels $P_{ij}(x)$ \cite{altpar} are given by
[$T_f=1/2$]
\begin{eqnarray}
  P_{gg}(x,\delta) &=&  2N \left[\frac{1}{x(1-x)}
   + x(1-x) -2\right] \Theta(1-x-\delta) \\
  & &{} +\left[2 N \log\delta + \frac{1}{2}\beta_0^L\right] \delta(1-x)
  \nonumber\\
  P_{qq}(x,\delta) &=&  C_F\, \frac{1+x^2}{1-x} \,\Theta(1-x-\delta)
  + C_F\left[2\log\delta +\frac{3}{2}\right] \delta(1-x) \\
  P_{gq}(x) &=& C_F\,\frac{1 + (1-x)^2}{x}\\
  P_{qg}(x) &=& T_f\left[x^2 + (1-x)^2\right].
\end{eqnarray}
The parameter $\delta$ is related to the IR cut-off parameter $\Delta$
through relations of the form $\delta=\Delta/(s+t_x)$ or $\delta =
\Delta/(s+u_x)$, as can be read off from Eq.~(\ref{defreduced}) by
solving the $\delta$-functions in the LO distributions. For the mass
factorization of the collinear divergences related to gluon radiation,
only the diagonal splitting functions $\Gamma_{ii}$ are required. [In
SUSY-QCD, additional splitting functions are realized in the final-state 
distributions at very high energies. For the sake of completeness
they are collected in Appendix~\ref{AP_kernels}.]

After performing the mass factorization in this way, the final results
for the virtual corrections plus gluon radiation are free of collinear
divergences.

We will use the standard \MS mass-factorization scheme in which most
of the experimentally determined parton densities have been
parametrized. The transition to the \DR scheme is non-trivial and
involves a careful matching for gluon-initiated heavy-particle
production. [This was first observed for top production within
standard QCD \cite{ggtt} and is not related to supersymmetry aspects.]

\subsection{Final States with an Additional Massless Quark}

In this subsection we shall discuss the partonic reactions that can
only be realized in next-to-leading order of the SUSY-QCD perturbative
expansion. These reactions involve final states with an additional
massless (anti)quark. In such reactions, explicit particle poles show
up inside the allowed phase space, requiring the careful isolation of
on-shell squark and gluino production and a subsequent subtle
subtraction procedure of the poles.

\subsubsection{Crossing}

The reactions that involve the radiation of a massless (anti)quark
only contribute at NLO:
\begin{alignat}{5} 
  &g &+&\qb_j &\,\longrightarrow\,& \sq_k &+&\sqb_l 
  &+& \qb_i\label{hard7}\\[0.2cm]
  &q_i &+&g &\,\longrightarrow\,& \sq_k &+&\sqb_l 
  &+& q_j\label{hard8}\\[0.2cm]
  &q_i &+&g   &\,\longrightarrow\,& \sq_i &+&\sq_j 
  &+& \qb_j\label{hard9}\\[0.2cm]
  &g &+&\qb_i &\,\longrightarrow\,& \gl   &+&\gl 
  &+& \qb_i\label{hard10}\\[0.2cm]
  &q_i &+&g &\,\longrightarrow\,& \gl   &+&\gl 
  &+& q_i\label{hard11}\\[0.2cm]
  &g &+&g     &\,\longrightarrow\,& \sq_i &+&\gl 
  &+& \qb_i\label{hard12}\\[0.2cm]
  &q_i &+&\qb_j     &\,\longrightarrow\,& \sq_k &+&\gl 
  &+& \qb_l\label{hard13}\\[0.2cm]
  &q_i &+&q_j     &\,\longrightarrow\,& \sq_k &+&\gl 
  &+& q_l.\label{hard14}
\end{alignat}
Again, the momenta of the initial-state partons are denoted by $k_1,
k_2$, while those of the particles in the final states are denoted by
$p_1, p_2$, and $k_3$.  In Fig.~\ref{fig:feynhard2} we give a few
selected Feynman diagrams for (a) the squark--antisquark--quark final
state and (b) the gluino--gluino--quark final state.
\begin{figure}[b]
  \begin{center}
   \vspace*{-2.5cm}
    \hspace*{-0.5cm}
    \epsfig{file=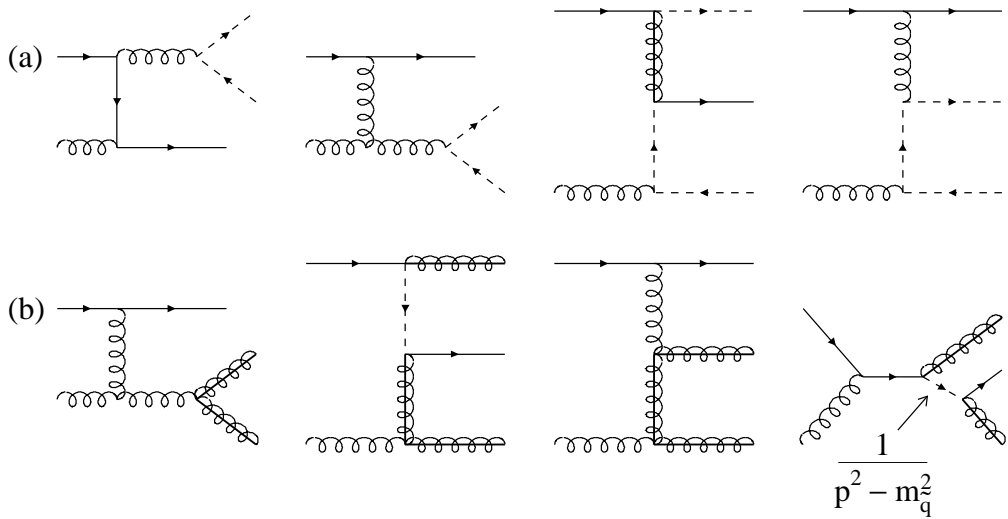,width=17cm}
    \vspace*{-2.5cm}
  \end{center}
  \caption{Selected set of Feynman diagrams for subprocesses that
    involve additional massless quarks in the final state.
    Squark--antisquark production (a), and gluino--gluino production
    (b).}
  \label{fig:feynhard2}
\end{figure}

In fact, only the matrix element for process (\ref{hard13}) requires a
new calculation. The squared matrix elements for the other reactions
can be related to subprocesses (\ref{hard1}), (\ref{hard3}),
(\ref{hard4}), (\ref{hard6}), and \emph{a posteriori} (\ref{hard13})
by means of crossing. This involves the interchange of the particles
with either the momenta $k_1$ and $k_3$ ($1 \leftrightarrow 3$) or
$k_2$ and $k_3$ ($2 \leftrightarrow 3$).

The ($1 \leftrightarrow 3$) crossing corresponds to the replacement
$k_1 \leftrightarrow -k_3$. In terms of the kinematical invariants
of Eq.~(\ref{hardkin}), this is equivalent to the exchanges
\begin{equation}
  s \leftrightarrow t' \qquad
  s_3/s_{3g} \leftrightarrow u_1/u_g \qquad
  s_4/s_{4g} \leftrightarrow u_7/u_{7g},
\end{equation}
whereas the other invariants are not affected. In view of the
different sign of the quark momentum inside the spinor sum, the
resulting squared matrix elements have to be multiplied by a factor
$(-1)$.

Analogously, the ($2 \leftrightarrow 3$) crossing corresponds to the
replacement $k_2 \leftrightarrow -k_3$. In terms of the kinematical
invariants of Eq.~(\ref{hardkin}), this is equivalent to
\begin{equation}
  s \leftrightarrow u' \qquad
  s_3/s_{3g} \leftrightarrow t_1/t_g \qquad
  s_4/s_{4g} \leftrightarrow u_6/u_{6g},
\end{equation}
whereas the other invariants are not affected. Again the resulting
squared matrix elements have to be multiplied by a factor $(-1)$.

The double-differential distributions are defined by
Eq.~(\ref{defhard}). In parallel to real-gluon radiation, the
integrand is cast into the appropriate form by means of partial
fractioning before the angular integrals are performed.  Since no IR
divergences are generated, the split-up into soft and hard regimes is
not needed.

However, the distributions contain initial-state collinear
singularities, which can be removed by means of mass factorization
[Eq.~(\ref{defreduced})].  The non-diagonal splitting functions are
needed, together with the LO distributions with one gluon more or one
gluon less in the initial state. At this point the difference in the
degrees of freedom for the gluons and quarks in $n$ dimensions starts
to play a role, leading to finite contributions to the reduced
cross-sections.

\subsubsection{On-Shell Intermediate Squark/Gluino States}
\label{subtract}

Quarks in the final state can be decay products of on-shell squarks
($\sq\to\gl q$) if $\ms > \mg$, or of on-shell gluinos ($\gl\to\sqb q$) if
$\mg > \ms$. Since these processes occur at lowest order and since the
branching ratios for the decays are large, these channels are the
dominant production channels for quarks in the final state. If the
squarks and gluinos are off shell, cf.~Fig.~\ref{fig:feynhard2}, the
production cross-section is suppressed by the strong coupling $\as$
and classified as a higher-order correction. However, by inspecting
Fig.~\ref{fig:feynhard2}, it is obvious that some of the higher-order
amplitudes are smoothly connected with the Born amplitudes. Formally
they are marked by singularities, $1/(p^2-\ms^2)$ or $1/(p^2-\mg^2)$,
if the momentum flow approaches the squark or gluino mass. These
problems can easily be solved by introducing the non-zero widths of
squarks/gluinos and regularizing in this way the higher-order
amplitudes. After subtracting the Breit--Wigner pole contributions from
the higher-order diagrams, which are already accounted for by the Born
diagrams, the cross-section for the production of squarks and gluinos
is defined properly and no double-counting occurs.

To exemplify this procedure we restrict ourselves to the case in which
the squarks are heavier than the gluinos so that the `stable' final
states, with respect to SUSY-QCD, are 2-gluino final states. This
example exhibits the full scope of subtleties inherent in the
regularization and subtraction procedures. [The singularities
generated in the other case in which the gluinos decay to squarks, are
treated in a similar way.] To be specific, we consider the subprocess
$qg\to\gl\gl q$.

The last diagram of Fig.~\ref{fig:feynhard2}b gives rise to a
particle pole if the squark momentum approaches the $\ms$ mass shell;
the other diagrams correspond to continuum $\gl\gl$ production. The
pole is regularized by introducing the non-zero squark width $\Gs$,
substituting for the propagator the Breit--Wigner form
\begin{equation}
  \frac{1}{p^2-\ms^2} \to  \frac{1}{p^2-\ms^2 +i\ms\Gs}.
\end{equation}
Denoting the on-shell resonance contribution to the matrix element, defined 
for $p^2=\ms^2$, by ${\cal M}_{res}$, and the sum of the
off-shell resonance contribution and the gluino continuum contribution
by ${\cal M}_{rem}$, the squared matrix element can be decomposed into
the resonance part and a remainder,
\begin{equation}
  |{\cal M}|^2 =   |{\cal M}_{res}|^2 
  + 2\,\text{Re}\left[{\cal M}_{res}^\ast{\cal M}_{rem} \right]   
  + |{\cal M}_{rem}|^2. 
  \label{eq:res-cont}
\end{equation}
Integrating over the entire phase space of the $\gl\gl q$ final state,
the resonance contribution to the cross-section represents the
$\sq\gl$ Born cross-section [including the branching ratio
$\Gamma(\sq\to q\gl)/\Gs$], while the remainder is to be attributed to
the ${\cal O}(\as)$ higher-order corrections to $\gl\gl$ 
production\footnote{Technical details on the separation of the resonance 
contribution are deferred to the Appendices.}.

The particle pole $1/(q^2-\ms^2)$ associated with the other
final-state gluino can be treated in a similar way. The presence of
complex masses in the $p$ and $q$ propagators gives rise to real
contributions from the interference of the two imaginary parts. This
requires a careful treatment of the angular integrations [see
Appendix~\ref{phasespace}]. Single poles in $|{\cal M}|^2$ of the form 
$1/(p^2-\ms^2)$ or
$1/(q^2-\ms^2)$, corresponding to configurations with only one
on-shell propagator, give rise to principal-value integrals, resulting
in finite contributions to the cross-sections.  In these contributions
we use a very small decay width for the squarks instead of the
physical width $\Gs$.  The influence of the actual size of the squark
width was found to be very small.

\medskip %
Adding the ${\cal O}(\as)$ corrections to the process $qg\to\sq\gl$,
the final result for the cross-section $\sigma(qg\to\gl\gl q)$
including all radiative corrections to ${\cal O}(\as)$, may therefore
be written as:
\begin{equation}
 \hspace*{-1.5cm} 
 \sigma(qg\to\gl\gl q) =   \sigma_{res}(qg\to\sq\gl\to\gl\gl q\,
           [\mathrm{LO+NLO}]) + \Delta\sigma(qg\to\gl\gl q\,[rem.]),
\end{equation}
with $\Delta\sigma$ denoting the interference term and the continuum
contribution in Eq.~(\ref{eq:res-cont}). By definition we will
attribute $\sigma_{res}$ to $\sq\gl$ production, but $\Delta\sigma$ to
$\gl\gl$ production.

\medskip %
For squark--antisquark and squark--squark final states,
similar procedures have to be followed if the gluinos are heavier than
the squarks. For squark--gluino final states the subtraction procedure is 
required for $\ms>\mg$ as well as for $\mg>\ms$.

\medskip %
After all singularities have been removed, we end up with well-defined
double-differen\-tial distributions for the irreducible
squark--antisquark, squark--squark, gluino--gluino, and squark--gluino
final states. These irreducible final states include only those
topologies in which the lightest of the coloured SUSY particles is not
produced in on-shell decays of the heavier particle.

\section{Results}

\subsection{Partonic Cross-Sections}
We first present the NLO SUSY-QCD results at the parton level for the
production of squarks and gluinos in quark and gluon collisions. To
classify the contributions it is convenient to decompose the partonic
cross-sections into scaling functions. In contrast with the
double-differential cross-sections, these scaling functions for the
total cross-sections can in general not be presented in analytic form.
Nevertheless, for two kinematical limits, at high energies and close
to the production threshold, compact analytical expressions can be
derived.

\subsubsection{Scaling Functions}
\label{sec:partonresults}
The partonic cross-sections can be calculated from the
double-differential distributions, discussed in the previous
subsections, by integration over the Mandelstam variables $t$ and
$s_4$. The exact boundaries for the integration can be found in
Appendix~\ref{phasespace}. For a detailed analysis of the partonic
cross-sections we introduce scaling functions
\begin{displaymath}
  \hat{\sigma}_{ij} = \frac{\as^2(Q^2)}{m^2} \left\{ f_{ij}^{B}(\eta,r)
  + 4 \pi \as(Q^2) \left[f_{ij}^{V+S}(\eta,r,r_t) + f_{ij}^{H}(\eta,r) +
  \bar{f}_{ij}(\eta,r) \log\left(\frac{Q^2}{m^2}\right) \right] \right\}
\end{displaymath} 
\begin{equation}
  \label{eq:scalingfun}
  \quad\eta = \frac{s}{4 m^2} -1 \qquad r =
  \frac{\mg^2}{\ms^2}\qquad r_t = \frac{m_t^2}{m^2}.
\end{equation}
The indices $i,j=g,q,\bar{q}$ indicate the partonic initial state of
the reaction.  As before, $m = (\sqrt{p_1^2}+\sqrt{p_2^2}\,)/2$ is the average 
mass of the produced particles.
The centre-of-mass energy of the partonic reaction $\sqrt{s}$ is
absorbed in the quantity $\eta$, which is better suited for analyzing
the scaling functions in the various regions of interest. Note that we
have identified the renormalization and factorization scales,
$Q=Q_R=Q_F$, properly justified in the next subsection.  For identical
particles in the final state, \emph{i.e.}~gluino pairs or squark pairs
with equal flavours and chiralities, we have taken into account the
statistical factor $1/2$. The scaling functions are divided into the
Born term $f^{B}$, the sum of virtual and soft-gluon corrections
$f^{V+S}$, the hard-gluon corrections $f^{H}$, and the scale-dependent
contributions $\bar{f}$. In this context it should be noted that the
$\log^i (\Delta/m^2)$ terms ($i=1,2$) are removed from the soft-gluon
corrections and added to the hard-gluon part.  The hard-gluon
corrections are therefore independent of the cut-off for $\Delta \ll
m^2$.

\medskip %
The scaling functions for squark--antisquark production are displayed
in Fig.~\ref{fig:sb_fun}. Unless stated otherwise, we use
$\ms=280$~GeV, $\mg=200$~GeV, and $m_t=175$~GeV as mass-parameter
input, representing an allowed mass configuration close to the present
exclusion boundaries. In LO the only possible initial states are
gluon--gluon (a) and quark--antiquark states [with equal (b) and
different (c) flavours].  The gluon--quark (d) channel is only realized
at NLO.

In Fig.~\ref{fig:ss_fun} we present the scaling functions for
squark-pair production. To lowest order, the final states are
generated exclusively in quark--quark collisions [with equal (a) and
different (b) flavours]. At NLO the gluon--quark (c) initial state
starts to contribute.

The scaling functions for gluino-pair production are displayed in
Fig.~\ref{fig:gg_fun}. In LO this final state can only be produced in
gluon--gluon (a) and quark--antiquark (b) reactions. Yet again, the
gluon--quark (c) initial state is possible at NLO. The adopted mass
configuration [with $\ms > \mg$] allows an on-shell intermediate
squark--gluino state, with subsequent decay of the squark into a gluino
and a massless quark. As discussed in Section~\ref{subtract}, the
associated singularity has to be subtracted in order to avoid double
counting.  After the subtraction is performed, a remaining, though
integrable, singularity shows up in the scaling function $f_{gq}^H$ at the 
threshold for squark--gluino production. This integrable singularity is
regularized by using a small non-zero squark width.

The scaling functions corresponding to gluino--squark production are
given in Fig.~\ref{fig:sg_fun}. Solely the gluon--quark (a) initial state 
contributes at LO. All other initial states, \emph{i.e.} gluon--gluon (b), 
quark--antiquark (c), and quark--quark (d), appear
only at NLO.  The singularities associated with on-shell
squark--(anti)squark intermediate states are subtracted, leaving
behind an integrable remnant.

The comparison of the various scaling functions reveals that
contributions involving at least one gluon in the initial state and at
least one gluino in the final state are dominant. This is a
straightforward consequence of the large colour charge of particles in
the adjoint representation.

\begin{figure}[p]
  \begin{center}
  \vspace*{-2.0cm}
  \hspace*{-0.5cm}
  \epsfig{file=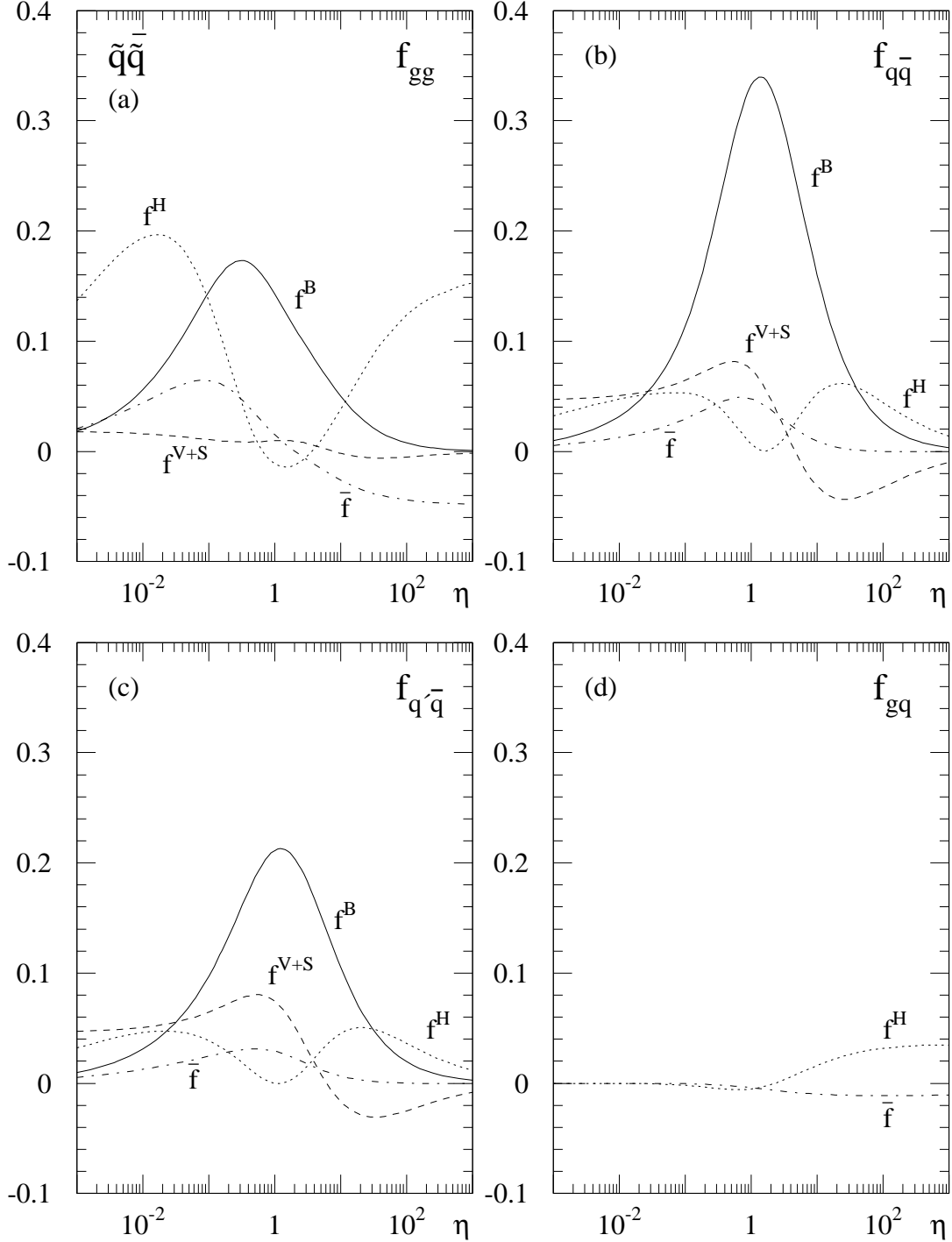,width=17cm}
  \vspace*{-2.0cm}
  \end{center}
  \caption[]{The scaling functions [Eq.~(\ref{eq:scalingfun})] for
    squark--antisquark production from (a) $gg$, (b) $q\bar{q}$, (c)
    $q'\bar{q}$, and (d) $gq$ initial states. Flavours and chiralities
    of the squarks are summed over.  Mass parameters: $\ms=280$~GeV,
    $\mg=200$~GeV, and $m_t=175$~GeV.}
  \label{fig:sb_fun}
\end{figure}
\begin{figure}[p]
  \begin{center}
  \vspace*{-2.0cm}
  \hspace*{-0.5cm}
  \epsfig{file=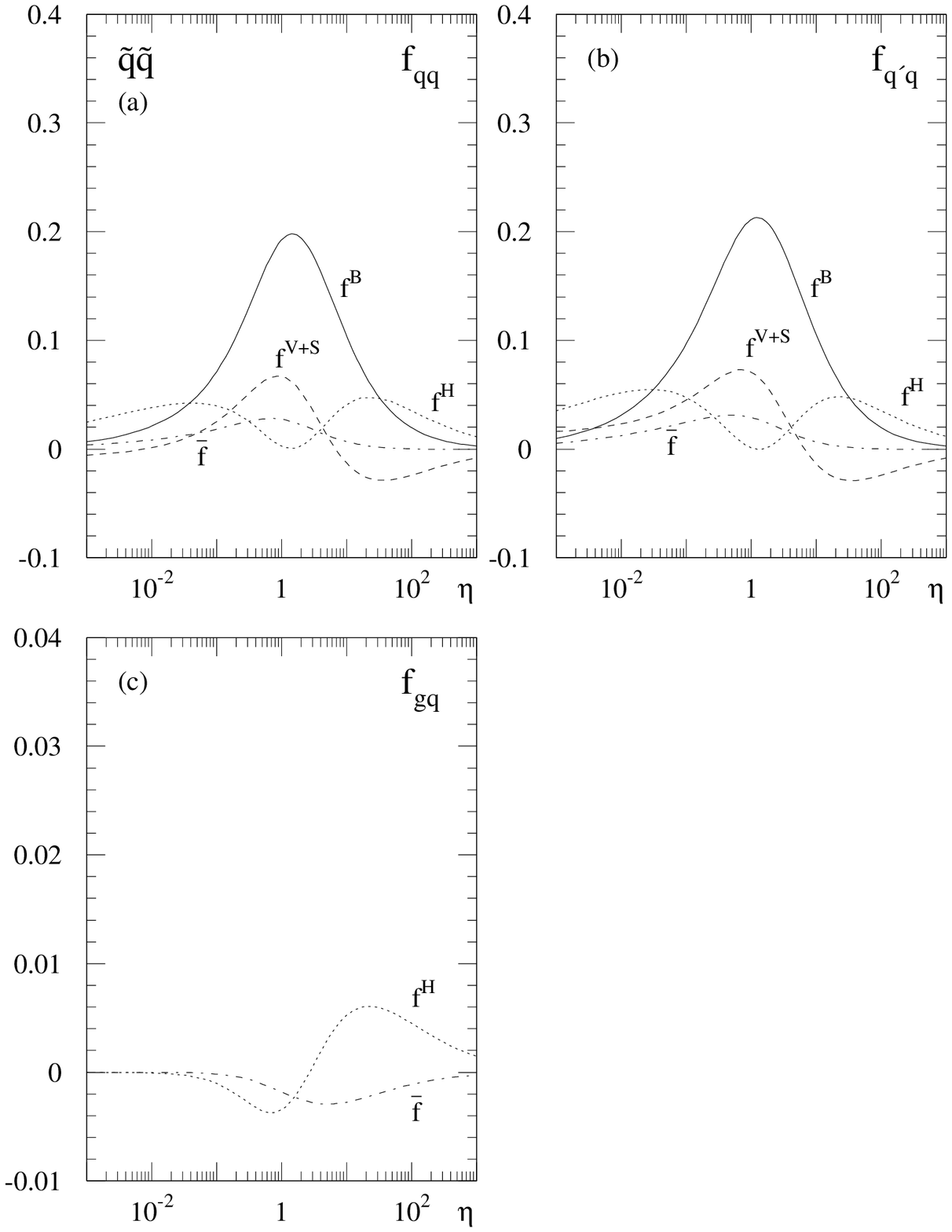,width=17cm}
  \vspace*{-2.0cm}
  \end{center}
  \caption[]{The scaling functions [Eq.~(\ref{eq:scalingfun})] for
    squark-pair production from (a) $qq$, (b) $q'q$, and (c) $gq$
    initial states. Flavours and chiralities of the squarks are summed
    over. Mass parameters: $\ms=280$~GeV, $\mg=200$~GeV, and
    $m_t=175$~GeV.}
  \label{fig:ss_fun}
\end{figure}
\begin{figure}[p]
  \begin{center}
  \vspace*{-2.0cm}
  \hspace*{-0.5cm}
  \epsfig{file=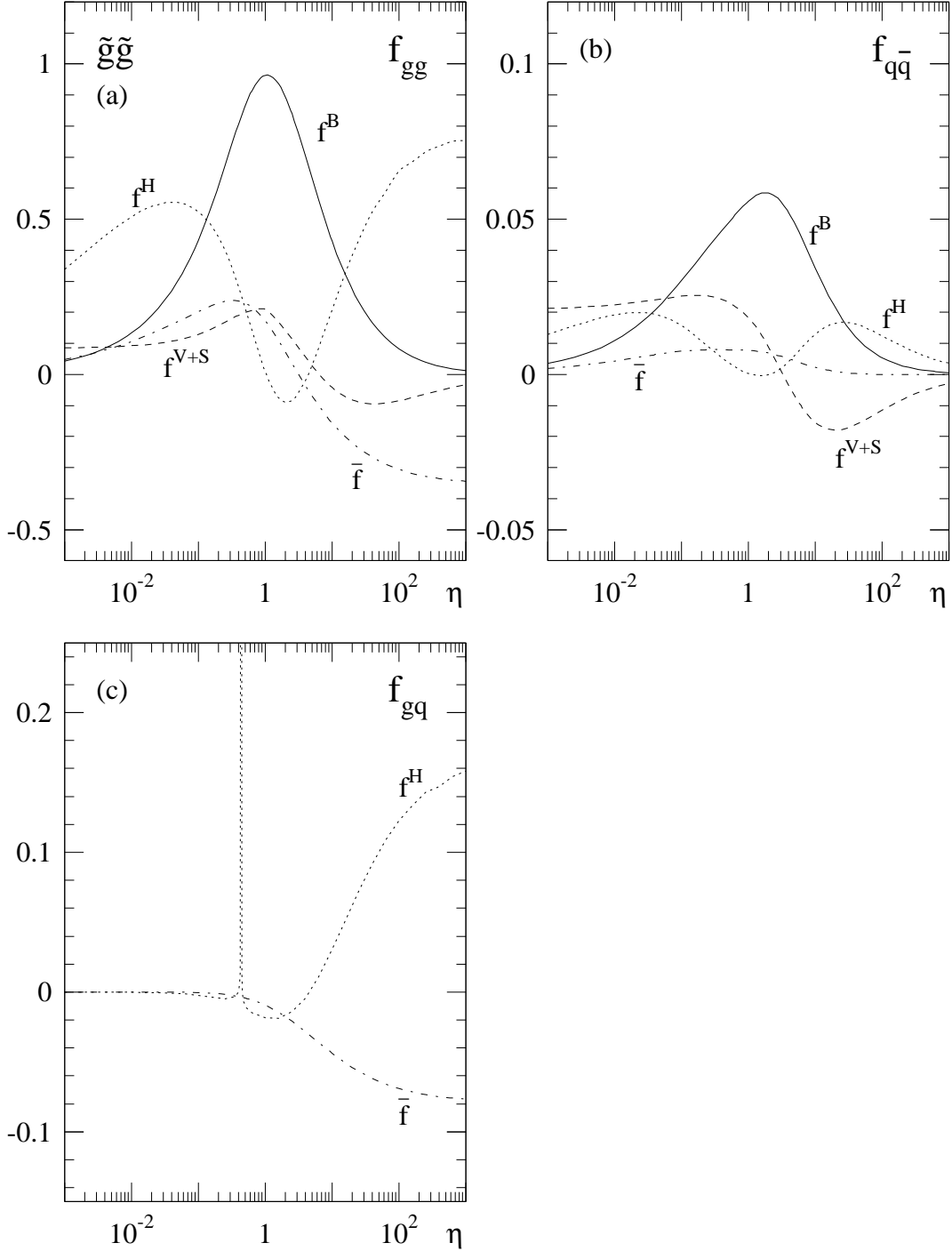,width=17cm}
  \vspace*{-2.0cm}
  \end{center}
  \caption[]{The scaling functions [Eq.~(\ref{eq:scalingfun})] for
    gluino-pair production from (a) $gg$, (b) $q\bar{q}$, and (c) $gq$
    initial states. The (integrable) singularity in (c) is the
    result of on-shell intermediate squark--gluino states; this singularity is
    regularized by using a small non-zero squark width. 
    Mass parameters: $\ms=280$~GeV, $\mg=200$~GeV, and $m_t=175$~GeV.}
  \label{fig:gg_fun}
\end{figure}
\begin{figure}[p]
 \begin{center}
 \vspace*{-2.0cm}
 \hspace*{-0.5cm}
 \epsfig{file=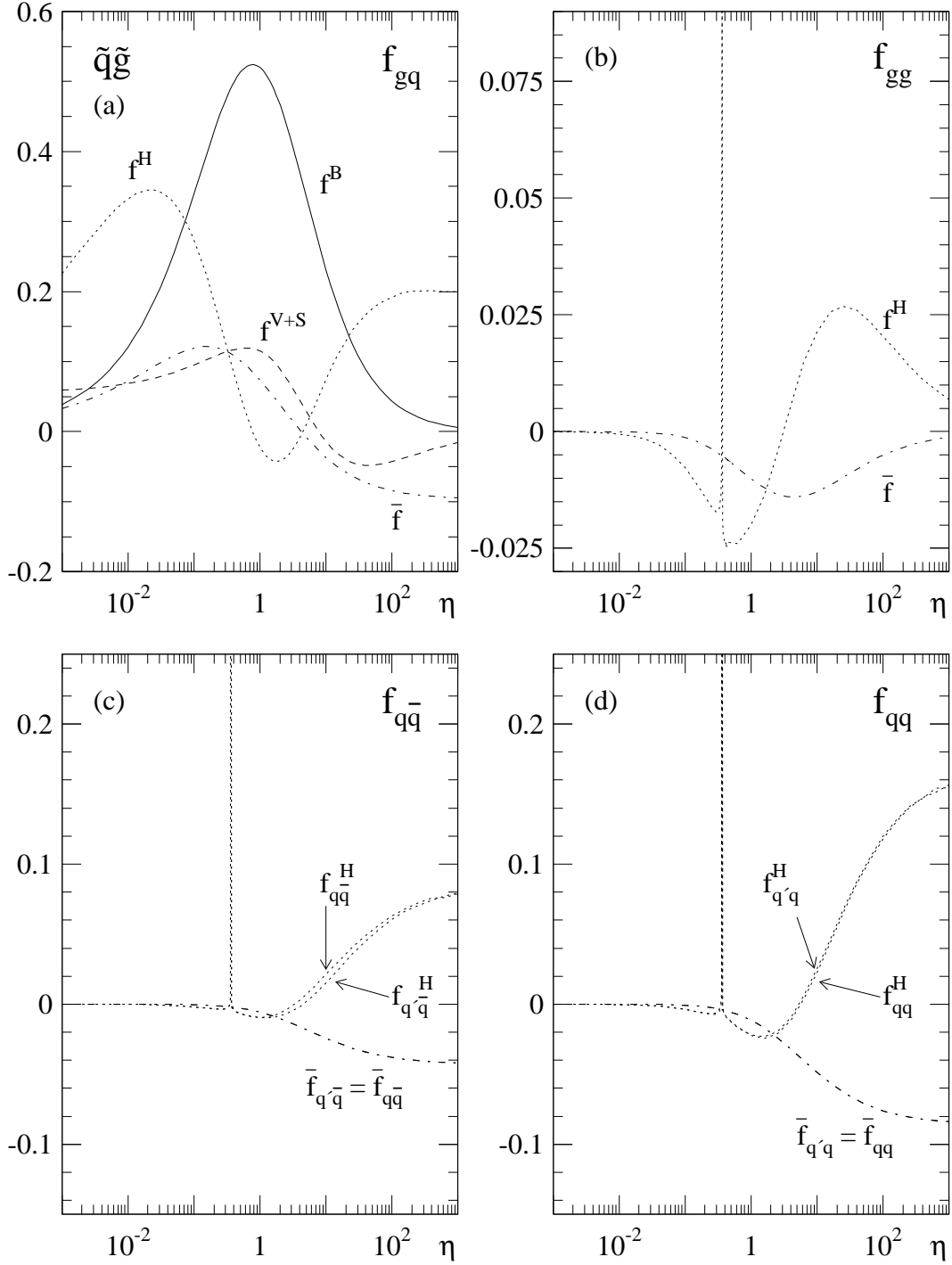,width=17cm}
 \vspace*{-2.0cm}
 \end{center}
  \caption[]{The scaling functions [Eq.~(\ref{eq:scalingfun})] for
    squark--gluino production from (a) $gq$, (b) $gg$, (c) $q\bar{q}$,
    and (d) $qq$ initial states. Flavours and chiralities of the
    squarks are summed over. The (integrable) singularities in (b) and
    (c) are caused by on-shell intermediate squark--antisquark states,
    in (d) by on-shell intermediate squark--squark states; these
    singularities are regularized by using a small non-zero squark width.
    Mass parameters: $\ms=280$~GeV, $\mg=200$~GeV, and $m_t=175$~GeV.}
  \label{fig:sg_fun}
\end{figure}

Noteworthy is the squark-mass dependence of the cross-section for
gluino-pair production from quark--antiquark annihilation. This
dependence is exemplified in Fig.~\ref{fig:gluinofun} for the scaling
functions, using the mass parameters $\mg =200$~GeV, as before, and
$\ms=175,200,225$~GeV. In the region of almost mass-degenerate squarks
and gluinos, the maximum of the LO scaling function $f^{B}$ decreases
with increasing squark mass.  By contrast, the virtual and soft
corrections $f^{V+S}$ increase for small and intermediate energies [$
\eta \lesssim 10$], as is also evident from Fig.~\ref{fig:gg_fun}b. The
hard-gluon corrections $f^{H}$ are nearly independent. The ratio of
the higher-order correction over the lowest-order cross-section will
therefore vary rapidly for gluino-pair production in the range where
gluino and squark masses are of the same order.

\begin{figure}[t]
  \begin{center}
  \vspace*{-2.0cm}
  \hspace*{-0.5cm}
   \epsfig{file=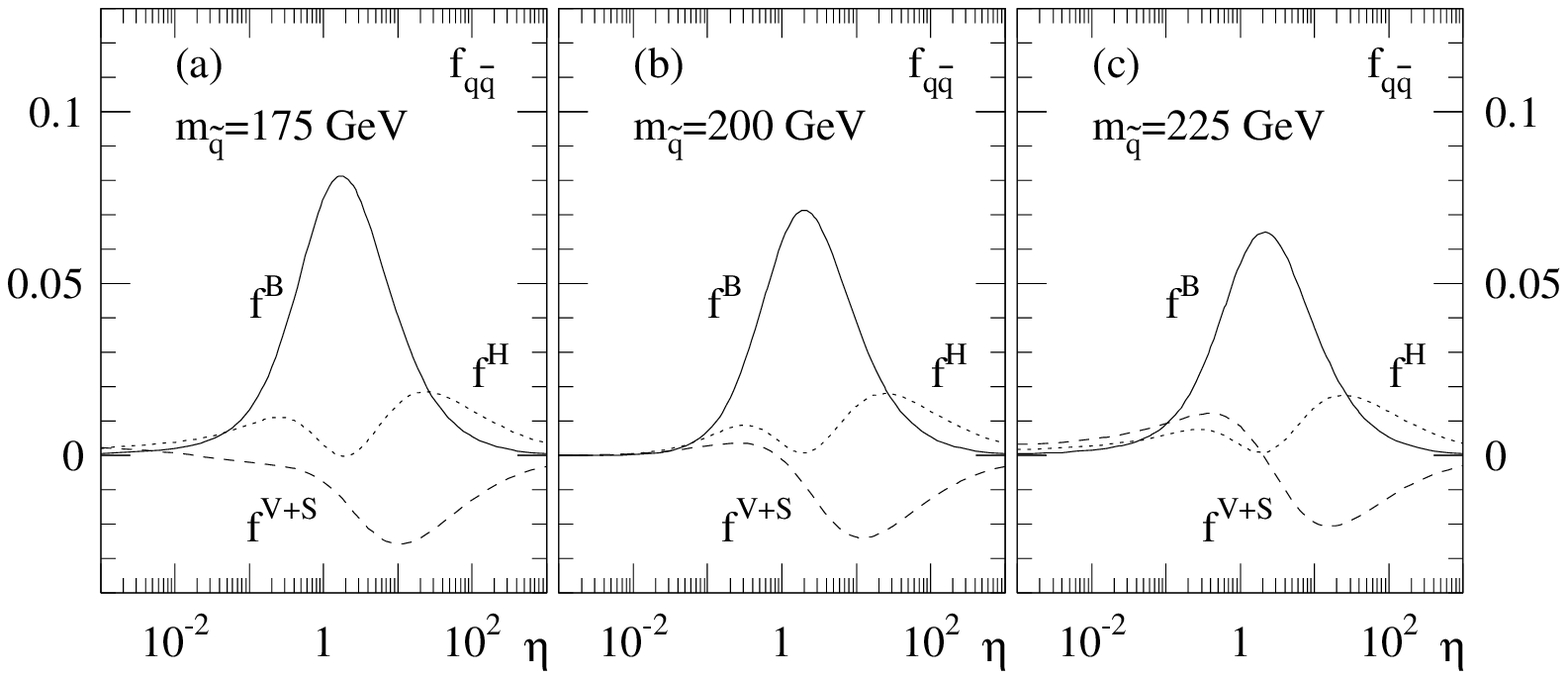,width=17cm}
  \vspace*{-2.0cm}
  \end{center}
  \caption[]{The variation of the scaling functions
    [Eq.~(\ref{eq:scalingfun})] for $q\bar{q}\to\gl\gl$ over a squark-mass
    interval around $\ms=\mg$: $\mg = 200$~GeV and (a) $\ms = 175$~GeV, 
    (b) $\ms = 200$~GeV, (c) $\ms = 225$~GeV.}
  \label{fig:gluinofun}
\end{figure}

\subsubsection{Threshold Region}

The energy region near the production threshold is the base for an
important part of the contributions to the hadronic cross-sections.
This region is characterized by the small velocity $\beta$ of the
produced heavy particles in their centre-of-mass system [$\beta \ll
1$].  Two sources of large corrections can be identified in this
threshold region, the leading terms of which can be calculated
analytically at NLO. These analytical expressions provide powerful
checks of the numerically integrated NLO corrections for arbitrary
mass parameters.

First of all, the exchange of (long-range) Coulomb gluons between the
slowly moving massive particles in the final state [see
Fig.~\ref{fig:coulomb}a] leads to a singular correction factor $\sim
\pi\as/\beta$, which compensates the LO phase-space suppression factor
$\beta$ \cite{somm}. The scaling function $f^{V+S}$ therefore tends to
a non-zero constant at threshold. It should be noted, however, that
the screening due to the non-zero lifetimes of the squarks/gluinos
reduce this effect considerably.

\begin{figure}[t]
  \begin{center}
    \vspace{-1.5cm}
    \hspace*{-1cm}
    \epsfig{file=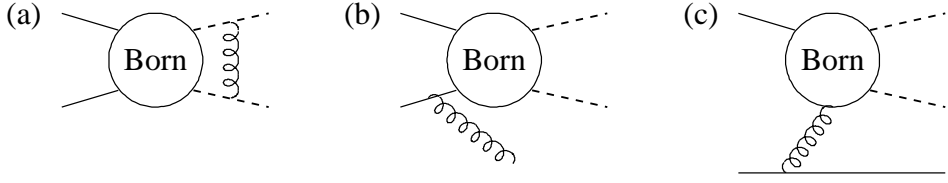,width=15cm}
    \vspace{-1.2cm}
  \end{center}
  \caption{Generic diagrams leading to (a) the Coulomb singularity,
    (b) the large threshold logarithms, and (c) the high-energy
    plateau. The solid external lines represent gluons/(anti)quarks, the
    dashed ones respresent gluinos/(anti)squarks.}
  \label{fig:coulomb}
\end{figure}

Secondly, as a result of the strong energy dependence of the
cross-sections near threshold, large positive ``soft'' corrections
$\sim \log^i(\beta^2)$ ($i=1,2$) are observed in the initial-state
gluon-radiation contribution [see Fig.~\ref{fig:coulomb}b]. They can
in fact be resummed \cite{logbeta,logbeta2}. In our analysis we will,
however, stick to the strict NLO corrections.

Near threshold the scaling functions can be expanded in $\beta$,
leading to the following analytical expressions [suppressing $\Theta(\beta)$]:

\paragraph{Squark--Antisquark:}
\begin{small}
\begin{alignat}{2}
  \\[-0.7cm]
  &f_{gg}^B =  \frac{7 n_f \pi \beta}{192} &\qquad
  &f_{q\bar{q}}^B = \frac{4 \pi \beta \ms^2 \mg^2}{9(\ms^2 +\mg^2)^2}\\
  &f_{gg}^{V+S}  =  f_{gg}^{B} \frac{11}{336 \beta} &\qquad
  &f_{q\bar{q}}^{V+S} =  f_{q\bar{q}}^{B} \frac{7}{48 \beta} \nonumber\\
  &f_{gg}^{H} =  f_{gg}^{B} \left[ \frac{3}{2\pi^2}\log^2(8\beta^2)
  -\frac{183}{28\pi^2} \log(8\beta^2) \right] &\qquad
  &f_{q\bar{q}}^{H}  =  f_{q\bar{q}}^{B} \left[
  \frac{2}{3\pi^2}\log^2(8\beta^2) -\frac{11}{4\pi^2} \log(8\beta^2) 
  \right] \nonumber\\
  &\bar{f}_{gg} = -f_{gg}^B \frac{3}{2\pi^2}\log(8\beta^2)  &\qquad
  &\bar{f}_{q\bar{q}} = -f_{q\bar{q}}^B \frac{2}{3\pi^2}\log(8\beta^2).
  \nonumber
\end{alignat}
\end{small}

\noindent For squark--antisquark final states the scaling functions for equal
and different flavours are identical ($f_{q'\bar{q}}=f_{q\bar{q}}$)
near threshold.

\paragraph{Squark--Squark:}
\begin{small}
\begin{alignat}{2}
  \\[-0.7cm]
  &f_{qq}^B  = \frac{8 \pi \beta \ms^2 \mg^2}{27(\ms^2 +\mg^2)^2}&\qquad
  &f_{q'q}^B = \frac{4 \pi \beta \ms^2 \mg^2}{9(\ms^2 +\mg^2)^2}\\
  &f_{qq}^{V+S}  =  f_{qq}^B \frac{1}{24\beta}&\qquad
  &f_{q'q}^{V+S} = f_{q'q}^B \frac{1}{24\beta} \nonumber\\
  &f_{qq}^{H}  =  f_{qq}^B \left[ \frac{2}{3\pi^2}\log^2(8\beta^2)
  -\frac{7}{2\pi^2} \log(8\beta^2) \right] &\qquad
  &f_{q'q}^{H} =  f_{q'q}^B \left[ \frac{2}{3\pi^2}\log^2(8\beta^2)
  -\frac{19}{6\pi^2} \log(8\beta^2) \right]\nonumber\\
  &\bar{f}_{qq}  =  -f_{qq}^B \frac{2}{3\pi^2}\log(8\beta^2)&\qquad
  &\bar{f}_{q'q} =  -f_{q'q}^B \frac{2}{3\pi^2}\log(8\beta^2).
  \nonumber
\end{alignat}
\end{small}

\paragraph{Gluino--Gluino:}
\begin{small}
\begin{alignat}{2}
  \\[-0.7cm]
  &f_{gg}^B =  \frac{27 \pi \beta}{64} &\qquad
  &f_{q\bar{q}}^B = \frac{\pi \beta}{3}\left( \frac{\mg^2-\ms^2}{\ms^2
    +\mg^2}\right)^2  \\
  &f_{gg}^{V+S}  =  f_{gg}^{B} \frac{1}{16 \beta} &\qquad
  &f_{q\bar{q}}^{V+S} =  f_{q\bar{q}}^{B} \frac{3}{16 \beta} \nonumber\\
  &f_{gg}^{H} =  f_{gg}^{B} \left[ \frac{3}{2\pi^2}\log^2(8\beta^2)
  -\frac{29}{4\pi^2} \log(8\beta^2) \right] &\qquad
  &f_{q\bar{q}}^{H}  =  f_{q\bar{q}}^{B} \left[
  \frac{2}{3\pi^2}\log^2(8\beta^2) -\frac{41}{12\pi^2} \log(8\beta^2) 
  \right] \nonumber\\
  &\bar{f}_{gg} = -f_{gg}^B \frac{3}{2\pi^2}\log(8\beta^2)  &\qquad
  &\bar{f}_{q\bar{q}} = -f_{q\bar{q}}^B \frac{2}{3\pi^2}\log(8\beta^2).
  \nonumber
\end{alignat}
\end{small}

\noindent The LO and NLO threshold cross-sections for gluino-pair
production from quark--antiquark annihilation vanish if the squarks
and gluinos are mass degenerate. This follows from the destructive
interference between the three LO diagrams. In
Fig.~\ref{fig:gluinofun} this phenomenon is clearly visible.

\paragraph{Squark--Gluino:}
\begin{small}
\begin{alignat}{1}
  \\[-0.7cm]
  & f_{qg}^{B} = \frac{\pi\beta}{(\ms+\mg)^3}\left\{
  \frac{2}{9}\ms\mg^2 +\frac{1}{2}\ms^2\mg +\frac{1}{2}\ms^3 \right\}
  \\
  & f_{qg}^{V+S} = \frac{\pi}{(\ms+\mg)^3} \left\{-\frac{1}{192}\ms\mg^2
  +\frac{3}{64}\ms^3 \right\} \nonumber\\
  & f_{qg}^{H} = \frac{f_{qg}^B}{\pi^2}
  \left\{\frac{13}{12}\log^2(8\beta^2)
    +\frac{13}{6}\log(8\beta^2)\log\left(\frac{4\ms\mg}{(\ms+\mg)^2}\right)
  \right\} \nonumber\\
  & \qquad +\frac{\beta}{\pi(\ms+\mg)^3}\log(8\beta^2) \left\{
  -\frac{529}{432}\ms\mg^2 -\frac{65}{24}\ms^2\mg -\frac{121}{48}\ms^3
\right\} \nonumber\\
 & \bar{f}_{qg} = - f_{qg}^B \frac{13}{12\pi^2}\log(8\beta^2) \qquad
 \beta = \sqrt{1 - \frac{4 \ms \mg}{s
     -(\ms-\mg)^2}}~. 
  \nonumber
\end{alignat}
\end{small}

\noindent Note that for squark--gluino production the leading
corrections factorize only partly in terms of the LO cross-section.
This in contrast to the other production processes, which involve
equal-mass final-state particles. 

\subsubsection{High-Energy Region}

At high energies the NLO partonic cross-sections can asymptotically
approach a non-zero constant, rather than scaling with $1/s$ as the LO
cross-sections:
\begin{alignat}{2}
\hat\sigma &\sim \frac{\as^2}{s}   &\qquad &\text{(LO)} \\
\hat\sigma &\sim \frac{\as^3}{m^2} &\qquad &\text{(NLO)}.
\end{alignat}
This is caused by almost on-shell, soft gluons in space-like
propagators, associated with hard-gluon/quark radiation [see
Fig.~\ref{fig:coulomb}c]. 
Since the splitting probabilities $q\to qg$ and $g\to gg$ are
scale-invariant \emph{mod.}~logarithms, the size of the NLO
cross-section is set by the centre-of-mass energy of the subprocess induced by
the virtual gluon (marked ``Born'' in Fig.~\ref{fig:coulomb}c). For the 
dominant contributions this centre-of-mass energy is of the order of the 
squark/gluino masses, \emph{i.e.}~not far above the threshold.
It should be noted that these high-energy plateaus only have a marginal 
influence on the hadronic cross-sections, as the main part of the contributions
originates from the partonic energy region near the production threshold.

Exploiting the factorization in the transverse gluon momentum at high
energies \cite{ktfac}, the high-energy scaling functions can be
determined analytically.

\paragraph{Squark--Antisquark:}
\begin{alignat}{3}
  \\[-0.7cm]
  f_{gg}^H  &=&  \frac{2159}{4320\pi} &\qquad 
 f_{gq}^H  &=&  \frac{2159}{19440\pi} \\
 \bar{f}_{gg} &=&  -\frac{11}{72\pi} &\qquad
 \bar{f}_{gq} &=&  -\frac{11}{324\pi}. \nonumber
\end{alignat}
The ratio of the $f_{gg}$ and $f_{gq}$ scaling functions is given by
$2 N:C_F = 9:2$. This ratio corresponds to the probability of
emitting a soft gluon from a gluon [$\sim N$, 2 sources] or a quark
($\sim C_F$, 1 source).

\paragraph{Squark--Squark:}%
The squark-pair production cross-section does not exhibit a
high-energy plateau, since no space-like gluon-exchange diagrams are
possible at NLO.

\paragraph{Gluino--Gluino:}
\begin{alignat}{3}
  \\[-0.7cm]
 f_{gg}^H & = & \frac{1949}{800\pi} &\qquad 
 f_{gq}^H & = & \frac{1949}{3600\pi} \\
 \bar{f}_{gg} &=&  -\frac{177}{160\pi} &\qquad
 \bar{f}_{gq} &=&  -\frac{59}{240\pi}. \nonumber
\end{alignat}
The ratio of the $f_{gg}$ and $f_{gq}$ scaling functions is given
again by $2 N:C_F = 9:2$.

\paragraph{Squark--Gluino:}
\begin{alignat}{6}
  \\[-0.7cm]
 f_{qg}^H      &=&  \frac{517}{864\pi} &\qquad 
 & f_{q\bar{q}}^H = & f_{q'\bar{q}}^H  &=&  \frac{517}{1944\pi} &\qquad
 & f_{qq}^H = & f_{q'q}^H        &=  \frac{517}{972\pi} \\
 \bar{f}_{qg}        &=&  -\frac{5}{18\pi} &\qquad
 & \bar{f}_{q\bar{q}} =& \bar{f}_{q'\bar{q}} &=&  -\frac{10}{81\pi} &\qquad
 & \bar{f}_{qq} = & \bar{f}_{q'q}       &=  -\frac{20}{81\pi}. 
 \nonumber
\end{alignat}
These (simple) high-energy limits are derived for equal squark and gluino
masses; the results for arbitrary masses are very complicated
and are therefore not given explicitly. The ratio of the
$f_{qg}$, $f_{q\bar{q}}$, and $f_{qq}$ scaling functions is given by
$N:C_F:2C_F=9:4:8$, irrespective of the precise values for the squark
and gluino masses.

\subsection{Hadronic Cross-Sections}
\label{sec:hadronsigmas}

Finally we discuss in this subsection the hadronic cross-sections for
the production of squarks and gluinos. The analyses are performed for
the Fermilab $\ppb$ collider Tevatron with a centre-of-mass energy of
$\sqrt{S}=1.8$~TeV, and for the CERN $pp$ collider LHC with a
centre-of-mass energy of $\sqrt{S}=14$~TeV. In analogy to the
experimental analyses, we consider the following four hadronic
production processes
\begin{eqnarray}
  p\bar p / pp & \rightarrow & \tilde q \bar{\tilde q} \label{ssbprod}
                                   \\[0.2cm]
  p\bar p / pp & \rightarrow & \sq\sq ,\sqb\sqb \label{ssprod}\\[0.2cm]
  p\bar p / pp & \rightarrow & \tilde g \tilde g \label{ggbprod}\\[0.2cm]
  p\bar p / pp & \rightarrow & \sq\gl ,\sqb\gl. \label{sgprod}
\end{eqnarray}
As before the chiralities and flavours of the squarks
(\emph{e.g.}~$\tilde{u}_L, \tilde{d}_R$) are implicitly summed over.
Yet stop production is not taken into account; these final states will
be analyzed in a forthcoming report \cite{stop}.  The charge-conjugate
final states are now properly taken into account in the reactions
(\ref{ssprod}) and (\ref{sgprod}). From now on we will refer to these
two hadronic reactions, for simplicity, as $\sq\sq$ (squark--squark)
and $\sq\gl$ (squark--gluino) production, respectively.

Various aspects of the above reactions will be presented in detail.
First of all, the dependence of the cross-sections on the
renormalization/factorization scale and parton densities is
investigated. Then the NLO corrections are studied for a default
choice of scale and parton densities. We discuss the NLO effects on
the differential distributions with respect to the rapidity $y$ and
transverse momentum $p_t$ of one of the outgoing particles. Finally we
describe the NLO effects on the total cross-sections and their
implications on the experimental search for squarks and gluinos.

\subsubsection{Scale and Parton-Density Dependence}

The total hadronic cross-sections are obtained by convoluting the
partonic cross-sections with the relevant parton densities [in the
proton or antiproton]:
\begin{equation}
  \sigma(S,Q^2) = \sum_{i,j=g,q,\bar{q}} \int_\tau^1 dx_1
  \int_{\tau/x_1}^1 dx_2 \, f_{i}^{h_1} (x_1,Q^2) \, f_{j}^{h_2} (x_2,Q^2)
  \, \hat{\sigma}_{ij}(x_1 x_2 S,Q^2) \Big|_{\tau = 4 m^2/S}
  \label{eq:sigscale} 
\end{equation}
The partons $i$ and $j$ carry fractions $x_1$ and $x_2$ of the original 
momenta of the hadrons $h_1$ ($=p$) and $h_2$ ($=\bar{p}/p$), respectively. 
The integrations in Eq.~(\ref{eq:sigscale}) are performed numerically, using 
the Monte Carlo integration routine VEGAS \cite{vegas}. The parton densities 
$f_{i}^h$ have been extracted from a large variety of experiments and are
available in various parametrizations. To estimate the associated
uncertainty, we compare the results for a sample of three different
parametrizations.

\medskip %
As a first step, we present for reactions
(\ref{ssbprod})--(\ref{sgprod}) the dependence of the total
cross-section on the renormalization and factorization scale
$Q=Q_R=Q_F$ in Figs.~\ref{fig:sigscale1} and \ref{fig:sigscale2}. We
have checked that the separate variation of the factorization scale
$Q_F$ and the renormalization scale $Q_R$ leads to a variation of the
next-to-leading order cross-section (roughly) within the band
generated by $Q$.  We can therefore keep the discussion to this
simplified case without loss of generality. The scale is restricted to
$Q\lesssim 1$~TeV, as the parton densities are not available beyond
this value.

The results for the Tevatron are given in Fig.~\ref{fig:sigscale1},
using the mass parameters $\ms=280$~GeV, $\mg=200$~GeV, and
$m_t=175$~GeV. For a consistent comparison of LO and NLO results, we
calculate all quantities [$\as(Q_R^2)$, the parton densities, and the
partonic cross-sections] in LO and NLO, respectively. In LO we use the
parton densities of GRV94 \cite{GRV}\footnote{For charm and bottom
  quarks we use the earlier GRV distributions \cite{grvalt}.} and
CTEQ3 \cite{CTEQ} with the corresponding QCD couplings. At NLO this is
compared with the parton densities of GRV94 \cite{GRV}, CTEQ3
\cite{CTEQ}, and MRS(A') \cite{MRS}.  In LO the scale dependence is
steep and monotonic. Changing the scale from $Q=2m$ to $Q=m/2$, the
cross-section increases by 100--120\%. 
In NLO the scale dependence is strongly reduced, to about
40--50\% in this interval. At the same time the cross-section is
significantly enhanced at the central scale ($Q=m$).  The uncertainty
originating from the various parametrizations of the parton densities
in NLO amounts to $\lesssim 10\%$ at the central scale.  An exception
is the squark--squark cross-section, which is dependent on the badly
determined sea-quark distribution.

The results for the LHC are given in Fig.~\ref{fig:sigscale2}, using
the mass parameters $\ms=600$~GeV, $\mg=500$~GeV, and $m_t=175$~GeV.
Here, too, a strong reduction of the scale dependence is observed by
taking into account the NLO corrections, as well as a clear
enhancement of the cross-section at the central scale. In the interval
between $Q=m$ and $Q=m/2$ the LO cross-section increases by about
35\%, whereas the variation for the NLO results is reduced to only
5--10\%. At the LHC the dependence on the factorization scale is very
weak and the residual scale dependence is dominated by $\as$ [in
contrast with the Tevatron where fairly large $x$ values in the parton
distributions give rise to a stronger factorization-scale dependence].
The uncertainty due to different parametrizations of the parton
densities in NLO amounts to $\lesssim 13\%$ at the central scale. This
is a result of the prominent role played by the gluon densities at the
LHC.

In conclusion, for all four reactions and for both hadron colliders
the scale dependence is reduced by a factor of 2.5--4 when the
theoretical predictions are improved by taking into account the
next-to-leading order SUSY-QCD corrections. Even a broad and shallow
maximum develops at scales near one third of the central scale.  In
the following we will adopt GRV94 parton densities and take $Q=m$ as
the default scale; this results in a \emph{conservative} estimate of
the cross-sections.

\begin{figure}[p]
  \begin{center}
  \vspace*{-2.5cm}
  \hspace*{-0.5cm}
  \epsfig{file=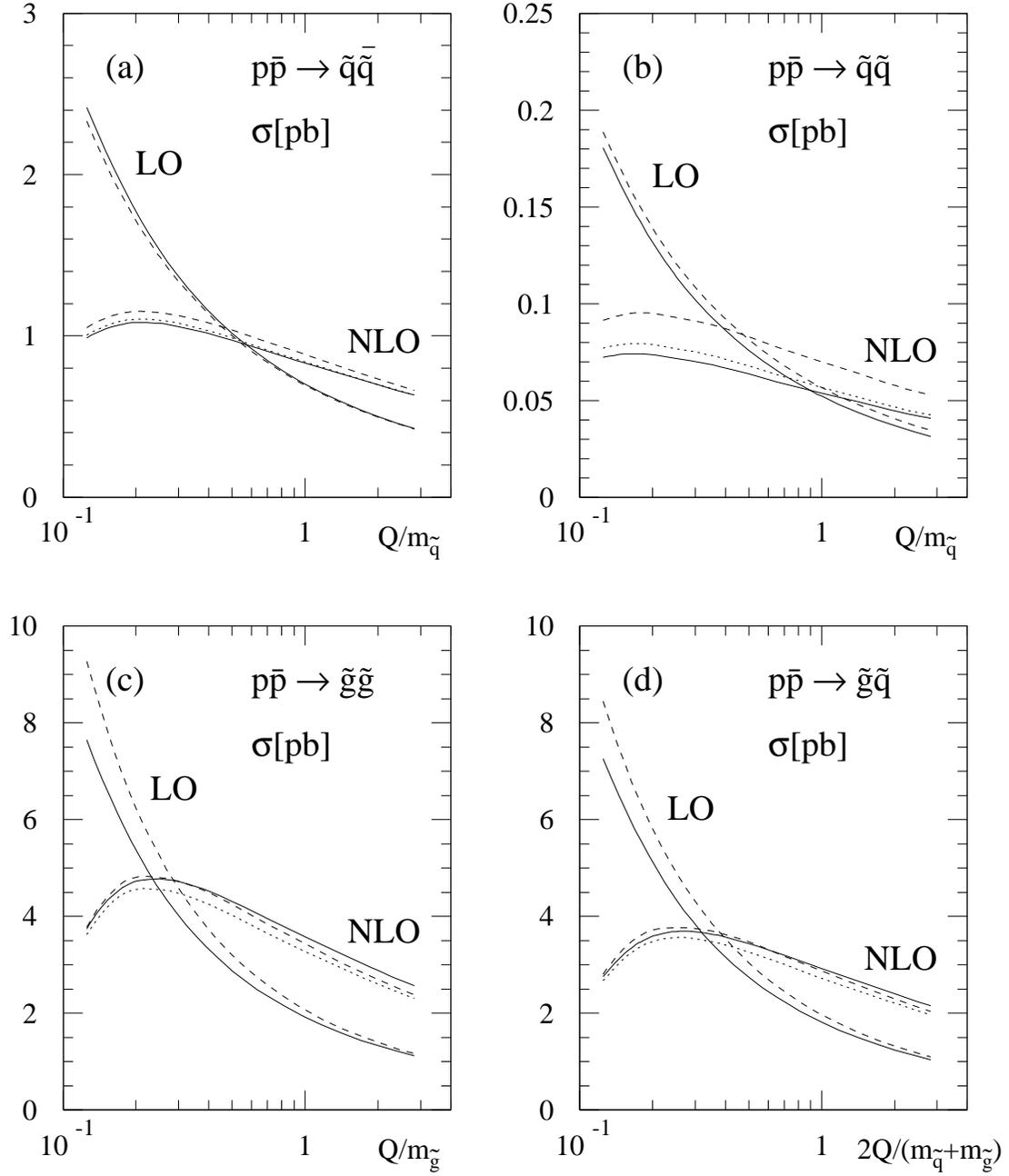,width=17cm}
  \vspace*{-2cm}
  \end{center}
  \caption[]{The dependence on the renormalization/factorization scale
    $Q$ of the LO and NLO cross-sections for (a) squark--antisquark,
    (b) squark--squark, (c) gluino--gluino, and (d) squark--gluino
    production at the Tevatron ($\sqrt{S}=1.8$~TeV).  Parton
    densities: GRV94 (solid), CTEQ3 (dashed), and MRS(A') (dotted).
    Mass parameters: $\ms=280$~GeV, $\mg=200$~GeV, and $m_t=175$~GeV.}
  \label{fig:sigscale1}
\end{figure}

\begin{figure}[p]
  \begin{center}
    \vspace*{-2.5cm}
    \hspace*{-0.5cm}
  \epsfig{file=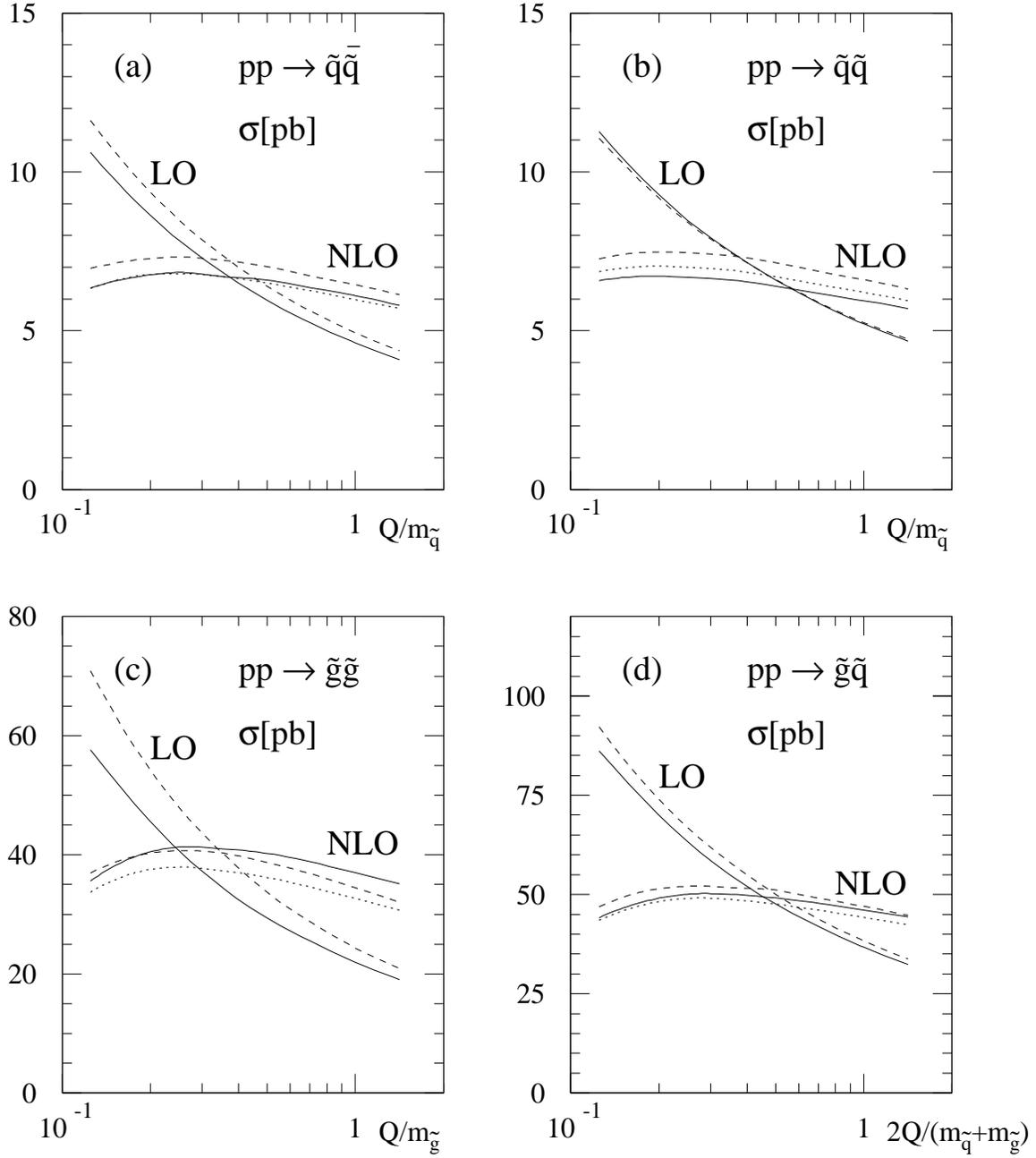,width=17cm}
  \vspace*{-2cm}
  \end{center}
  \caption[]{The dependence on the renormalization/factorization scale
    $Q$ of the LO and NLO cross-sections for (a) squark--antisquark,
    (b) squark--squark, (c) gluino--gluino, and (d) squark--gluino
    production at the LHC ($\sqrt{S}=14$~TeV).  Parton densities:
    GRV94 (solid), CTEQ3 (dashed), and MRS(A') (dotted).  Mass
    parameters: $\ms=600$~GeV, $\mg=500$~GeV, and $m_t=175$~GeV.}
  \label{fig:sigscale2}
\end{figure}

\pagebreak

\subsubsection{Differential Distributions: Transverse Momentum and
  Rapidity} 

The hadronic differential distributions are obtained by convoluting
the partonic double-differential distributions with the relevant
parton densities. We consider the distributions with respect to the
transverse momentum $p_t$ and rapidity $y$ of one of the outgoing
squarks/gluinos.  The corresponding double-differential hadronic
distribution is given by\footnote{The preferred scales $Q$ are the
  transverse masses; however, for convenience we have chosen the
  particle masses, as for the total cross-sections.}
\begin{equation}
  \label{eq:sigdiff}
  \frac{d^2\sigma}{dp_t dy} = 2 p_t S \sum_{i,j=g,q,\bar{q}}
  \int_{x_1^-}^1 dx_1 \int_{x_2^-}^1 dx_2 \, x_1 f_{i}^{h_1}
  (x_1,Q^2)\, x_2 f_{j}^{h_2} (x_2,Q^2)
  \,\frac{d^2\hat{\sigma}_{ij}(x_1 x_2 S,Q^2)}{dt\, du}.
\end{equation}
The definitions of $p_t$ and $y$ and of the integration boundaries can
be found in Appendix~\ref{phasespace}. Squarks and antisquarks are not
distinguished in the final state. So, for $\sq\sqb$, $\sq\sq$, and
$\gl\gl$ final states we have to add the distributions with respect to
both final-state particles. The rapidity distributions, as presented in
Figs.~\ref{fig:tevy} and \ref{fig:lhcy}, are defined as the sum of the
contributions of positive and negative rapidity.  The distributions
are normalized to unity.

In Fig.~\ref{fig:tevpt} the normalized $p_t$ distribution is given for
the Tevatron. The input mass parameters are $\ms=280$~GeV,
$\mg=200$~GeV, and $m_t=175$~GeV. For the scale $Q$ and for the parton
densities we take the default settings [$Q=m$ and GRV94]. In order to
perform a consistent comparison of LO and NLO results, the parton
densities are used with the associated values for $\as$. In
Fig.~\ref{fig:tevy} the normalized $y$ distribution is shown.  The
corresponding distributions for the LHC can be found in
Figs.~\ref{fig:lhcpt} and \ref{fig:lhcy}, elaborated for
$\ms=600$~GeV, $\mg=500$~GeV, and $m_t=175$~GeV.

\medskip%
The normalized $p_t$ distributions are hardly affected by the
transition from LO to NLO. In total, the NLO corrections render the
distributions a little softer. This is caused by the fact that for
high-energetic massive particles the probability of losing energy
through radiation is large. The normalized rapidity spectra in LO and
NLO are identical for all practical purposes.

In conclusion, the properly normalized distributions of the squarks
and gluinos in transverse momentum and rapidity are described quite
well by the lowest-order approximation. 

\begin{figure}[p]
  \begin{center}
  \vspace*{-2.5cm}
  \hspace*{-1cm}
  \epsfig{file=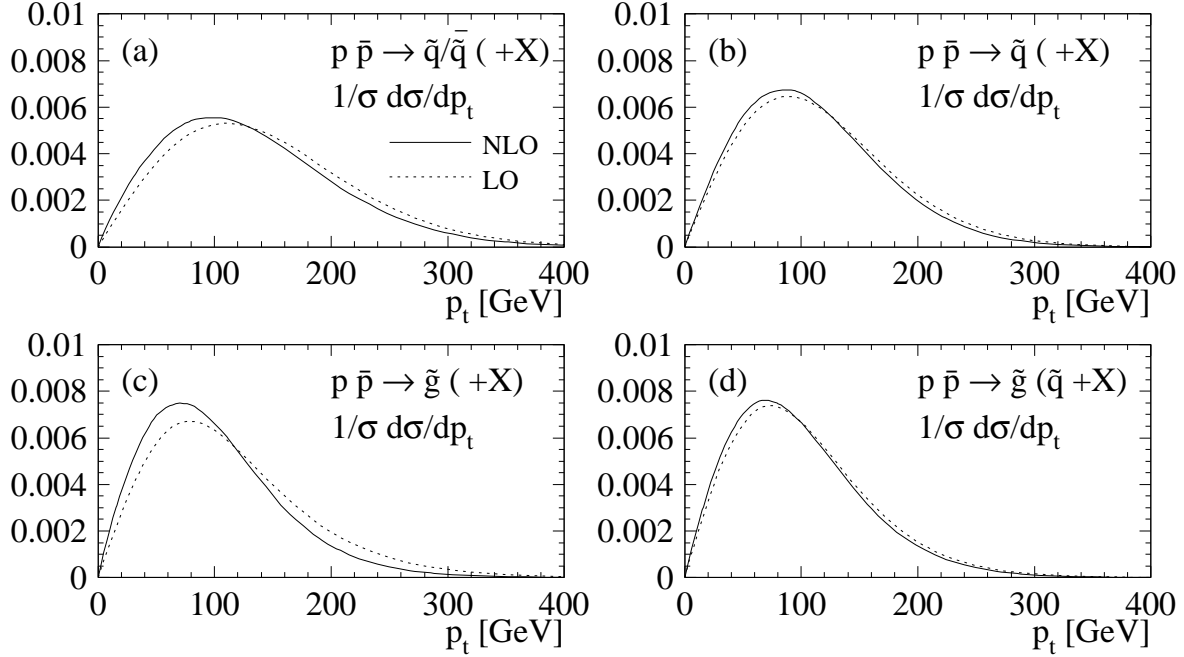,width=18cm}
  \vspace*{-2.3cm}
  \end{center}
  \caption[]{Normalized transverse-momentum distributions in LO
    (dotted) and NLO (solid) at the Tevatron ($\sqrt{S}=1.8$~TeV).
    Parton densities: GRV94, with scale $Q=m$; mass parameters:
    $\ms=280$~GeV, $\mg=200$~GeV, and $m_t=175$~GeV.}
  \label{fig:tevpt}
\end{figure}

\begin{figure}[p]
  \begin{center}
  \vspace*{-2.5cm}
  \hspace*{-1cm}
  \epsfig{file=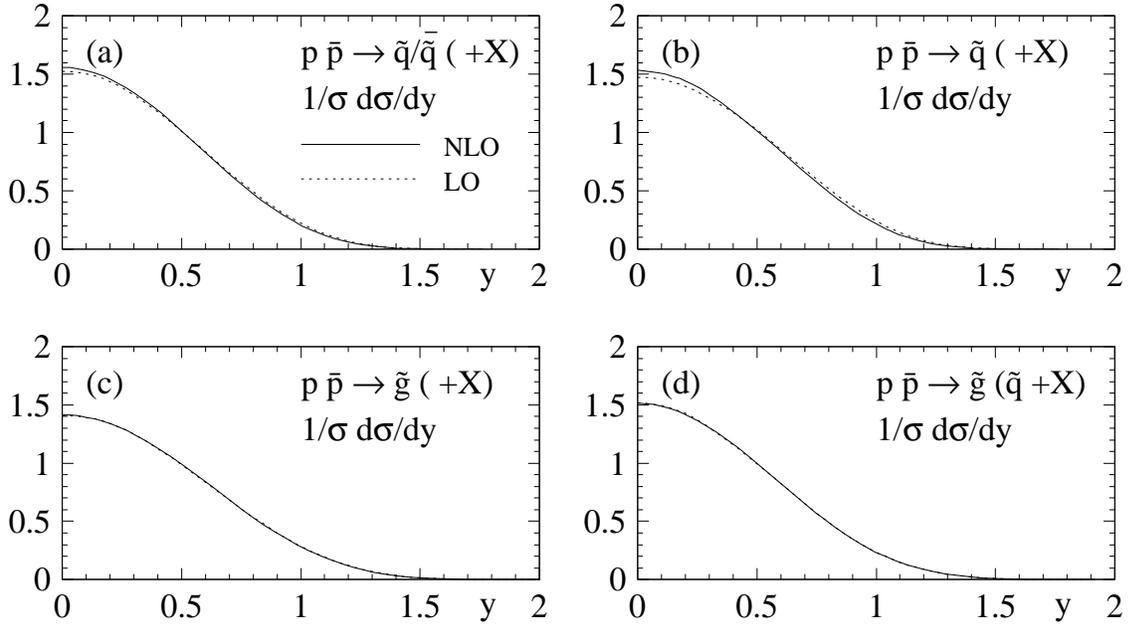,width=18cm}
  \vspace*{-2.3cm}
  \end{center}
  \caption[]{Normalized rapidity distributions in LO (dotted) and 
    NLO (solid) at the Tevatron ($\sqrt{S}=1.8$~TeV). Parton
    densities: GRV94, with scale $Q=m$; mass parameters:
    $\ms=280$~GeV, $\mg=200$~GeV, and $m_t=175$~GeV.}
  \label{fig:tevy}
\end{figure}

\begin{figure}[p]
  \begin{center}
  \vspace*{-2.5cm}
  \hspace*{-1cm}
  \epsfig{file=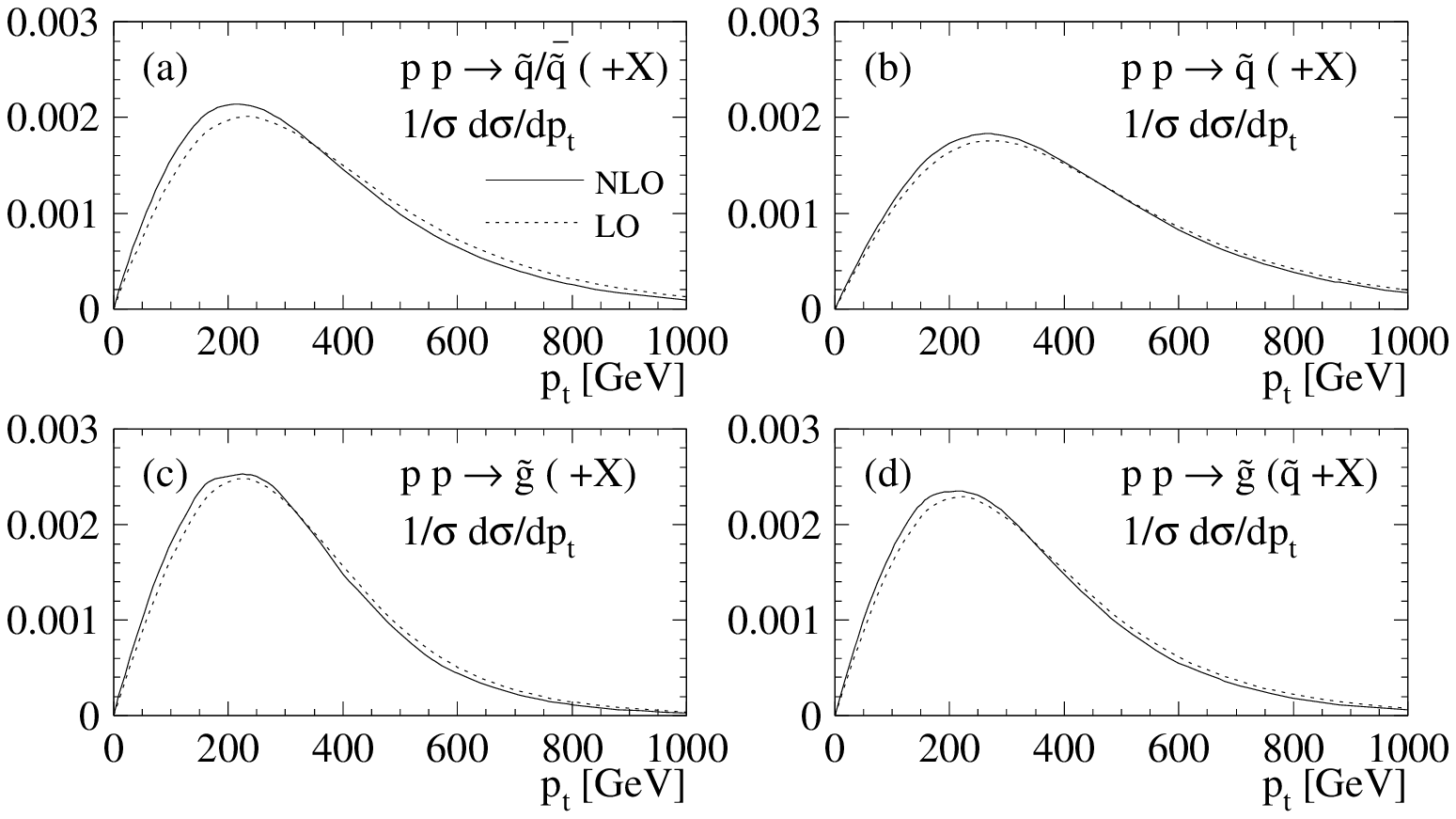,width=18cm}
  \vspace*{-2.3cm}
  \end{center}
  \caption[]{Normalized transverse-momentum distributions in LO
    (dotted) and NLO (solid) at the LHC ($\sqrt{S}=14$~TeV). Parton
    densities: GRV94, with scale $Q=m$; mass parameters:
    $\ms=600$~GeV, $\mg=500$~GeV, and $m_t=175$~GeV.}
  \label{fig:lhcpt}
\end{figure}

\begin{figure}[p]
  \begin{center}
  \vspace*{-2.5cm}
  \hspace*{-1cm}
  \epsfig{file=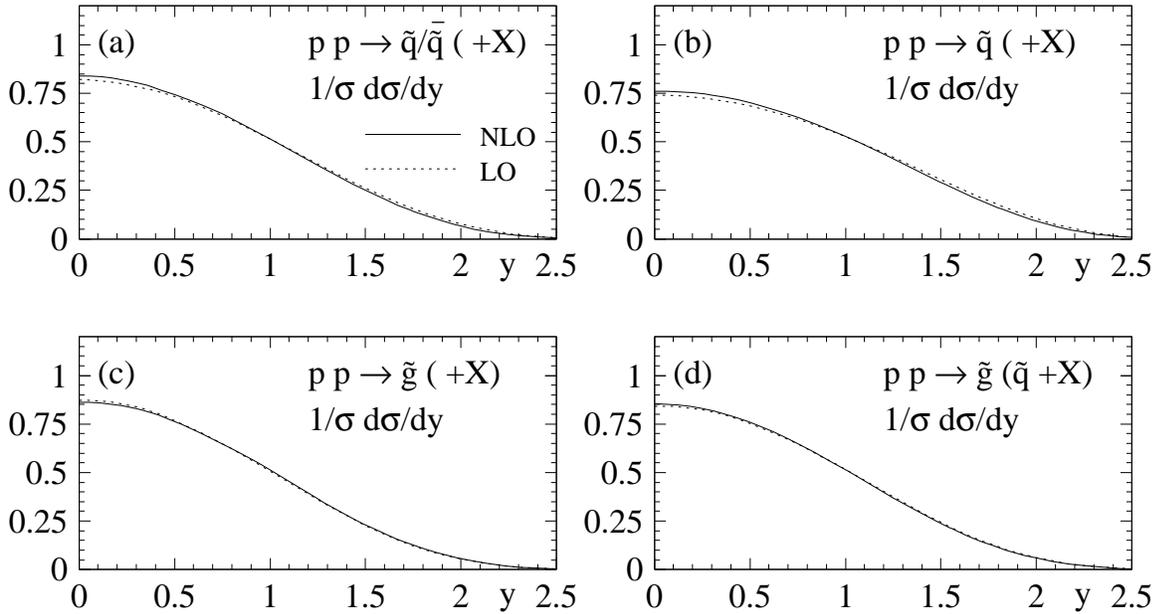,width=18cm}
  \vspace*{-2.3cm}
  \end{center}
  \caption[]{Normalized rapidity distributions in LO (dotted) and
    NLO (solid) at the LHC ($\sqrt{S}=14$~TeV). Parton densities:
    GRV94, with scale $Q=m$; mass parameters: $\ms=600$~GeV,
    $\mg=500$~GeV, and $m_t=175$~GeV.}
  \label{fig:lhcy}
\end{figure}

\pagebreak

\subsubsection{Total Cross-Sections for Squark and Gluino Production}
\paragraph{$K$-factors:} To facilitate the quantitative comparison of
LO and NLO cross-sections we define the ratio
\begin{equation}
  \label{kfacdef} 
  K = \sigma_{NLO}/ \sigma_{LO}, 
\end{equation}
usually referred to as the $K$-factor. For consistency, the
cross-section $\sigma_{LO}$ ($\sigma_{NLO}$) is calculated for all
entries taken at leading (next-to-leading) order, \emph{i.e.}~couplings, 
parton densities, and parton cross-sections.

In Fig.~\ref{fig:tevkfac} we present the $K$-factors at the Tevatron
for the reactions (\ref{ssbprod})--(\ref{sgprod}).  For the scale $Q$
and the parton densities we take the default settings [$Q=m$ and
GRV94].  In Fig.~\ref{fig:tevkfac}a  the $K$-factors are displayed
for a gluino mass $\mg=200$~GeV and for squark masses in the range
150--400~GeV, whereas in Fig.~\ref{fig:tevkfac}b  the role of the
squarks and gluinos is interchanged. The corrections strongly depend 
on the process. With the exception of the squark-pair production
process, which is rather unimportant at the Tevatron, all processes
are subject to large, positive corrections between $+10\%$ and
$+90\%$. The $K$-factors for squark final states are almost
mass-independent.  By contrast, the $K$-factors for the (dominant)
final 
states that involve at least one gluino exhibit a strong mass
dependence. The particularly strong mass dependence of the gluino-pair
$K$-factors for almost mass-degenerate squarks and gluinos is a direct
consequence of the phenomena described in
Section~\ref{sec:partonresults} for the scaling functions, since a large
part of the hadronic cross-section originates from the
quark--antiquark channel. For a fixed gluino mass and increasing
squark masses the LO cross-section decreases, whereas the virtual
corrections increase. This leads to the observed increase of the
$K$-factor. If the squark mass is kept fixed and the gluino mass is
increased, the reverse is observed.

\medskip%
In Fig.~\ref{fig:lhckfac} the $K$-factors are presented for the LHC.
The input is the same as before, except for the fact that we consider
fixed ratios of the squark and gluino masses: $\ms/\mg = 0.8, 1.2,
1.6\,\text{and}\, 2$ [Figs.~\ref{fig:lhckfac} a -- d ].  Again, the
corrections are positive and in general large, between $+5\%$ and
$+90\%$. The mass dependence and the absolute size of the $K$-factors
for squark final states are moderate. By contrast, for final states
involving gluinos the corrections are substantially larger and exhibit
a strong mass dependence. The influence of the squark mass on the
gluino-pair cross-section is less pronounced than for the Tevatron,
since the gluon--gluon initial state yields the dominant
contributions.

\begin{figure}[p]
  \begin{center}
  \vspace*{-2.5cm}
  \hspace*{-0.5cm}
  \epsfig{file=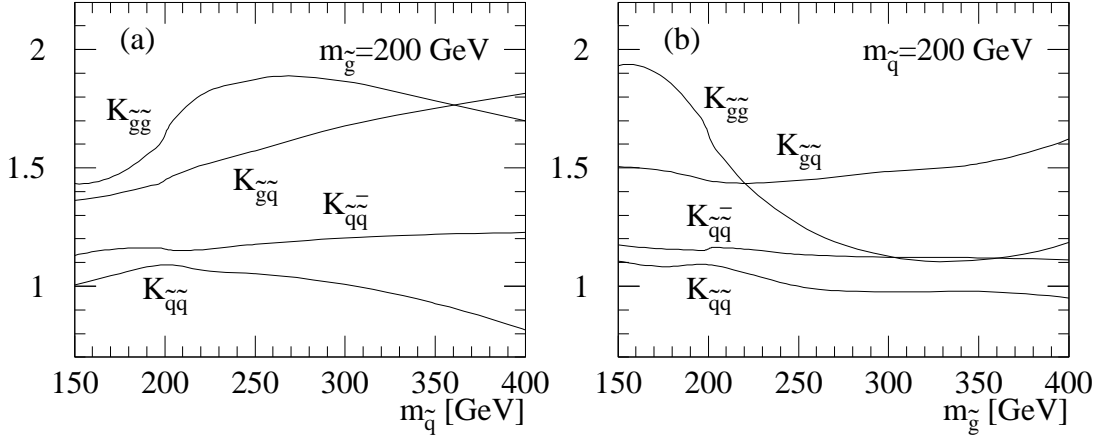,width=17cm}
  \vspace*{-2.cm}
  \end{center}
  \caption[]{The $K$-factors [Eq.~(\ref{kfacdef})] for the Tevatron 
    ($\sqrt{S}=1.8$~TeV). Parton densities: GRV94, with scale $Q=m$;
    mass parameters: $m_t=175$~GeV and (a) $\mg=200$~GeV, (b) $\ms=200$~GeV.}
  \label{fig:tevkfac}
\end{figure}

\begin{figure}[p]
  \begin{center}
  \vspace*{-2.8cm}
  \hspace*{-0.5cm}
  \epsfig{file=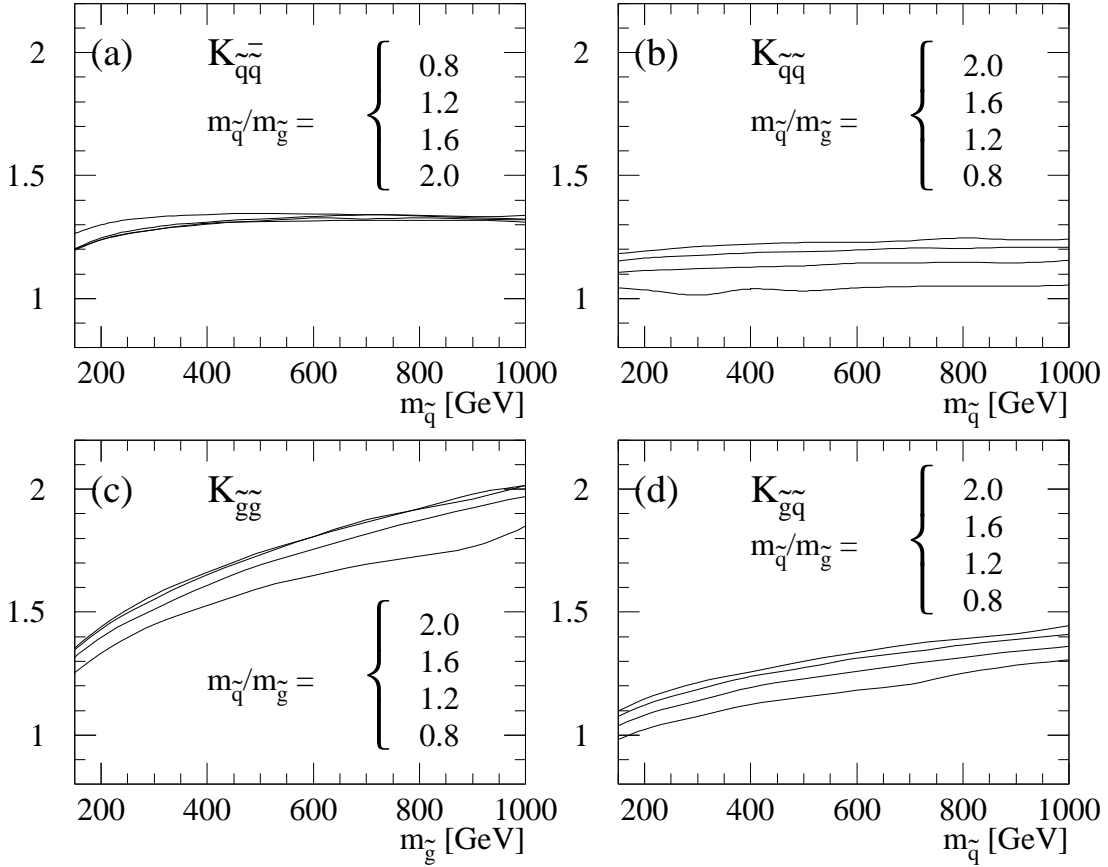,width=17cm}
  \vspace*{-2.0cm}
  \end{center}
  \caption[]{The $K$-factors [Eq.~(\ref{kfacdef})] for the LHC 
    ($\sqrt{S}=14$~TeV). Parton densities: GRV94, with scale $Q=m$;
    top-quark mass: $m_t=175$~GeV.}
  \label{fig:lhckfac}
\end{figure}

\paragraph{Total cross-sections:}
The absolute size of the total hadronic cross-sections plays a crucial
role in the experimental analyses. As long as no squarks and gluinos
are discovered, the exclusion limits of the masses are derived by
comparing the experimental data with the expected rates based on the
theoretical predictions of the cross-sections. If squarks and gluinos
are discovered, the masses of the particles will be determined
experimentally by the same method.

In Fig.~\ref{fig:tevtot} the total cross-sections are given for the
Tevatron. The NLO results are based on the default settings:
renormalization/factorization scale $Q=m$ and GRV94 parton densities.
These cross-sections are compared with the LO parametrizations adopted
in the experimental analyses \cite{cdf,d0} until recently [EHLQ parton
densities, with $Q$ equal to the partonic centre-of-mass
energy\footnote{The scale choice $Q=\sqrt{s}$ is only legitimate in LO;
  in NLO a hadronic scale, \emph{e.g.}~a particle mass, is required
  by the renormalization group.}].  Over the full mass range covered
by the Tevatron, the net effect of the NLO corrections is to raise the
derived lower mass bounds by $+10$~GeV to $+30$~GeV.

\medskip%
In Fig.~\ref{fig:lhctot} the total cross-sections are given for the
LHC, using the default settings and a representative range of squark
and gluino masses.  Over the full mass range covered by the LHC, the
cross-sections are increased by the NLO corrections, leading to a
shift in the associated particle masses in the range between $+10$~GeV
and $+50$~GeV.

\begin{figure}[p]
  \begin{center}
  \vspace*{-2.5cm}
  \hspace*{-0.5cm}
  \epsfig{file=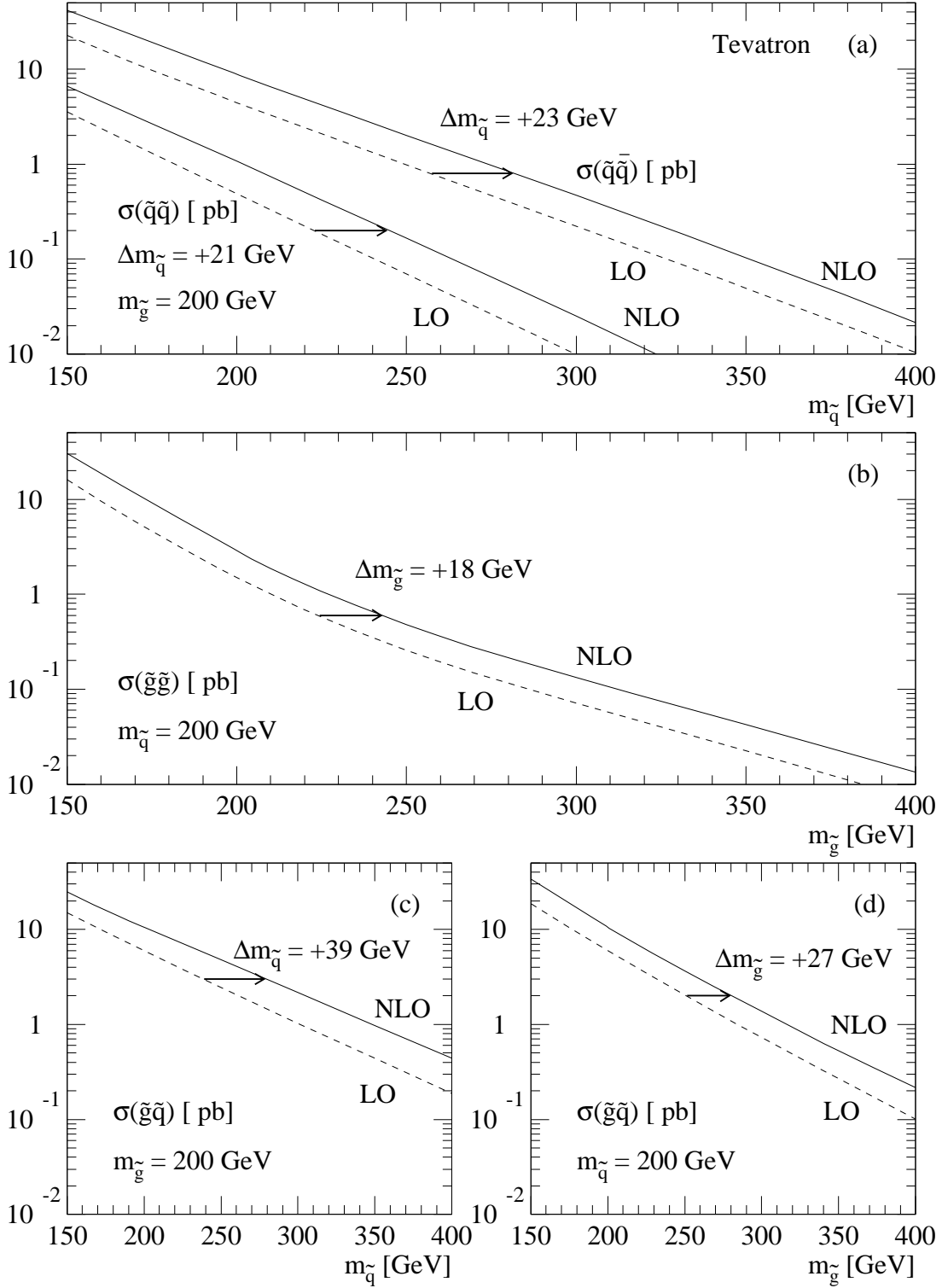,width=17cm}
  \vspace*{-2.cm}
  \end{center}
  \caption[]{The total cross-section for the Tevatron
    ($\sqrt{S}=1.8$~TeV). NLO (solid): GRV94 parton densities, with
    scale $Q=m$; compared with LO (dashed): EHLQ parton densities, at
    the scale $Q=\protect\sqrt{s}$.}
  \label{fig:tevtot}
\end{figure}

\begin{figure}[p]
  \begin{center}
  \vspace*{-2.5cm}
  \hspace*{-0.5cm}
  \epsfig{file=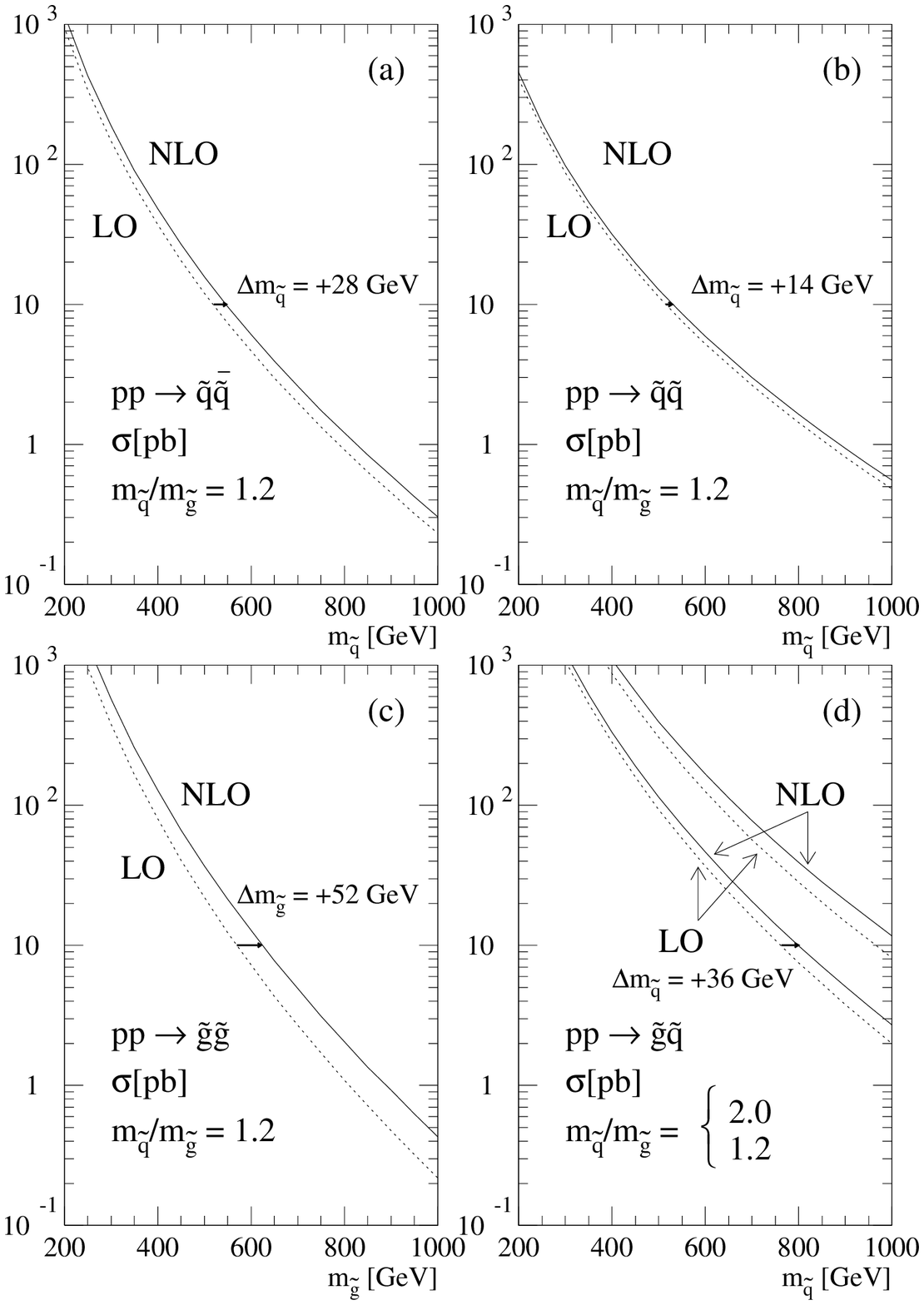,width=17cm}
  \vspace*{-2cm}
  \end{center}
  \caption[]{The total cross-section for the LHC ($\sqrt{S}=14$~TeV).
    NLO (solid) compared with LO (dotted). Parton densities: GRV94, with
    scale $Q=m$.}
  \label{fig:lhctot}
\end{figure}

\subsubsection{Implications for Experimental Searches}

The precise knowledge of the cross-sections at next-to-leading order
SUSY-QCD has a profound impact on the experimental analyses:

\emph{(i)} The renormalization/factorization scale dependence is
reduced by roughly a factor of 2.5--4 compared with leading-order
calculations, and the theoretical predictions of the cross-sections
are stable.  Taking for the renormalization/factorization scale $Q$
the average mass $m$ of the produced particles results in a
conservative estimate for the cross-sections at NLO.

\emph{(ii)} The NLO corrections are large and positive at the central
scale $Q=m$. The NLO corrections must therefore be included in the
analyses to obtain adequate theoretical predictions for the total
cross-sections, as required for deriving experimental mass bounds or
measuring the squark and gluino masses.

\emph{(iii)} The shape of the differential distributions in
transverse momentum and rapidity of one of the outgoing squarks or
gluinos is hardly affected by the NLO corrections. 

\emph{(iv)} The NLO cross-sections raise the present lower mass bounds
for squarks and gluinos derived from Tevatron data by $+10$~GeV to
$+30$~GeV.

\newpage
\section{Conclusions and Outlook}

In this report we have presented the next-to-leading order SUSY-QCD
corrections for the production cross-sections of squarks and gluinos
at the hadron colliders Tevatron and LHC. By reducing the scale
dependence of the cross-sections considerably, the quality of the
theoretical predictions is substantially improved compared with the
lowest-order calculations. The NLO cross-sections provide a solid
basis for experimental analyses of squark and gluino mass
bounds/measurements at hadron colliders.

So far the calculations have been performed for mass-degenerate
squarks associated with the five light quark flavours. Generally,
small mass differences between the $L$ and $R$ squark states for a
given flavour are suggested by supergravity-inspired parametrizations
of low-energy supersymmetry. Due to the large top--Higgs Yukawa
coupling, however, the assumption of mass degeneracy is expected to
be strongly broken for the two stop states $\tilde{t}_1$ and
$\tilde{t}_2$, mixtures of the $L$ and $R$ chirality states
$\tilde{t}_L$ and $\tilde{t}_R$.  In these cases the NLO SUSY-QCD
cross-sections require the extension of the theoretical analysis to
different left- and right-handed couplings of the quarks to squarks
and gluinos, adding to the complexity of the calculation in a
non-trivial way.  For final-state stop particles this analysis is in
progress and will be completed in due time.

\section*{Note}
The Fortran codes of the NLO cross-sections can be obtained from
hoepker\,@@\,x4u2.desy.de, spira\,@@\,cern.ch, 
or wimb\,@@\,lorentz.leidenuniv.nl.

\section*{Acknowledgements}
We have benefited from discussions with K.I.~Hikasa, M.~Kr{\"a}mer,
and W.L.~van Neerven. Special thanks go to our experimental colleagues
S.~Lammel and M.~Paterno for valuable comments on the Tevatron data
and useful suggestions for experimentally convenient parametrizations
of the theoretical results worked out in this report.

\newpage
\appendix

\section{Fermion-Flow Diagrams in SUSY-QCD}
\label{feynman_rules}
The field-theoretic components of supersymmetric QCD are
quarks/squarks and gluons/gluinos. The Majorana character of gluinos
renders the evaluation of Feynman diagrams involving these particles
somewhat cumbersome. However, a simple prescription has recently been
proposed in Ref.~\cite{majo}, which allows an easy and fail-safe
evaluation of the diagrams. It involves the definition of a continuous 
fermion-flow line, which in general does not coincide with the flow of
the fermion number [given by the direction of the Dirac propagator
line]. The amplitude must be evaluated along the fermion-flow
according to the standard rules. In this way also
fermion-number-violating processes like $qq\to\sq\sq$ can be described
in a straightforward fashion.

This method is equivalent to the usual evaluation of SUSY-QCD diagrams
if the analytic form of the propagators and vertices involving
fermions is adjusted properly. The analytic expressions associated
with these propagators and vertices are given explicitly in
Fig.~\ref{fig:majo}. Indices of the fundamental/adjoint representation
of colour $SU(3)$ are denoted by $i,j$/$a$--$c$, the generators of the
fundamental representation by $t^a=\lambda^a/2\,\ (a=1,...,8)$, and
the structure constants by $f^{abc}$. The index $\mu$ is a Lorentz
index, $\alpha$ and $\beta$ are Dirac indices. The coupling constants
$g_s$ and $\ghat$ are the gauge and Yukawa couplings, respectively.
The fermion-flow is defined in the diagrams by an additional directed
line, parallel to the propagators.  All other SUSY-QCD vertices and
propagators are not affected by defining the fermion-flow lines. The
flavour of the quarks and squarks are conserved in the SUSY-QCD
interactions.

\begin{figure}[t]
  \begin{center}
  \vspace*{-0.5cm}
  \hspace*{-1cm}
  \epsfig{file=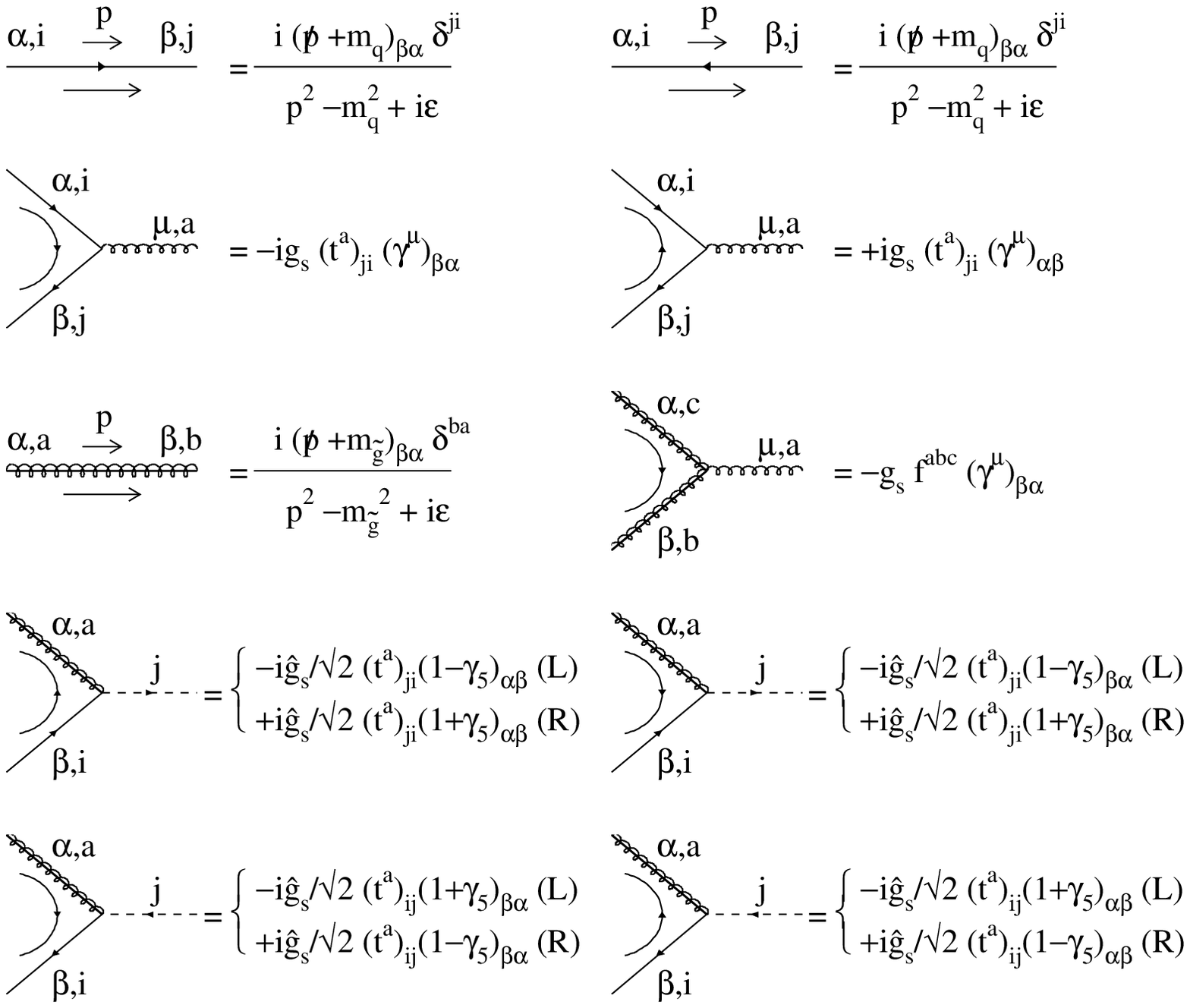,width=17cm}
  \vspace*{-1.2cm}
  \end{center}
  \caption[]{Analytic expressions for the SUSY-QCD propagators and vertices 
    that involve a specific fermion-flow, defined by the line parallel to
    the propagators.}
  \label{fig:majo}
\end{figure}

\section{Kinematics and Phase Space}
\label{phasespace}
The kinematics and phase space for the production processes of a pair
of squarks or a pair of gluinos are the same as for top--antitop
production \cite{ggtt}. However, for the production of squark--gluino
final states, the masses of the heavy particles are different. The
kinematical and phase-space relations must therefore be derived for
this more general case. Identifying squark and gluinos masses, we
recover the formulae for squark and gluino pairs.

\subsection{Partonic Processes}
We shall discuss the partonic process
\begin{equation}
  q(k_1) + g(k_2) \longrightarrow \sq(p_1) + \gl(p_2) 
  \,\,\left[ +g(k_3) \right]
\end{equation}
for illustration. All particles are on shell, \emph{i.e.}~$k_1^2=
k_2^2= 0$, $p_1^2=\ms^2$, $p_2^2=\mg^2$ and $k_3^2=0$. The kinematics of
the radiative process is characterized by the invariant variables
introduced in Eq.~(\ref{hardkin}); they are related by
energy--momentum conservation:
\begin{alignat*}{2}
  s_4 &= s + t_g + u_1 \qquad\qquad\qquad & s_3 &= s + u_6 + u_7 \\
  s_5 &= s + t' + u'   \qquad\qquad\qquad & u_6 &= -s -t_g -t'   \\
  u_7 &= -s -u_g -u'.   
\end{alignat*}
The total partonic cross-section is obtained by integrating the
double-differential cross-section
\begin{eqnarray}
  \label{totpartdef1}
  \sihat &=& \int_{t_g^-}^{t_g^+} dt_g \int_0^{s_{4}^{max}(t_g)} ds_{4}\, 
  \frac{d^2\sihat}{dt_g\,ds_4} 
\end{eqnarray}
within the limits
\begin{eqnarray*}
  t_g^\pm &= &{}-\frac{s+\mg^2-\ms^2}{2} 
  \pm\frac{1}{2}\sqrt{(s-\mg^2-\ms^2)^2 -4\mg^2 \ms^2} \\
  s_{4}^{max}(t_g) &=& s + t_g +\mg^2 -\ms^2 + \frac{\mg^2
    s}{t_g}. 
\end{eqnarray*}
The invariant energy of the $\sq g$ subsystem in the final state is
characterized by the variable $s_4$, the square of the momentum
transfer from the gluon (quark) in the initial state to the gluino in the
final state by $t=t_g+\mg^2$ \,($u=u_g+\mg^2$).

The double-differential cross-section in Eq.~(\ref{totpartdef1}) can
be derived by integrating the general four-fold differential
cross-section over the angles $\theta_i$ defined in the centre-of-mass
frame of the $\sq g$ subsystem. In this particular frame the
$n$-dimensional momenta are given by
\begin{eqnarray}
  k_1 &=& (w_1,0,...,0,0,|p|\sin\psi,|p|\cos\psi-w_2)\\
  k_2 &=& (w_2,0,...,0,0,0,w_2)\nonumber\\
  k_3 &=& (w_3,0,...,0,w_3\sin\theta_1\sin\theta_2,
  w_3\sin\theta_1\cos\theta_2,w_3\cos\theta_1)\nonumber  \\
  p_1 &=& (E_1,0,...,0,-w_3\sin\theta_1\sin\theta_2,
  -w_3\sin\theta_1\cos\theta_2,-w_3\cos\theta_1)\nonumber  \\
  p_2 &=& (E_2,0,...,0,0,|p|\sin\psi,|p|\cos\psi). \nonumber
\end{eqnarray}
The energies $w_i, E_i$, the momentum $|p|$, and the angle $\psi$ are
related to the $\theta_i$-independent invariants $s,\,t_g,\,u_g$, and $s_4$ 
introduced earlier:
\begin{alignat}{3}
  \\[-0.7cm]
  w_1 &= \frac{s+u_g}{2\sqrt{s_4+\ms^2}}& \qquad 
  w_2 &= \frac{s+t_g}{2\sqrt{s_4+\ms^2}}& \qquad 
  w_3 &= \frac{s_4}{2\sqrt{s_4+\ms^2}} \nonumber\\
  E_1 &= \frac{s_4 +2\ms^2}{2\sqrt{s_4+\ms^2}} &\qquad 
  E_2 &= {}-\frac{t_g +u_g +2\mg^2}{2\sqrt{s_4+\ms^2}}& \qquad \\
  |p| &= \frac{\sqrt{(t_g+u_g)^2-4\mg^2s}}{2\sqrt{s_4+\ms^2}} &\qquad 
  \cos\psi &= \frac{t_g s_{4g} -s(u_g
    +2\mg^2)}{(s+t_g)\sqrt{(t_g+u_g)^2-4\mg^2s}}. \nonumber
\end{alignat}
Using these relations, all the invariant variables defined in
Eq.~(\ref{hardkin}) can be expressed in terms of
$s,t_g,u_g,s_4,\theta_i$. For certain combinations of invariants, it is
convenient to define the $z$-axis with respect to $k_1$ or $p_2$
(instead of $k_2$). Writing the $n$-dimensional angular part of the
phase-space element as $d\Omega_n = \sin^{1-2\eps}(\theta_1)\,
d\theta_1\, \sin^{-2\eps}(\theta_2) \,d\theta_2$, the
double-differential cross-section is obtained from the matrix element
${\cal M}^R$ in the following way:
\begin{eqnarray}
  \label{defhardint}
  s^2\,\frac{d^2\sigma^R}{dt_g\,ds_4} &=& K_{ij}\,\frac{ S_\eps^2
    \mu^{2\eps}}{2\Gamma(1-2\eps)}
  \left[\frac{t_g u_g-\mg^2s}{\mu^2 s}\right]^{-\eps} 
  \Theta(t_g u_g-\mg^2s)\,\Theta(s-[\ms+\mg]^2) \nonumber \\[1mm]
                                      & &
  \times\,\frac{(s_4)^{1-2\eps}}{(s_4+\ms^2)^{1-\eps}}\,\Theta(s_4)\int
  d\Omega_n \sum |{\cal M}^R|^2. \qquad
\end{eqnarray}
In the limit $\ms=\mg$ the well-known expressions of Ref.~\cite{ggtt}
are recovered for the equal-mass case. The $\Theta$-functions
represent the requirements of positive energies and $\cos^2\psi \le
1$. They translate into the integration boundaries of the $s_4$ and
$t_g$ integrals in Eq.(\ref{totpartdef1}).

As described in Section~\ref{subtract}, on-shell resonance production
of intermediate particles is possible. For squark--gluino production,
this must be analyzed carefully in the kinematical range where the
three-particle final state $\gl\sq\qb$ approaches the two-particle
final state $\gl\gl$ (if $\mg > \ms$) or $\sq\sqb$ (if $\ms > \mg$).
This gives rise to singularities in $1/s_3$ or $1/s_{4g}$, as is
evident from Fig.~\ref{fig:hardsub2}, if the momentum flow approaches
the $\sq$ or $\gl$ mass shells.

\begin{figure}[t]
  \begin{center}
  \vspace*{-1.5cm}
  \hspace*{-1cm}
  \epsfig{file=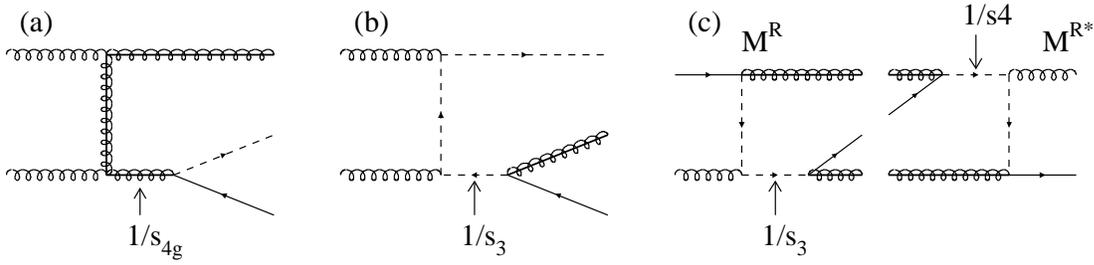,width=16cm}
  \vspace*{-2.0cm}
  \end{center}
  \caption{Examples for on-shell intermediate resonance states: 
    (a) gluino-pair intermediate state for squark--gluino final
    states; (b) squark--antisquark intermediate state for
    squark--gluino final states; and (c) squark--gluino intermediate
    state for gluino-pair final states.}
  \label{fig:hardsub2}
\end{figure}

After exchanging the order of the integrations, the total
cross-section corresponding to the diagram in
Fig.~\ref{fig:hardsub2}a can be written as
\begin{equation}
  \sihat = \int_0^{s_{4}^+} ds_{4}
  \int_{t_g^-(s_{4})}^{t_g^+(s_{4})} dt_g\,
  \frac{d^2\sihat}{dt_g\,ds_{4}}
  \equiv \int_0^{s_{4}^+} ds_{4}\, \frac{f(s_{4g})}{s_{4g}^2 }
\end{equation}
\begin{displaymath}
  s_4^+ = s +\mg^2 -\ms^2 -2\sqrt{s\mg^2} \qquad
  t_g^\pm(s_4) = {}-\frac{s -s_{4g}}{2} \pm
  \frac{1}{2}\sqrt{(s-s_{4g})^2 -4s\mg^2}.
\end{displaymath}
The singularity in $s_{4g}$ for $\mg>\ms$ can be regularized by
inserting the non-zero gluino width and introducing the Breit--Wigner
form, \emph{i.e.}~$1/[s_{4g}^2 +\mg^2\Gg^2]$, for the (absolute) square
of the propagator.

Inserting the identity $f(s_{4g}) = [f(s_{4g}) -f(0)] + f(0)$, the
part of the cross-section related to $f(0)$ corresponds to the (LO)
production of an on-shell intermediate gluino and subsequent (LO)
decay of this gluino into a squark and antiquark, since
\begin{displaymath}
  f(0)=\sihat^B_{\gl\gl}\,\frac{\mg\Gg}{\pi}\,
\frac{\Gamma^B_{\gl\to\sq}}{\Gg}
\end{displaymath}
and $\mg\Gg/[s_{4g}^2 +\mg^2\Gg^2] \to \pi\delta(s_{4g})$ for small
$\Gg$.  This part is already accounted for by the lowest-order
$\gl\gl$ cross-section and has to be subtracted from the
$\gl\sq\bar{q}$ cross-section.  After the formal expansion of
$f(s_{4g})$ around $s_{4g}=0$, the remaining part
\begin{equation}
  \Delta\sihat = \int_0^{s_{4}^+} ds_4\, 
                 \frac{f(s_{4g}) -f(0)}{s_{4g}^2+\mg^2\Gg^2} 
\end{equation}
is finite and well defined as a principal-value integral, since
$s_{4g}/[s_{4g}^2+\mg^2\Gg^2] \to {\cal P}(1/s_{4g})$ for small $\Gg$.

For $\ms>\mg$ also on-shell intermediate squark states will occur 
[see Fig.~\ref{fig:hardsub2}b]. This case is treated in analogy to
the previous example by regularizing the pole in $s_3$. The
integration over $s_3$ is hidden in the angular integrations so that
the technique must be described in some detail. To transform the
angular integration to the integration over $s_3$, we define a
reference frame in which $p_2$ defines the $z$-axis:
\begin{eqnarray}
  && \int_0^\pi d\theta_1 \sin \theta_1 \int_0^\pi d\theta_2\,
  \frac{1}{s_3^2}\, ( A + B\cos\theta_1 +
  C\sin\theta_1\cos\theta_2)^{-l} =\\ &&
  = \int_{s_{3g}^-(s_4)}^{s_{3g}^+(s_4)} ds_{3g} \int_0^\pi d\theta_2\,
  \frac{2(s_4+\ms^2)}{s_4 \sqrt{(s-s_{4g})^2 -4\mg^2 s}}
  \,\frac{1}{s_3^2}\, ( A + B\cos\theta_1 +
  C\sin\theta_1\cos\theta_2)^{-l} \nonumber
\end{eqnarray}
\vspace*{-0.3cm}
\begin{eqnarray*}
  && s_{3g} = \frac{s_4(s-s_4-\ms^2-\mg^2)}{2(s_4+\ms^2)} - \cos\theta_1\,
  \frac{s_4\sqrt{(s-s_{4g})^2 -4\mg^2 s}}{2(s_4+\ms^2)} \nonumber\\ 
  && s_{3g}^\pm(s_4) = \frac{s_4}{2(s_4+\ms^2)}\, \left[ s
  -s_4 -\ms^2 -\mg^2\pm \sqrt{(s-s_{4g})^2 -4\mg^2 s}\;\right]. \nonumber
\end{eqnarray*}
The integrals over $\theta_2$ are given by
\begin{alignat*}{2}
  & \int_0^\pi d\theta_2 \,\left[ X + Y\cos\theta_2 \right]^2  = 
  \pi\left[X^2 +\frac{1}{2} Y^2\right] 
  \qquad & 
  & \int_0^\pi d\theta_2 \,\frac{1}{\left[ X + Y\cos\theta_2 \right]^2} 
  =  -\,\frac{\pi X}{\left[X^2-Y^2\right]^{\frac{3}{2}}}\\
  & \int_0^\pi d\theta_2 \,\left[ X + Y\cos\theta_2 \right] = 
  \pi X 
  \qquad &
  & \int_0^\pi d\theta_2 \,\frac{1}{X + Y\cos\theta_2}  =  
  -\,\frac{\pi}{\sqrt{X^2-Y^2}} \\
  & \int_0^\pi d\theta_2  = \pi 
\end{alignat*}
for $X = A + B\cos\theta_1, \quad Y=C\sin\theta_1, \quad X< 0,
\quad X^2 > Y^2$. Using this formalism, the total partonic
cross-section can finally be rewritten as 
\begin{eqnarray} \label{eq:s3g_int}
  \sihat &=& \int_0^{s_4^{+}} ds_4 \int_{t_g^-(s_4)}^{t_g^+(s_4)} dt_g
  \int_{s_{3g}^-(s_4)}^{s_{3g}^+(s_4)} ds_{3g} \int_0^\pi d\theta_2\, 
  \frac{d^4\sihat}{ds_4\, dt_g \, ds_{3g}\, d\theta_2} \nonumber\\
  &=&
  \int_0^{s_{3g}^{+}} ds_{3g} \int_{s_4^-(s_{3g})}^{s_4^+(s_{3g})} 
  ds_4 \int_{t_g^-(s_4)}^{t_g^+(s_4)} dt_g \int_0^\pi d\theta_2\,
  \frac{d^4\sihat}{ds_{3g}\, ds_4\, dt_g \, d\theta_2} \\[0.3cm]
  s_4^\pm(s_{3g}) &=& \frac{s_{3g}}{2(s_{3g}+\mg^2)}\,
  \left[ s  -s_{3g} -\ms^2 -\mg^2\pm \sqrt{(s-s_3)^2 -4\ms^2 s}\;\right]
  \nonumber \\
  s_{3g}^{+} &=& s +\ms^2 -\mg^2 -2\sqrt{s\ms^2}~. \nonumber
\end{eqnarray}
The singularity in $s_3$ for $\ms > \mg$ can be regularized by
introducing the non-zero squark width in the propagator. The
separation of the lowest-order $\sq\sqb$ contribution and the
integration of the residual (principal-value) integral can be carried
out in the same way as in the previous case.

For identical particles in the final state, gluino or squark pairs,
singularities can occur in both  $s_4$ and $s_3$. The existence of 
interference terms [see Fig.~\ref{fig:hardsub2}c], involving both types of 
singularities, demands some special care. Let us consider the gluino-pair 
production process as an example. In that case the interference terms can be 
treated by substituting 
\begin{eqnarray}
  \frac{1}{s_4 -i\epsilon} &=& {\cal P}\left(\frac{1}{s_4}\right) 
  +i\pi  \delta(s_4)\\
  \frac{1}{s_3 +i\epsilon} &\to& \frac{1}{a +i\epsilon +b\cos\theta_1}
  = {\cal P}\left(\frac{1}{a+b\cos\theta_1}\right)
  -i\pi\delta(a+b\cos\theta_1). 
\end{eqnarray}
The simultaneous singularities $\,(s_4 -i\epsilon)^{-1}(s_3
+i\epsilon)^{-1}\,$ do not require a subtraction procedure, since
these configurations are kinematically suppressed.  However, if $\ms >
\mg$ and $s > (\ms+\mg)^2$ in our example, the product of the two
non-zero imaginary parts gives rise to a non-zero contribution to the
cross-section.

\subsection{Hadronic Differential Cross-Sections}
For the hadroproduction of mixed squark--gluino final states,
\begin{equation}
  h_1(K_1) + h_2(K_2) \longrightarrow \sq(p_1) +\gl(p_2)
  \,\,[+g(k_3)/q(k_3)/\bar{q}(k_3)],
\end{equation}
we introduce the following hadronic variables:
\begin{alignat}{2}
  \\[-0.7cm]
  S   &= (K_1 + K_2)^2 &\\
  T_1 &= (K_2 -p_2)^2 -\ms^2 &\qquad  T_{g} &= (K_2 -p_2)^2 -\mg^2 
  \nonumber\\
  U_1 &= (K_1 -p_2)^2 -\ms^2 &\qquad  U_{g} &= (K_1 -p_2)^2 -\mg^2.
  \nonumber
\end{alignat}
The double-differential hadronic cross-section is given by the
integration of the parton cross-sections over the parton densities:
\begin{multline}
  \frac{d^2\sigma}{dT_g dU_g}(S,T_g,U_g,Q^2)
  = \\
  = \sum_{i,j=g,q,\bar{q}}
  \int_{x_1^-}^1 dx_1 \int_{x_2^-}^1 dx_2 \, x_1 f_{i}^{h_1} (x_1,Q^2)\,
  x_2 f_{j}^{h_2} (x_2,Q^2)
  \,\frac{d^2\hat{\sigma}_{ij}(s,t_g,u_g,Q^2)}{dt_g\, du_g} 
\end{multline}
\begin{displaymath}
   x_1^- = \frac{-T_g-\mg^2 +\ms^2}{S+U_g}
  \qquad x_2^- = \frac{-x_1 U_g -\mg^2 +\ms^2}{x_1 S + T_g}.
\end{displaymath}
Since the parton $i$ has been assigned to the hadron $h_1$, we can
express the invariant parton variables in terms of the hadronic
variables:
\begin{displaymath}
  s = x_1 x_2 S \qquad t_g = x_2 T_g \qquad u_g = x_1 U_g.
\end{displaymath}
The integration over $x_2$ can be transformed into the
integration over $s_4$,
\begin{equation}
  \int_{x_1^-}^1 dx_1 \int_{x_2^-}^1 dx_2 = 
  \int_{x_1^-}^1 dx_1 \int_0^{s_4^{\ast}}
  \,\frac{ds_4}{x_1 S +T_g} 
\end{equation}
\begin{displaymath}
   s_4^{\ast} = x_1(S+U_g)+T_g+\mg^2-\ms^2
  \qquad x_2 = \frac{s_4 -x_1 U_g -\mg^2 +\ms^2}{x_1 S +T_g}.
\end{displaymath}
The integral over $s_4$ splits into contributions from soft and hard
gluons.  Both the soft and hard contributions contain terms that
depend logarithmically on the cut-off parameter $\Delta$. The soft
terms of that type are mapped into the hard contributions, so that the
scaling functions are independent of $\Delta$ and well defined in the
limit $\Delta\to 0$.

The differential cross-section in transverse momentum and rapidity of
the produced gluino can be obtained from the double-differential
cross-section by the following transformation:
\begin{equation}
    \frac{d^2\sigma}{dp_t \, dy} = 2 p_t S \,
    \frac{d^2\sigma}{dT_g \, dU_g} 
\end{equation}
\begin{equation}
  p_t^2 = \frac{T_g\,U_g}{S} -\mg^2=\frac{t_g\,u_g}{s}
  -\mg^2 \qquad 
  y = \frac{1}{2}\Lg{T_g}{U_g}.
\end{equation}
[In general, transverse momentum and rapidity are defined for the
particle carrying the momentum $p_2$.] The transverse mass of the
gluino is given by the quantity $\sqrt{p_t^2 + \mg^2}$~. We define the
rapidity spectrum by the sum of the contributions from positive and
negative values\footnote{Since squarks and antisquarks are not
  discriminated in the experimental analyses, we have not
  distinguished them in the final states in the numerical discussion.
  For $\sq\sqb$, $\sq\sq$, and $\gl\gl$ final states we have therefore
  added the distributions with respect to both final-state particles.
  These $y$ distributions are symmetric under $y\to -y$. Note
  that an extra factor $1/2$ has to be inserted when using these
  (summed) distributions to obtain the total cross-section.}. For
identical particles, the spectrum is defined for positive rapidity
only, accounting in this way for the statistical factor $1/2$ in those
cases.  Integrating over transverse momentum and rapidity, the total
cross-section is reproduced:
\begin{equation}
    \label{deftotal2}
  \sigma(S,Q^2) = \int_0^{p_t^{max}(0)} dp_t 
  \int_{0}^{y^{max}(p_t)} dy\, \frac{d^2\sigma}{dp_t dy}
  = \int_{0}^{y^{max}(0)} dy \int_0^{p_t^{max}(y)} dp_t\,
  \frac{d^2\sigma}{dp_t dy}
\end{equation}
\begin{eqnarray*}
   p_t^{max}(y) & = &
  \frac{1}{2\sqrt{S}\,\text{cosh}y}\sqrt{\left(S+\mg^2-\ms^2\right)^2 
  -4\mg^2 S \,\text{cosh}^2y} \\[1mm]
  y^{max}(p_t) & = & 
  \text{arccosh}\left(\frac{S+\mg^2-\ms^2}{2\sqrt{S(p_t^2+\mg^2)}}\right).
\end{eqnarray*}

\medskip %
In order to reproduce the total hadronic cross-section
Eq.(\ref{eq:sigscale}), special care must be exercised in calculating
the hadronic differential distributions for the resonance
contributions:

\paragraph{\boldmath $1/s_{4g}$ singularity:}
For the subtraction it is convenient to write the double-differential
hadronic cross-section in the form
\begin{eqnarray}
  \frac{d^2\sigma}{dT_g dU_g} & = & 
  \int_{\tau}^1 dx_1 \int_{\tau/x_1}^1 dx_2\,f_i^{h_1}(x_1,Q^2)\, 
  f_j^{h_2}(x_2,Q^2) \nonumber \\
                              & &
  \times\,\int_{0}^{s_4^+} ds_4 \int_{t_g^-(s_4)}^{t_g^+(s_4)} dt_g \,
  \frac{d^2 \hat\sigma_{ij}}{ds_4 dt_g}\,
  \delta\left(T_g - \frac{t_g}{x_2}\right)\,
  \delta\left(U_g - \frac{u_g}{x_1} \right) \nonumber\\
                              & \equiv &
  \int_{\tau}^1 dx_1 \int_{\tau/x_1}^1 dx_2\,f_i^{h_1}(x_1,Q^2)\, 
  f_j^{h_2}(x_2,Q^2)\,\int_{0}^{s_4^+} ds_4 \,\frac{g(s_{4g})}{s_{4g}^2}.
\end{eqnarray}
The integration boundaries are the ones defined in the previous subsection
on the partonic processes (and $\tau = (\ms+\mg)^2/S$).
The fact that the arguments of the $\delta$-functions should have zeros inside
the allowed integration region results in additional conditions on $x_i,\,T_g$,
and $U_g$. As we have opted for a subtraction of on-shell intermediate states 
at the parton level, the subtracted double-differential hadronic distribution
is given by  
\begin{eqnarray}
  \frac{d^2\Delta\sigma}{dT_g dU_g} & = &  
  \int_{\tau}^1 dx_1 \int_{\tau/x_1}^1 dx_2\,
  f_i^{h_1}(x_1,Q^2)\,f_j^{h_2}(x_2,Q^2)
  \int_0^{s_{4}^+} ds_{4} \,\frac{g(s_{4g})-g(0)}{s_{4g}^2 +\mg^2\Gg^2}
\end{eqnarray}
after regularization by means of the non-zero gluino width.
From this it is clear that upon integration over the hadronic variables $T_g$ 
and $U_g$ the subtraction for the total hadronic cross-section is reproduced.

\paragraph{\boldmath $1/s_3$ singularity:}
In this case we start off by applying Eq.(\ref{eq:s3g_int}) and write
\begin{eqnarray}
  \frac{d^2\sigma}{dT_g dU_g} & = & 
  \int_{\tau}^1 dx_1 \int_{\tau/x_1}^1 dx_2\,f_i^{h_1}(x_1,Q^2)\, 
  f_j^{h_2}(x_2,Q^2) \\
                              & &
  \times\,\int_0^{s_{3g}^+} ds_{3g} \int_{s_4^-(s_{3g})}^{s_4^+(s_{3g})} 
  ds_4 \int_{t_g^-(s_4)}^{t_g^+(s_4)} dt_g \,
  \frac{d^3 \hat\sigma_{ij}}{ds_{3g} ds_4 dt_g}\,
  \delta\left(T_g - \frac{t_g}{x_2}\right)\,
  \delta\left(U_g - \frac{u_g}{x_1} \right). \nonumber
\end{eqnarray}
Now the regularization by means of the non-zero squark width and
subsequent subtraction of the resonance contribution can be performed
in the usual way.

\paragraph{\boldmath Mixed $1/s_3$ and $1/s_4$ singularities for identical
  final-state particles:} After exchanging the order of the
integrations, the singularities in the propagators can easily be
isolated. They can be treated as discussed in the subsection on
partonic processes, giving rise to contributions from the two non-zero
imaginary parts.

\newpage
\section{Splitting Functions}
\label{AP_kernels}
In this appendix we list all next-to-leading order SUSY-QCD splitting
functions. The splitting functions involving massless partons [quarks
and gluons] are known from the standard QCD evolution equations
\cite{split1,split2,altpar}. The splitting functions involving massive
coloured SUSY particles are realized in final-state distributions at
very high energies \cite{split3}. We present all functions for the
momentum fraction $x$ in the restricted range $0<x<1$, excluding in
this way the end-point singularities in $P_{aa}$. The various
end-point singularities can be derived from quark conservation, 
$\int_0^1 dx\,[ P_{qq}(x) + P_{\sqL q}(x) + P_{\sqR q}(x) ] = 0$ 
and $\int_0^1 dx\,[ P_{q\sq}(x) + P_{\sq\sq}(x) ] = 0$, 
and from momentum conservation, $\int_0^1 dx\,x \sum_a P_{ab}(x)=0$. 
[As usual, $C_F=4/3$, $N=3$, and $T_f=1/2$.]

\begin{alignat*}{2}
\intertext{\bf Quarks and gluons:}
  q &\to q\,(+g)   \qquad & P_{qq}(x) &= C_F\,\frac{1 + x^2}{1-x} \\
  q &\to g\,(+q)   \qquad & P_{gq}(x) &= C_F\,\frac{1 + (1-x)^2}{x} \\
  g &\to q\,(+\qb) \qquad & P_{qg}(x) &= T_f\left[x^2 + (1-x)^2\right] \\
  g &\to g\,(+g)   \qquad & P_{gg}(x) &= 2\cdot N \left[\frac{1}{x(1-x)}
                                         +x(1-x) -2\right] \\
\intertext{\bf Squarks and gluons:}
  \sq &\to \sq\,(+g)    \qquad & P_{\sqL\sqL}(x) &= P_{\sqR\sqR}(x) = 
                                                    2\,C_F\,\frac{x}{1-x} \\
  \sq &\to g\,(+\sq)    \qquad & P_{g\sqL}(x)    &= P_{g\sqR}(x) = 
                                                    2\,C_F\,\frac{1-x}{x} \\
  g   &\to \sq\,(+\sqb) \qquad & P_{\sqL g}(x)   &= P_{\sqR g}(x) =
                                                    T_f\,x(1-x) \\
\intertext{\bf Gluinos and gluons:}
  \gl &\to \gl(\,+g)   \qquad & P_{\gl\gl}(x) &= N\,\frac{1 + x^2}{1-x} \\
  \gl &\to g\,(+\gl)   \qquad & P_{g\gl}(x)   &= N\,\frac{1 + (1-x)^2}{x} \\
  g   &\to \gl\,(+\gl) \qquad & P_{\gl g}(x)  &= 2\cdot\frac{1}{2}\,N
                                                 \left[x^2 + (1-x)^2\right] \\
\intertext{\bf Squarks, gluinos and quarks:}
  \sq & \to  q\,(+\gl)        \qquad & P_{q\sqL}(x)   &= P_{q\sqR}(x) = C_F \\
  \sq & \to \gl\,(+ q)        \qquad & P_{\gl\sqL}(x) &= P_{\gl\sqR}(x) = C_F 
                                                         \\
  \gl & \to \sq\,(+\qb)       \qquad & P_{\sqL\gl}(x) &= P_{\sqR\gl}(x) = 
                                                         \frac{1}{2}\,T_f\,x \\
  \gl & \to \qb\,(+\sqL/\sqR) \qquad & P_{\qb\gl}(x)  &= 2\cdot\frac{1}{2}\,
                                                         T_f\,(1-x) \\
  q   & \to \sq\,(+\gl)       \qquad & P_{\sqL q}(x)  &= P_{\sqR q}(x) =
                                                         \frac{1}{2}\,C_F\,x \\
  q   & \to \gl\,(+\sqL/\sqR) \qquad & P_{\gl q}(x)   &= 2\cdot\frac{1}{2}\,
                                                         C_F \,(1-x)
\end{alignat*}
The splitting functions for the charge-conjugate transitions are identical.

Using the notation of Section~\ref{massfac}, the end-point singularities in
$P_{aa}$ are given by $C_F\,[2\log\delta + 1]\,\delta(1-x)$ for $P_{qq}$ and 
$P_{\sq\sq}$, and $[2N\log\delta + \beta_0/2]\,\delta(1-x)$ for $P_{gg}$ and 
$P_{\gl\gl}$.
Note that in Section~\ref{massfac} the end-point singularities for $P_{gg}$ 
and $P_{qq}$ have been derived for the case in which massive particles are 
decoupled.

\frenchspacing
 \newcommand{\zp}[3]{{\sl Z. Phys.} {\bf #1} (19#2) #3}
 \newcommand{\np}[3]{{\sl Nucl. Phys.} {\bf #1} (19#2)~#3}
 \newcommand{\pl}[3]{{\sl Phys. Lett.} {\bf #1} (19#2) #3}
 \newcommand{\pr}[3]{{\sl Phys. Rev.} {\bf #1} (19#2) #3}
 \newcommand{\prl}[3]{{\sl Phys. Rev. Lett.} {\bf #1} (19#2) #3}
 \newcommand{\prep}[3]{{\sl Phys. Rep.} {\bf #1} (19#2) #3}
 \newcommand{\fp}[3]{{\sl Fortschr. Phys.} {\bf #1} (19#2) #3}
 \newcommand{\nc}[3]{{\sl Nuovo Cimento} {\bf #1} (19#2) #3}
 \newcommand{\ijmp}[3]{{\sl Int. J. Mod. Phys.} {\bf #1} (19#2) #3}
 \newcommand{\ptp}[3]{{\sl Prog. Theo. Phys.} {\bf #1} (19#2) #3}
 \newcommand{\sjnp}[3]{{\sl Sov. J. Nucl. Phys.} {\bf #1} (19#2) #3}
 \newcommand{\cpc}[3]{{\sl Comp. Phys. Commun.} {\bf #1} (19#2) #3}
 \newcommand{\mpl}[3]{{\sl Mod. Phys. Lett.} {\bf #1} (19#2) #3}
 \newcommand{\cmp}[3]{{\sl Commun. Math. Phys.} {\bf #1} (19#2) #3}
 \newcommand{\jmp}[3]{{\sl J. Math. Phys.} {\bf #1} (19#2) #3}
 \newcommand{\nim}[3]{{\sl Nucl. Instr. Meth.} {\bf #1} (19#2) #3}
 \newcommand{\el}[3]{{\sl Europhysics Letters} {\bf #1} (19#2) #3}
 \newcommand{\ap}[3]{{\sl Ann. of Phys.} {\bf #1} (19#2) #3}
 \newcommand{\jetp}[3]{{\sl JETP} {\bf #1} (19#2) #3}
 \newcommand{\jetpl}[3]{{\sl JETP Lett.} {\bf #1} (19#2) #3}
 \newcommand{\acpp}[3]{{\sl Acta Physica Polonica} {\bf #1} (19#2) #3}
 \newcommand{\vj}[4]{{\sl #1~}{\bf #2} (19#3) #4}
 \newcommand{\ej}[3]{{\bf #1} (19#2) #3}
 \newcommand{\vjs}[2]{{\sl #1~}{\bf #2}}
 \newcommand{\hep}[1]{{hep--ph/}{#1}}
 \newcommand{\desy}[1]{{DESY-Report~}{#1}}

\end{document}